\def\unredoffs{\hoffset-.14truein\voffset-.2truein}
\def\redoffs{\voffset=-.45truein\hoffset=-.21truein}
\def\speclscape{}
\newbox\leftpage \newdimen\fullhsize \newdimen\hstitle \newdimen\hsbody
\tolerance=1000\hfuzz=2pt
\catcode`\@=11 
\def\bigans{b }
\def\answ{b }
\ifx\answ\bigans\message{(This will come out unreduced.}
\magnification=1200\unredoffs\advance\vsize by .5truein
\advance\voffset by .25truein\baselineskip=16pt plus 2pt minus 1pt
\hsbody=\hsize \hstitle=\hsize 
\else\message{(This will be reduced.} \let\l@r=L
\magnification=1000\baselineskip=16pt plus 2pt minus 1pt \vsize=6.75truein
\redoffs\advance\voffset by .25truein 
\hstitle=8truein\hsbody=4.75truein\fullhsize=10truein\hsize=\hsbody
\output={\ifnum\pageno=0 
  \shipout\vbox{\speclscape{\hsize\fullhsize\makeheadline}
    \hbox to \fullhsize{\hfill\pagebody\hfill}}\advancepageno
  \else
  \almostshipout{\leftline{\vbox{\pagebody\makefootline}}}\advancepageno
  \fi}
\def\almostshipout#1{\if L\l@r \count1=1 \message{[\the\count0.\the\count1]}
      \global\setbox\leftpage=#1 \global\let\l@r=R
 \else \count1=2
  \shipout\vbox{\speclscape{\hsize\fullhsize\makeheadline}
      \hbox to\fullhsize{\box\leftpage\hfil#1}}  \global\let\l@r=L\fi}
\fi
\hfuzz=2pt
\vfuzz=4pt
\pretolerance=5000
\tolerance=5000
\parskip=0pt plus 1pt
\parindent=16pt
\font\fourteenrm=cmr10 scaled \magstep2
\font\fourteeni=cmmi10 scaled \magstep2
\font\fourteenbf=cmbx10 scaled \magstep2
\font\fourteenit=cmti10 scaled \magstep2
\font\fourteensy=cmsy10 scaled \magstep2

\font\sans=cmssbx10
\def\bss#1{\hbox{\sans #1}}
\font\bdi=cmmib10
\def\bi#1{\hbox{\bdi #1\/}}
\font\eightrm=cmr8
\font\eighti=cmmi8
\font\eightbf=cmbx8
\font\eightit=cmti8
\font\eightsl=cmsl8
\font\eightsy=cmsy8
\font\srm=cmr7
\font\sbf=cmbx7
\font\sixrm=cmr6
\font\sixi=cmmi6
\font\sixsy=cmsy6

\def\tenpoint{\def\rm{\fam0\tenrm}%
  \textfont0=\tenrm \scriptfont0=\sevenrm
                      \scriptscriptfont0=\fiverm
  \textfont1=\teni  \scriptfont1=\seveni
                      \scriptscriptfont1=\fivei
  \textfont2=\tensy \scriptfont2=\sevensy
                      \scriptscriptfont2=\fivesy
  \textfont3=\tenex   \scriptfont3=\tenex
                      \scriptscriptfont3=\tenex
  \textfont\itfam=\tenit  \def\it{\fam\itfam\tenit}%
  \textfont\slfam=\tensl  \def\sl{\fam\slfam\tensl}%
  \textfont\bffam=\tenbf  \scriptfont\bffam=\sevenbf
                            \scriptscriptfont\bffam=\fivebf
                            \def\bf{\fam\bffam\tenbf}%
  \normalbaselineskip=20 truept
  \setbox\strutbox=\hbox{\vrule height14pt depth6pt
width0pt}%
  \let\sc=\eightrm \normalbaselines\rm}
\def\eightpoint{\def\rm{\fam0\eightrm}%
  \textfont0=\eightrm \scriptfont0=\sixrm
                      \scriptscriptfont0=\fiverm
  \textfont1=\eighti  \scriptfont1=\sixi
                      \scriptscriptfont1=\fivei
  \textfont2=\eightsy \scriptfont2=\sixsy
                      \scriptscriptfont2=\fivesy
  \textfont3=\tenex   \scriptfont3=\tenex
                      \scriptscriptfont3=\tenex
  \textfont\itfam=\eightit  \def\it{\fam\itfam\eightit}%
  \textfont\bffam=\eightbf  \def\bf{\fam\bffam\eightbf}%
  \normalbaselineskip=16 truept
  \setbox\strutbox=\hbox{\vrule height11pt depth5pt width0pt}}
\def\fourteenpoint{\def\rm{\fam0\fourteenrm}%
  \textfont0=\fourteenrm \scriptfont0=\tenrm
                      \scriptscriptfont0=\eightrm
  \textfont1=\fourteeni  \scriptfont1=\teni
                      \scriptscriptfont1=\eighti
  \textfont2=\fourteensy \scriptfont2=\tensy
                      \scriptscriptfont2=\eightsy
  \textfont3=\tenex   \scriptfont3=\tenex
                      \scriptscriptfont3=\tenex
  \textfont\itfam=\fourteenit  \def\it{\fam\itfam\fourteenit}%
  \textfont\bffam=\fourteenbf  \scriptfont\bffam=\tenbf
                             \scriptscriptfont\bffam=\eightbf
                             \def\bf{\fam\bffam\fourteenbf}%
  \normalbaselineskip=24 truept
  \setbox\strutbox=\hbox{\vrule height17pt depth7pt width0pt}%
  \let\sc=\tenrm \normalbaselines\rm}
\def\today{\number\day\ \ifcase\month\or
  January\or February\or March\or April\or May\or June\or
  July\or August\or September\or October\or November\or
December\fi
  \space \number\year}
\newcount\yearltd\yearltd=\year\advance\yearltd by -1900
\newcount\secno      
\newcount\subno      
\newcount\subsubno   
\newcount\appno      
\newcount\tableno    
\newcount\figureno   
\def\title#1
   {\nopagenumbers\vglue.in\hsize=\hstitle
   {\baselineskip=24pt
    \pretolerance=10000    
    \fourteenpoint\centerline{\bf #1}}
    \pageno=0}
\def\author#1
  {\medskip
    \centerline{\bf #1}\bigskip}
\def\address#1
   {\bigskip
    \centerline{\sl #1}}
\def\shorttitle#1
   {\vfill
    \noindent \rm Short title: {\sl #1}\par
    \medskip}
\def\pacs#1
   {\noindent \rm PACS number(s): #1\par
    \medskip}
\def\jnl#1
   {\noindent \rm Submitted to: {\sl #1}\par
    \medskip}
\def\date
   {\noindent \today\par
   \supereject\global\hsize=\hsbody
   \footline={\hss\tenrm\folio\hss}}
\def\beginabstract
   {\bigskip\noindent  {\bf Abstract:} \rm}
\def\endabstract
   {\par\bigskip}
\def\contents
   {{\noindent
    \bf Contents
    \par}
    \rightline{Page}}
\def\entry#1#2#3
   {\noindent
    \hangindent=20pt
    \hangafter=1
    \hbox to20pt{#1 \hss}#2\hfill #3\par}
\def\subentry#1#2#3
   {\noindent
    \hangindent=40pt
    \hangafter=1
    \hskip20pt\hbox to20pt{#1 \hss}#2\hfill #3\par}

\def\ack
   {\vskip-\lastskip
    \vskip36pt plus12pt minus12pt
    \bigbreak
    \noindent{\bf Acknowledgments\par}
    \nobreak
    \bigskip
    \noindent}

\def\Section#1
   {\vskip0pt plus.1\vsize\penalty-250
    \vskip0pt plus-.1\vsize\vskip24pt plus12pt minus6pt
    \subno=0 \subsubno=0
    \global\advance\secno by 1
    \centerline{\bf \the\secno. #1}
    \indent}
\def\Subsection#1
   {\vskip-\lastskip
    \vskip24pt plus12pt minus6pt
    \bigbreak
    \global\advance\subno by 1
    \subsubno=0
    \centerline{\sl \the\secno.\the\subno. #1}
    \nobreak
    \indent}
\def\appendice#1#2
   {\vskip0pt plus.1\vsize\penalty-250
    \vskip0pt plus-.1\vsize\vskip24pt plus12pt minus6pt
    \subno=0
    \centerline{\bf Appendix #1. #2}
    \indent}
\def\subappendice#1#2
   {\vskip-\lastskip
    \vskip36pt plus12pt minus12pt
    \bigbreak
    \global\advance\subno by 1
    \centerline{\sl #1\the\subno. #2}
    \nobreak
    \indent}
\def\Figcaption#1#2
   {\noindent {\sbf Figure #1.} \rm#2\par
    \medskip}
\def\ackapref#1#2#3
   {\noindent
    \hangindent=20pt
    \hangafter=1
    #1\  #2\hfill #3\par}
\def\tabcaption#1#2
   {\bigskip\smallskip\noindent {\sbf Table #1.} \rm#2\smallskip}

\def\references
     {\vfill\eject
     {\noindent \bf References\par}
      \parindent=0pt
      \bigskip}
\def\refjl#1#2#3#4
   {\hangindent=16pt
    \hangafter=1
    \rm #1
   {\frenchspacing\sl #2
    \bf #3}
    #4\par}
\def\refbk#1#2#3
   {\hangindent=16pt
    \hangafter=1
    \rm #1
   {\frenchspacing\sl #2}
    #3\par}
\def\numrefjl#1#2#3#4#5
   {\parindent=40pt
    \hang
    \noindent
    \rm {\hbox to 30pt{\hss #1\quad}}#2
   {\frenchspacing\sl #3\/
    \bf #4}
    #5\par\parindent=16pt}
\def\numrefbk#1#2#3#4
   {\parindent=40pt
    \hang
    \noindent
    \rm {\hbox to 30pt{\hss #1\quad}}#2
   {\frenchspacing\sl #3\/}
    #4\par\parindent=16pt}
\def\dash{---{}---}
\def\frac#1#2{{#1 \over #2}}

\def\d{{\rm d}}
\def\e{{\rm e}}
\def\i{\ifmmode{\rm i}\else\char"10\fi}

\def\boldrule#1{\vbox{\hrule height1pt width#1}}
\def\medrule#1{\vbox{\hrule width#1}}

\catcode`\@=11
\def\ind{\hbox to 5pc{}}
\def\eq(#1){\hfill\llap{(#1)}}

\def\deqn#1{\displ@y\halign{\hbox to \displaywidth
    {$\@lign\displaystyle##\hfil$}\crcr #1\crcr}}
\def\indeqn#1{\displ@y\halign{\hbox to \displaywidth
    {$\ind\@lign\displaystyle##\hfil$}\crcr #1\crcr}}
\def\indalign#1{\displ@y \tabskip=0pt
  \halign to\displaywidth{\ind$\@lign\displaystyle{##}$\tabskip=0pt
    &$\@lign\displaystyle{{}##}$\hfill\tabskip=\centering
    &\llap{$\@lign##$}\tabskip=0pt\crcr
    #1\crcr}}
\catcode`\@=12
\def\si#1#2{\ifx\answ\bigans#1\else#2\fi}
\def\sips#1#2{\ifx\answ\bigans\epsfxsize=#1\else\epsfxsize=#2\fi}
\def\lpsn#1{\hfill LPSN-\number\yearltd-LT#1}
\def\boite#1{\hbox to .08\hsbody{\hfil#1\hfil}}
\def\Boite#1#2{\hbox to #1{\hfil #2\hfil}}
\def\sst{\scriptstyle}

\def\sn{\mathop{\rm sn}\nolimits}
\def\J{\mathop{\rm J}\nolimits}
\def\P{\mathop{\rm P}\nolimits}
\def\arcosh{\mathop{\rm arcosh}\nolimits}
\def\F{\mathop{\rm F}\nolimits}
\def\demi{\mathop{1\over 2}\nolimits}
\def\max{\mathop{\rm max}\nolimits}
\def\Nabla{\mathop{\nabla \hskip-8pt\nabla 
\hskip-8.2pt\nabla \hskip-8.4pt\nabla \hskip-8.6pt\nabla
\hskip-8.8pt\nabla }\nolimits}

 \def\Tr{{\rm Tr}}
\def\JPA{J. Phys. A: Math. Gen.}
\def\JPC{J. Phys. C: Solid State Phys.}     

\def\PR{Phys. Rev.}
\def\PRL{Phys. Rev. Lett.}

\overfullrule=0pt
\def\head{\eightsl Inhomogeneous systems}
\input epsf

\headline{cond-mat/9312077 \lpsn{3}}\si{\vglue3.5cm}{\vglue2.5cm}
\title{Inhomogeneous systems with~unusual~critical~behaviour}
\footnote{}{\tenit Accepted for publication in Advances in Physics.}
\author{F Igl\'oi\footnote\dag{\tenit 
igloi@power.szfki.kfki.hu}, 
I Peschel\footnote\ddag{\tenit peschel@aster.physik.fu-berlin.de}
and L Turban\footnote\S{\tenit turban@lps.u-nancy.fr} }
{\sl\bigskip\centerline{\dag\  Research Institute for Solid State Physics,
P. O. Box 49}
\centerline{H-1525 Budapest 114, Hungary}
\medskip\centerline{\ddag\  Fachbereich Physik, Freie Universit\" at
Berlin, Arnimallee 14}
\centerline{D-14195 Berlin, Germany}
\medskip\centerline{\S\  Laboratoire de Physique du Solide, URA CNRS 155,
Universit\'e de Nancy I, BP239}
\centerline{F-54506 Vand\oe uvre l\`es Nancy cedex, France}}
\beginabstract
The phase transitions and critical properties of two types
of inhomogeneous systems are reviewed. In one case, the
local critical behaviour results from
the particular shape of the system. Here scale-invariant
forms like wedges or cones are considered as well as general
parabolic shapes. In the other case the system contains 
defects, either narrow ones in the form of lines or stars, 
or extended ones where the couplings deviate 
from their bulk values according to power laws. In each
case the perturbation may be irrelevant, marginal or relevant.
In the marginal case one finds local exponents which depend
on a parameter. In the relevant case unusual stretched
exponential behaviour and/or local first order transitions
appear. The discussion combines mean field theory, scaling
considerations, conformal transformations and
perturbation theory. A number of examples are Ising models for
which exact results can be obtained. Some walks and polymer
problems are considered, too.
\endabstract
\vfill\date
\eject
\headline{\hfil\head}
\contents
\entry{1.}{Introduction}{3}
\entry{2.}{Planar surfaces}{5}
\subentry{2.1.}{Mean field theory}{\si{5}{6}}
\subentry{2.2.}{Correlation functions and critical 
profiles}{\si{9}{11}}
\subentry{2.3.}{Surface exponents}{\si{12}{14}}
\entry{3.}{Wedges, corners and cones}{\si{14}{17}}
\subentry{3.1.}{Mean field theory}{\si{15}{18}}
\subentry{3.2.}{Ising models}{\si{17}{21}}
\subentry{3.3.}{Other systems}{\si{19}{23}}
\subentry{3.4.}{Conformal results}{\si{20}{24}}
\entry{4.}{Parabolic shapes}{\si{23}{27}}
\subentry{4.1.}{Mean field theory}{\si{23}{28}}
\subentry{4.2.}{Conformal and scaling results}{\si{25}{30}}
\subentry{4.3.}{Ising model}{\si{27}{32}}
\subentry{4.4.}{Other systems}{\si{28}{33}}
\entry{5.}{Extended defects at surfaces}{\si{31}{37}}
\subentry{5.1.}{Ising model, general results}{\si{31}{37}}
\subentry{5.2.}{Ising model boundary magnetization}{\si{33}{39}}
\subentry{5.3.}{Conformal invariance for the marginal case}{\si{35}{42}}
\subentry{5.4.}{Related problems}{\si{36}{44}}
\subentry{5.5.}{Scaling considerations for relevant
inhomogeneities}{\si{38}{46}}
\entry{6.}{Narrow line defects in the bulk}{\si{40}{48}}
\subentry{6.1.}{Mean field theory and scaling}{\si{40}{49}}
\subentry{6.2.}{Ising model with a defect line}{\si{41}{50}}
\subentry{6.3.}{Conformal invariance, star-like defects}{\si{45}{54}}
\entry{7.}{Extended defects in the bulk}{\si{47}{58}}
\subentry{7.1.}{Ising model, conformal results}{\si{48}{58}}
\entry{8.}{Radially extended defects}{\si{50}{60}}
\subentry{8.1.}{Scaling considerations}{\si{50}{61}}
\subentry{8.2.}{Ising model}{\si{50}{61}}
\subentry{8.3.}{Conformal considerations}{\si{52}{63}}
\entry{9.}{Conclusion}{\si{53}{65}}
 
\ackapref{Acknowledgments}{}{\si{54}{66}}
\ackapref{Appendix A.}{Scaling and
conformal invariance}{\si{55}{67}}
\subentry{A.1.}{Scaling and critical exponents}{\si{55}{67}}
\subentry{A.2.}{Conformal invariance}{\si{57}{70}}
\ackapref{Appendix B.}{Transfer matrices}{\si{60}{73}}
\subentry{B.1.}{Row transfer matrix}{\si{61}{74}}
\subentry{B.2.}{Corner transfer matrix}{\si{63}{77}}
\subentry{B.3.}{Rescaling}{\si{65}{79}}
\ackapref{Appendix C.}{Perturbation theory for extended
defects}{\si{66}{81}}
\subentry{C.1.}{General}{\si{66}{81}}
\subentry{C.2.}{Relevance-irrelevance criterion}{\si{67}{82}}
\subentry{C.3.}{Marginal behaviour}{\si{68}{84}}
\ackapref{References}{}{\si{71}{87}}
\vfill\eject
\Section{Introduction}
In all fields of physics homogeneous systems have the
simplest properties and thus play a particular r\^ole.
This also holds with respect to phase transitions and
critical phenomena. The general behaviour, the universal
features near second-order transitions and the 
universality classes in this case are well-known [1]. On
the other hand, all real systems are inhomogeneous in one
way or another. This may affect their properties in
different ways. Enhanced couplings in a finite region
of a ferromagnetic system will result in a locally 
stronger magnetic order. For the case of competing
interactions the effects of an inhomogeneity may be more
complicated in detail but will still be of a local and
quantitative nature. If, however, the perturbation has
sufficient extent, the character of the phase transition
and the critical behaviour may change.
 
One example for this situation is well-known: a system 
with a free surface. To obtain it from a homogeneous 
system one has to cut an infinite number of bonds. The case
of planar surfaces was first treated for the
two-dimensional Ising model [2,~3] and
subsequently studied in great detail [4, 5]. It was found
that, connected with the surface, there is a set of
critical exponents with values different from those in the
bulk. The modified critical behaviour is seen in a
boundary layer which has a width of the order of the
correlation length.
 
The planar surface, however, is only the simplest example
showing such an effect. A~number of other inhomogeneous
systems display similar features. It is then of interest
to collect and present these cases, to see relations
between them and to discuss the basic aspects and
mechanisms. This is the aim of the following review.
 
The systems to be discussed are mainly classical spin
systems with short-range ferromagnetic interactions and,
occasionally, walks and polymers. The inhomogeneities have
a regular, non-random nature and the examples fall
into two groups. One consists of systems in various
geometrical forms with free boundaries described by
algebraic curves, the other contains models with defect
lines or extended inhomogeneities with algebraically
varying strength. In both cases unusual and nonuniversal
critical behaviour is found if the perturbation is
sufficiently effective. Then the local exponents depend
continuously on a parameter or the behaviour changes
altogether from power laws to streched exponentials. In
the second group of systems also local ordering at or
above the bulk transition temperature may occur. The
necessary condition for these effects always involves
scaling dimensions of the undisturbed system.
 
There are homogeneous systems which also show such
features. An example is the eight-vertex model [6]. In
this case, however, all exponents vary in a similar way
and one can obtain universal values by measuring
the temperature via the correlation length. The
inhomogeneous systems considered here have a richer
structure since local and bulk quantities appear
simultaneously. In those cases where the effects are
related to the geometry they also are of a rather general
nature since non-universal behaviour occurs for
any simple isotropic system and in all dimensions.
 
The mentioned phenomena can often already be seen in a
mean field treatment which is therefore given in various
places. More generally, renormalization and scaling
considerations allow to classify the perturbations
and the way they act. There also is a rather general
method to calculate the change of critical exponents via
perturbation theory. A majority of the examples are
two-dimensional systems and for them two other methods
are available. Conformal invariance makes detailed
predictions at the critical point and has been
used in various cases although its validity in 
inhomogeneous systems is not necessarily guaranteed. Exact
calculations for various Ising systems, however, have
always confirmed conformal results. Moreover they allow to
obtain a complete picture by giving results for all
temperatures. The main tool here are transfer matrices and
the two-dimensional problems are thereby related to certain
quantum spin chains. All these topics are treated in the
various sections and in three appendices. The latter
contain some background material and some more technical 
aspects.
 
Certain types of inhomogeneity will not be considered here
although they may lead to interesting effects. Random
systems, by their very nature, demand special methods and
are still under investigation with respect to
their critical properties. For reviews we refer to [7, 8].
In the case of wetting phenomena one is concerned with the
properties of interfaces which are created by appropriate
boundary conditions and which may be influenced
by inhomogeneities. This is also a field in itself and has
already been reviewed [9--11]. Finally, layered systems 
with a periodic structure [12--14] will also be omitted.
 
Some of the effects to be discussed should be observable
in experiments but the article deals with the theoretical
aspects. Nevertheless, it is not aimed at the specialist.
The notation has thus been chosen as simple and coherent as
possible and therefore differs from the usual one in some
places.
\Section{Planar surfaces}
At a free surface missing bonds reduce the order and, as
mentioned before, the critical behaviour is modified
locally. The loss of full translational invariance
leads to a decay of correlations which is different
parallel and perpendicular to the surface. In dimensions
higher than two, enhanced surface couplings may induce
surface order above the bulk critical temperature. The
local critical behaviour can also be influenced by the
surface geometry or through the introduction of long-range
surface induced perturbations. These last
two points will be considered in the next sections.
 
The easiest approach to the problem at hand makes use of
mean field theory which often brings a good qualitative
description of the phenomena and becomes exact above the
upper critical dimension where the fluctuation effects are
negligible. Since a detailed account of planar surface
critical behaviour within this frame can be found
elsewhere [15, 4] we limit ourselves to a brief study of 
the bulk transition and the ordinary surface transition for
illustrative purpose. This will serve as an introduction to the mean
field treatment of the other, less usual, inhomogeneous 
systems which are considered in the next sections.
 
There also exists a number of recent reviews dealing
with surface critical behaviour beyond mean field theory
[4, 5, 16, 17]. As a consequence, only selected results
concerning the structure of correlations, critical 
profiles and surface exponents are discussed in this
section.

\Subsection{Mean field theory}
In Landau mean field theory the system is treated in a
continuum description. The total free energy $F$ of a  
sytem with a volume $(V)$ limited by a surface $(S)$ is 
written as the sum of bulk and surface contributions
which  are functionals of the order parameter $m(\bi r)$ which is
nonvanishing in the ordered phase 
$$
F[m]=\int_{(V)}f_b[m]\ \d V+\int_{(S)}f_s[m]\ \d S.
\eqno(2.1)
$$

Let us consider for simplicity an Ising system with a 
scalar order parameter  $m$ and a free energy which
in zero external field  is even in $m$, i.e. symmetric
under order parameter reversal. Near a second order
transition, the order parameter is small   and the bulk
free energy density $f_b[m]$ is written as an
expansion in the order parameter and its gradient, limited
to the following terms 
$$ 
f_b[m]=f_b[0]+\demi C(\Nabla
m)^2+\demi Am^2+{1\over 4} Bm^4-hm.\eqno(2.2)
$$
The second one, with $C>0$, gives the extra energy
associated with a spatial variation of the order parameter,
$A\sim T-T_c$ measures the deviation from the critical
temperature $T_c$, $B$ is a positive constant to ensure
stability below $T_c$ and $h$ is the bulk external field.

In the same way the surface free energy density is written
phenomenologically as
$$
f_s[m]=f_s[0]+\demi C{m^2\over\Lambda}-h_sm
\eqno(2.3)
$$
where $m$ is the value of the order parameter on $(S)$. 
The quantity  $\Lambda$, with the dimension of a length, is
called the  extrapolation length and can be deduced from
the microscopic surface and bulk interactions through a
mean field treatment of the microscopic Hamiltonian of the 
system [4]. 

The equilibrium value of the order parameter $m(\bi r)$
minimizes the free energy in (2.1). It may be obtained 
through a variational calculation by looking for the first
order change of the free energy, $\delta F[m]$, associated
with a deviation $\delta m(\bi r)$ of the order parameter
from its equilibrium value. Using (2.1--3), one obtains
$$\eqalign{
\delta F[m]=&\int_{(V)}\left[C\Nabla m\cdot\Nabla
\delta m+(Am+Bm^3-h)\delta m\right]\d V\cr
&+\int_{(S)}\left[{C\over\Lambda}
m-h_s\right]\delta m\ \d S.\cr}\eqno(2.4)
$$
The first term in the volume integral may be rewritten as
$$
C\Nabla m\cdot\Nabla \delta m=\Nabla\cdot 
(C\delta m\Nabla m)-C\nabla^2 m\ \delta m\eqno(2.5)
$$
and the contribution to (2.4) of the first term on the 
right can be transformed into a surface integral through
Gauss theorem.  Then
$$\eqalign{
\delta F[m]=&\int_{(V)}\left[-C\nabla^2 m+Am+Bm^3-h\right]
\delta m\ \d V\cr
&+\int_{(S)}C\left[-\bi n\cdot\Nabla m+
{m\over\Lambda}-{h_s\over C}\right]\delta m\ \d S\cr}
\eqno(2.6)
$$
where $\bi n$ is a unit vector normal to the surface and 
pointing inside the system. 

At equilibrium the first order variation of the free 
energy vanishes. The volume part gives the Ginzburg-Landau 
equation 
$$
C\nabla^2m(\bi r)=Am(\bi r)+Bm^3(\bi r)-h\eqno(2.7)
$$
governing the bulk equilibrium behaviour whereas the 
surface part provides the boundary condition
$$
\bi n\cdot\Nabla m(\bi r)={m(\bi r)\over\Lambda}-
{h_s\over C}.\eqno(2.8) 
$$

In the bulk the l.h.s. in (2.7) vanishes and the 
zero-field magnetization is given by
$$
m_b=\left\{\matrix{(-A/B)^{1/2}\sim t^{1/2}&T\leq T_c\cr
                  0                       &T>T_c\cr}\right.
\eqno(2.9)
$$
where $t\sim\mid\! A\!\mid$ is the reduced temperature.
The  spontaneous magnetization vanishes at $T_c$ as a
power of the reduced temperature with a mean field bulk
exponent $\beta\!=\! 1/2$.

The connected part of the order parameter correlation
function $$
<m(\bi r)m(\bi r')>_c=<m(\bi r)m(\bi r')>-<m(\bi r)><m(\bi
r')>={\delta m(\bi r)\over\delta h(\bi r')}\eqno(2.10)
$$
can be obtained through the introduction of a varying bulk external
field $h(\bi r)$ in~(2.2). This simply amounts to 
replacing $h$ in~(2.7) by $h({\bi r})$.
Taking a derivative with respect to $h(\bi r')$ and using 
the definition (2.10) leads to
$$
\left[-C\nabla_r^2+A+3Bm^2(\bi r)\right]<m(\bi r)m(\bi
r')>_c =\delta(\bi r-\bi r').\eqno(2.11)
$$
 
For constant $m(\bi r)$, which is the case in a homogeneous
system below $T_c$ or generally above $T_c$, the correlation
function is therefore proportional to the Green function
which satisfies
$$
\left(-\nabla_r^2+{1\over\xi^2}\right)G(\bi r,\bi
r')=\delta(\bi r-\bi r').\eqno(2.12) 
$$
Here
$$
\xi=\left\{\matrix{\sqrt{{C\over A}}&T>T_c\cr
                 \sqrt{-{C\over 2A}}&T<T_c\cr}\right.
\eqno(2.13)
$$
is the bulk correlation length diverging at the critical 
point with a mean field exponent $\nu=1/2$ which is the
same in both phases. Hence the critical $G$ is the Green
function of the Laplace equation which establishes a link
to electrostatic problems.
 
For a flat surface at $y=0$ without external surface 
field, the boundary condition (2.8) gives 
$$
{\d m\over\d y}={m\over\Lambda}.\eqno(2.14)
$$
When the surface interactions are not larger than in the
bulk, due to the missing couplings at the boundary,
the magnetization always increases from the surface
into the bulk and the extrapolation length is positive. 
The surface transition is then driven by the bulk and there
is an ordinary  surface transition at $T_c$. Since
$\Lambda\!\ll\!\xi$ near the bulk critical point, the
magnetization profile extrapolates to zero near the
surface and one may use Dirichlet boundary conditions,
i.e. $m=0$ on the surface, instead of (2.14). With $\hat
m=m/m_b$ and $h=0$, (2.7) and (2.9) lead to  
$$ 
{\d^2\hat m\over\d y^2}={A\over C}\hat m+{B\over
C}m_b^2\hat m^3=-{A\over C}(\hat m^3-\hat m).\eqno(2.15) 
$$
Multiplying both sides by $2\d\hat m/\d y$ and 
integrating, one obtains
$$
{\d \hat m\over \d y}={1-\hat m^2\over 2\xi}\eqno(2.16)
$$
where the integration constant has been chosen to give a 
vanishing slope at infinity and the form of the correlation
length in  (2.13) has been used. The solution of (2.16)
with $\hat m=1$ at infinity then is 
$$
m(y)=m_b\tanh{y\over 2\xi}.\eqno(2.17)
$$
In the vicinity of the surface, $m\sim t\ y$, so that the
magnetization grows linearly with the distance to the 
surface. Its temperature dependence is different from the
bulk one and involves a new exponent $\beta_s=1$ which is
the mean field surface exponent at the ordinary transition.
 
The form of the correlation function for the ordinary
transition is given by the solution of (2.12) which
satisfies Dirichlet boundary conditions at the surface.
To obtain this function for a half-space, one can start from
a $d$-dimensional slab limited by two surfaces at
$y=0$ and $y=L$. With $\bi r\!-\!\bi r'\!=\!{\bi
r}_\parallel\! +\!(y-y')\bi n$ and taking advantage of
translational invariance in the $d\!-\! 1$ transverse
directions, one may rewrite $G({\bi r}_\parallel,y,y')$ as
a Fourier expansion with components $G_{\bi k}(y,y')$
satisfying $$
\left(-{\partial^2\over\partial y^2}+\kappa^2\right)G_{\bi
k}(y,y')=\delta(y-y')\eqno(2.18)
$$
where $\kappa\!=\!\sqrt{k^2+\xi^{-2}}$. The solution can be
written as an eigenfunction expansion
$$
G_{\bi k}(y,y')={2\over L}\sum_{n=1}^\infty{\sin(n\pi
y/L)\sin(n\pi y'/L)\over\kappa^2+n^2\pi^2/L^2}.\eqno(2.19)
$$

For a semi-infinite system $L\rightarrow\infty$ and the sum
over $n$ can be replaced by an integral which is evaluated
using the method of residues so that finally [15]
$$
G({\bi
r}_\parallel,y,y')={1\over(2\pi)^{d-1}}\int\d^{d-1}k\ 
{\e^{\i\bi k\cdot{\bi r}_\parallel}\over 2\kappa}\left[
\e^{-\kappa\mid y-y'\mid}-\e^{-\kappa(y+y')}\right].
\eqno(2.20) $$
The two parts of $G$ are each correlation functions of
the infinite system, one between $\bi r$ and $\bi r'$
and the other between $\bi r$ and the image point of
$\bi r'$ relative to the boundary. At the critical point
they decay with power $(d-2)$. From this the surface-bulk
correlations follow by taking
$r_\parallel=0$, $y'$ fixed and $y\gg y'$, leading to
the asymptotic
decay $G(y)\sim y^{-(x+x_s)}\sim y^{-(d-1)}$. Here $x$
and $x_s$ are the bulk and surface (ordinary) scaling
dimensions of the magnetization (Appendix A.1). The bulk
behaviour is obtained taking $y$  and $y'\gg 1$ with $\mid
y-y'\mid\gg 1$. Then the decay exponent is
$2x=d-2$ so that $$
x={d-2\over 2}\qquad\qquad x_s={d\over 2}\qquad\qquad({\rm
ordinary}\quad{\rm transition}).\eqno(2.21)
$$
The scaling relations (A4, A7) $\beta\!=\!\nu x\!=\! 1/2$
and $\beta_s\!=\!\nu x_s\!=\! 1$ are satisfied by the
mean field exponents at the upper critical dimension
which, in this case, is  $d_c=4$.

Finally, consider the critical profile of the 
magnetization when its value at the surface is fixed at
$m(0)=m_0$. It follows from (2.7) with $A=0$ and $h=0$ so
that  
$$
{\d^2m\over\d y^2}={B\over C}m^3.\eqno(2.22)
$$
Following the same steps as above, the integration gives 
$$
m(y)=m_0\left[1+m_0\left({B\over 2C}\right)^{1/2}\!\! y
\right]^{-1}\eqno(2.23)
$$
with the asymptotic behaviour
$$
m(y)\sim y^{-1}.\eqno(2.24)
$$
The magnetization decays as a power of the distance to 
the surface, a behaviour which is also obtained beyond 
mean field theory as discussed in the next section.

If the surface couplings are sufficiently enhanced and/or
a surface field is present, $\Lambda$
becomes negative and the surface orders at a temperature
which is higher than the bulk critical one.
Then bulk ordering at $T_c$ induces some 
singularity in the surface behaviour. The associated 
critical point is the extraordinary transition. Increasing
$\Lambda$ decreases the temperature of the surface
transition until the surface and the extraordinary
transition meet at the special transition. A detailed
 treatment can be found in
references [4, 15]. 

\Subsection{Correlation functions and critical profiles}  
Conformal methods (Appendix A.2) can be used to make quite
general statements about critical systems. They 
generalize, at the critical point, the covariance under
global scale transformations which is at the basis of the
renormalization group, by introducing local scale
transformations with a varying dilatation factor $b(\bi
r)$. Such local dilatations are realized via conformal
transformations.  In any dimension the conformal group is
constructed by including the inversion $\bi r'=\bi r/r^2$
besides the usual uniform transformations: translation,
rotation and dilatation. 

Particularly useful in the study of surface properties is 
the special conformal transformation [18,16] 
$$ 
{\bi r'\over r'^2}={\bi r\over r^2}+\bi a\eqno(2.25) 
$$
which combines an inversion followed by a translation and 
a new inversion.
A semi-infinite system with a flat surface containing the 
origin is invariant under such a transformation when the
translation $\bi a$ is parallel to the surface. Using an
infinitesimal translation,  covariance under this
transformation determines the form of
the critical correlation functions [18] and allows a
determination of boundary induced profiles. Although the
transformation works in any dimension, only the
two-dimensional situation is discussed below.

Consider a critical system on a half-plane with a surface 
at $y=0$. In Cartesian coordinates (2.25) translates into 
$$
x'=x+\epsilon (x^2-y^2)\qquad\qquad y'=y+2\epsilon xy
\eqno(2.26)
$$
where the infinitesimal translation is 
$\bi a=(-\epsilon,0)$. Using complex notations, $z=x+\i y$,
(2.25) is rewritten as 
$$
z'=z+\epsilon z^2\eqno(2.27)
$$
and the local dilatation factor is 
$$
b(x)=\left\vert{\d z\over\d z'}\right\vert=1-2\epsilon x.
\eqno(2.28)
$$
 
Let $\psi (\bi r)$ be some local operator (energy density, order
parameter) with bulk scaling dimension $x$, i.e. 
transforming as $\psi (\bi r/b)=b^{x}\psi(\bi r)$ under a
global change of scale. Its two-point critical correlation
function
$G(x_1,x_2,y_1,y_2)=<\!~\psi(x_1,y_1)\psi(x_2,y_2)~\!>$, which
is a function of $x_1$ and $x_2$ through $u=x_1-x_2$ due to
translational invariance parallel to the surface, is
transformed into 
$$
G(u',y'_1,y'_2)=b(x_1)^{x}b(x_2)^{x}G(u,y_1,y_2)
\eqno(2.29) 
$$
according to (A14) under (2.26). A first-order expansion in $\epsilon$ leads
to the differential equation 
$$
(x_1^2-x_2^2-y_1^2+y_2^2)\ {\partial G\over\partial
u}+2x_1y_1{\partial G\over\partial y_1}+ 2x_2y_2{\partial
G\over\partial y_2}+2x(x_1+x_2)\ G=0.\eqno(2.30)
$$
Since $G$ depends on $x_1,\ x_2$ only through $u$, all the 
terms involving a factor $x_1+x_2$ sum up to zero and ones
obtains two partial differential equations
$$\eqalign {
u{\partial G\over\partial u}+y_1{\partial G\over\partial
y_1}+y_2{\partial G\over\partial y_2}+2xG&=0\cr
(y_2^2-y_1^2)\ {\partial G\over\partial u}+u\left(y_1
{\partial G\over\partial y_1}-y_2{\partial G\over\partial
y_2}\right)&=0.\cr}\eqno(2.31)
$$
The first one expresses the homogeneity of the correlation
function 
$$
G\left({u\over b},{y_1\over b},{y_2\over
b}\right)=b^{2x}G(u,y_1,y_2)\eqno(2.32)
$$
or, with $b=u$ and $\zeta_i=y_i/u$,
$$
G(u,y_1,y_2)=u^{-2x}G(1,\zeta_1,\zeta_2)=(u^2\zeta_1
\zeta_2)^{-x}\Xi(\zeta_1,\zeta_2).\eqno(2.33)
$$
Using this form, the second equation in (2.31) fixes the 
way spatial coordinates combine into a single scaling
variable $\rho$ in the scaling function $\Xi(\rho)$ so that,
finally, 
$$
G(x_1-x_2,y_1,y_2)=(y_1y_2)^{-x}\ \Xi
\left[{y_1y_2\over
(x_1-x_2)^2+(y_1-y_2)^2}\right].\eqno(2.34)
$$
The behaviour of the scaling function for small or large 
values of the argument can be
deduced from scaling considerations. For instance, with 
$y_1=y_2$ fixed and $\mid x_1-x_2\mid=\mid
u\mid\rightarrow\infty$, the surface-surface correlations
decay like $\mid u\mid^{-2x_s}$ where $x_s$ is the surface
scaling dimension of $\psi$. As a consequence 
$$ 
\Xi (\rho)\sim\rho^{x_s}\qquad\qquad\rho\rightarrow 0.
\eqno(2.35)
$$

In two dimensions the scaling function itself satisfies
a certain differential equation following from conformal
invariance and has been determined explicitly
for the Ising, Potts and $O(N)$ models [18, 16, 19, 20].

The same method applies to the determination of critical 
profiles when $\psi(\bi r)$ is some quantity with
nonvanishing average at the bulk critical point of a
semi-infinite system. This may be either the energy
density with free or fixed boundary conditions or the
order parameter with fixed boundary conditions. Since in
two dimensions the surface is one-dimensional, an
enhancement of surface interactions is  not sufficient to
maintain surface order at the bulk critical point.
Therefore the boundary variables must be fixed in order to
have a non-trivial order parameter profile.
 
Due to translational invariance parallel to the surface, the 
critical profile depends only on $y$. Following the same 
steps as above, a differential equation for $<\psi (y)>$ is
obtained, leading to the universal profile 
$$
<\psi (y)>\sim y^{-x}.\eqno(2.36)
$$
That (2.36) follows from conformal invariance was already 
noticed in a different way in [21, 22]. This algebraic form
was originally deduced from simple scaling
considerations [23]. In any number of dimensions,
$<\psi(r_\perp/b)>=b^{x}<\psi(r_\perp)>$ where $r_\perp$ is 
the distance to the surface, so that with $b=r_\perp$,  
$$ <\psi(r_\perp)>\sim r_\perp^{-x}.\eqno(2.37) $$

When $\psi$ is the order parameter, $x=\beta/\nu$ which
is equal to $1$ in mean field theory in agreement with 
Equation (2.24).

\Subsection{Surface exponents}
Some heuristic arguments have been used [24] to obtain
the values of surface exponents in arbitrary dimensions.
They apply to the ordinary and extraordinary transitions
for the energy density and to the extraordinary
transition for the order parameter. In two dimensions
the results are supported by independent conformal
arguments.

Consider a system inside a cube with $L^d$ interacting
spins and free boundaries. Increasing the size of the
system by $\delta L$ through the addition of a surface
layer, the partition function becomes
$$
Z_{L+\delta L}=\Tr \exp\left[-\beta ({\cal H}+\delta
{\cal H})\right]=
Z_L<\exp (-\beta\delta {\cal H})>_L\eqno(2.38)
$$
where $\delta{\cal H}$ is the energy change associated
with the extra layer.
To leading order in $L$ the free energy
$$
F_L=L^df_b+O(L^{d-1})\eqno(2.39)
$$
only involves the bulk free energy density
$f_b$. Its variation under an infinitesimal change of the
size is given by
$$
\delta F_L=-\beta^{-1}\ln <\exp (-\beta\delta
{\cal H})>_L\simeq <\delta {\cal H}>_L\simeq dL^{d-1}
f_b\delta L+O(L^{d-2})\eqno(2.40)
$$
so that
$$
\lim_{\delta L\rightarrow 0}{<\delta {\cal H}>_L\over
dL^{d-1}\delta L}
=f_b+O(L^{-1})\eqno(2.41)
$$
and the surface energy density $\varepsilon_s$ on the
l.h.s. scales  near the
bulk critical point like the bulk free energy density
$$
\varepsilon_s\sim t^{\nu x_s}\sim f_b\sim t^{2-\alpha }.
\eqno(2.42)
$$
Thus its scaling dimension $x_s$ at the ordinary
transition is equal to the dimensionality of the system
according to the Josephson hyperscaling relation
(A2). The same argument applies to the magnetization
and energy densities with fixed boundary conditions,
i.e. at the extraordinary transition. Then
$$
m_s\sim\varepsilon_s\sim t^{2-\alpha }\eqno(2.43)
$$
so that both scaling dimensions are equal to the
dimension $d$ of the system. For the surface energy
operator at the ordinary transition, this result was
first obtained by Dietrich and Diehl in the $O(N)$
model using renormalization
group methods and the short-distance expansion [25].
 
In two dimensions, any analytic function $w(z)$ on the 
complex plane provides a conformal transformation with a
local dilatation factor $b(z)=\mid\!\d z/\d w\!\mid$ 
(see~Appendix~A.2). Such a transformation can be used to deduce
surface exponents starting from the half-space critical 
profile. Under the conformal mapping $w\!=\!{L/\pi}\arcosh z$ 
the half-plane  $y\!>\!0$ is transformed into the
half-strip ($u\!>\!0$, $0\!<\!v\!<\!L$, $w\!=\!u\!+\!\i v$). The
dilatation factor is $b(z)\!=\!\pi/L\sqrt{\sinh^2(\pi
u/L)+\sin^2(\pi v/L)}$ and $y\!=\!\sinh(\pi u/L)\sin(\pi
v/L)$. The boundary-induced critical profile $<\psi(y)>$
(2.36) which has a bulk scaling dimension $x$, then
transforms into~[22]     
$$ \eqalign { 
<\psi(u,v)>&\sim b(z)^{x}y^{-x}\cr
          &\sim\left({\pi\over L}\right)^{x} \left[
\sinh^{-2}\left({\pi u\over L}\right)+\sin^{-2} \left({\pi
v\over L} \right)\right]^{x/2}\cr}\eqno(2.44)   
$$ 
in the half-strip. The profile in the new geometry can
also be written as an expansion in terms of the eigenstates of
the transfer operator ${\cal T}= \exp (-{\cal H})$ where
$\cal H$ is the strip Hamiltonian with appropriate
boundary conditions (Appendix B.1). Then   
$$ 
<\psi(u,v)>=\sum_nM_n<n\mid\psi (v)\mid 0>\exp\left[-
(E_n-E_0)u\right]\eqno(2.45) 
$$ 
where $E_0$ is the
ground-state energy of $\cal H$ and $M_n$ selects the
eigenstates $\mid n>$ compatible with the boundary
conditions at $u=0$. The smallest gap corresponding to a
nonvanishing matrix element gives the surface scaling
dimension $x_s$ of $\psi$ as in (A23). Since the large-$u$
expansion of the r.h.s. of (2.44) involves only powers of
$\exp\left( -2\pi u/L \right)$ the gaps in (2.45) are
multiples of $2\pi /L$. Assuming that $M_1$ is
non-nanishing, one may identify the surface scaling
dimension as 
$$ 
x_s=2\eqno(2.46)
$$
in agreement with previous results in the special case
$d$=$2$.
 
The considerations presented above suppose the
existence of a non-vanishing profile at the critical
point. As a consequence they cannot be applied to the
case of the magnetization at the ordinary surface
transition. As mentioned in the preceding section,
other conformal techniques can be used, which completely
determine the critical correlation functions
in the semi-infinite geometry. From these, the ordinary
surface exponents given in Table 2.1 for the $q$-state
Potts model and the $O(N)$ model have been identified
[18, 16].
{\medskip\par\begingroup\parindent=0pt\leftskip=1cm
\rightskip=1cm\parindent=0pt
\midinsert
\hfil\boldrule{0.5\hsbody}\hfil\par
\hfil $q$-state Potts model $(d=2)$\hfil\par
\hfil\medrule{0.5\hsbody}\hfil\par
\centerline{\boite{$q$}\boite{$0$}\boite{$1$}
\boite{$2$}\boite{$3$}\boite{$4$}}
\centerline{\boite{$x_s$}\boite{$0$}\boite{$1\over 3$}
\boite{$\demi$}\boite{$2\over 3$}\boite{$1$}}
\hfil\medrule{0.5\hsbody}\hfil\par
\hfil$O(N)$ model $(d=2)$\hfil\par
\hfil\medrule{0.5\hsbody}\hfil\par
\centerline{\boite{$N$}\boite{$-2$}\boite{$-1$}
\boite{$0$}\boite{$1$}\boite{$2$}}
\centerline{\boite{$x_s$}\boite{$1$}\boite{$13\over 16$}
\boite{$5\over 8$}\boite{$\demi$}\boite{$1\over 4$}}
\hfil\boldrule{0.5\hsbody}\hfil
\baselineskip=12truept
\tabcaption{2.1}{\srm Order parameter scaling dimension at the
ordinary surface transition
in the $\sst q$-state Potts model and the $\sst O(N)$ vector model.}
\endinsert
\endgroup
\par}
\Section{Wedges, corners and cones}
Systems in the form of a wedge show new features and
their study gives new insight into the influence of
geometry on critical behaviour. They were first
investigated by Cardy [26] using mean field theory
in various dimensions and an $\epsilon$-expansion
near $d=4$. Subsequently the corner geometry in $d=2$
was studied by accurate calculations, mainly
on Ising models, and by conformal mapping. The main
result is that the edge and corner exponents are
functions of the opening angle $\theta$ of the wedge.
The planar surface and its exponents can be viewed
as a simple special case.
 
This non-universal behaviour can be related to the
particular geometrical properties of a wedge, which
does not contain a length parameter and thus
is invariant under rescaling. The opening angle is
therefore a marginal variable in a renormalization
transformation and may enter into the expressions for
the exponents. The same will happen for all other
scale-invariant figures like arbitrary pyramids or
cones. Furthermore, the system does not have to fill
the figure, it may just cover the surface. For such a
situation there are even exact analytical results.
\Subsection{Mean field theory}
Within a continuum approximation, in the mean-field
approach one has to solve the same equations as in 
Section 2.1, but for a different geometry. The
order parameter correlation function is
proportional to the Green function (2.12) with Dirichlet
boundary conditions for the ordinary transition. 
This function also appears in electrostatics [27] and in
diffusion problems [28]. In the present context it was
discussed by Cardy [26] for wedges in $d$ dimensions.

For a wedge, one proceeds as for a slab and performs a
Fourier transformation in the $d\!-\! 2$ directions parallel
to the edge. In the remaining directions one uses
polar coordinates $\rho,\alpha$ where $\alpha$ is the
azimuthal angle in the wedge $(0\leq\alpha\leq\theta)$. 
The Fourier components $G_{\bi k}$ then satisfy
$$
\left(-{\partial^2\over\partial\rho^2}-{1\over\rho}{\partial
\over\partial\rho}-{1\over\rho^2}{\partial^2\over\partial
\alpha^2}+\kappa^2\right)G_{\bi k}(\rho,\alpha;\rho',\alpha')
={1\over\rho}\delta(\rho-\rho')\delta(\alpha-\alpha')\eqno(3.1)
$$
with $\kappa$ defined below (2.18). 
The solution can be written as an eigenfunction expansion and
takes the form
$$
G_{\bi k}(\rho,\alpha;\rho',\alpha')
={2\over\theta}\sum_{n=1}^\infty\int_0^\infty{\d\mu\ \mu
\over\kappa^2+\mu^2}{\J_{\nu}(\mu\rho)\J_{\nu}(\mu\rho')\sin\left({
n\pi\alpha\over\theta}\right)\sin\left({n\pi\alpha'\over
\theta}\right)}\eqno(3.2)
$$
where the $\J_{\nu}$ are Bessel functions and
$\nu=n\pi/\theta$. The continuous index $\mu$ is related
to the infinite extent of the system in the $\rho$-direction.
Using this expression one can find, for example, the critical
correlations parallel to the edge. Their asymptotic behaviour is
determined by the contributions of $n=1$ and small $\mu$. One obtains
$$ G(\bi r,\bi r')\sim{1\over\mid\bi r-\bi r'
\mid^{2x_e}}\eqno(3.3) 
$$ 
with the edge exponent
$$
x_e(\theta )={d-2\over 2}+{\pi\over\theta}.\eqno(3.4)
$$
The decay thus is always faster than in a homogeneous
system, where it would be given by the first term in
(3.4) as found in (2.21). This is an effect of the boundary
conditions. In an electrostatic picture, $G$ gives the
potential at ${\bi r}$ due to a point charge at ${\bi r'}$
and the effect comes from the induced charge of opposite
sign on the
(metallic) boundaries. For $\theta=\pi/n$, where $n$ is an
integer, it can be simulated by $(2n-1)$ image charges.
As the wegde becomes narrower, the effect becomes stronger.

To obtain the order parameter profile in the wedge one
would have to solve the non-linear Ginzburg-Landau equation
(2.7) with $h=0$. For the critical behaviour, however,
this is not necessary. The radial dependence has the
general scaling form $m_bf(\rho/\xi)$ and since $m$ is
small, the spatial dependence follows from the
linearized equation. This gives, for $\rho\ll\xi$
$$
m(\rho,\alpha)\sim m_b\left({\rho\over\xi}\right)^{\pi/\theta}
\sin\left({\pi\alpha\over\theta}\right)\eqno(3.5)
$$
Inserting the temperature
dependence of $\xi$ and $m_b$ then leads to the
exponent
$$
\beta_e(\theta )=\demi \left( 1+{\pi\over\theta }\right).
\eqno(3.6) 
$$ 
This result is valid in all dimensions. At
the upper critical dimension $d_c=4$ the scaling law
$\beta_e=\nu x_e$ as in (A7) is satisfied although
$\beta_e$ and $x_e$ are non-universal. For $\theta=\pi$
the surface exponent $\beta_s=1$ is recovered. For
smaller angles, the order parameter near the edge
vanishes with zero slope at the critical temperature.
This reflects the difficulty to maintain the ordered
state in such a geometry.

One can also treat a semi-infinite wedge where the third
planar boundary is perpendicular to the edge. This system
has a three-dimensional corner with two right angles and
the third one equal to $\theta$. The correlation function
then follows from the image method as
$$
G(\bi r,\bi r')=G_{\infty}(\bi r,\bi r')-G_{\infty}(\bi r,
\bi r'')\eqno(3.7)
$$
where $G_{\infty}$ is the result for the infinite wedge and
$\bi r''$ is the image point of $\bi r'$ with respect to
the third plane. For $\bi r'$ near the corner, the correlations 
parallel
to the edge then decay with a power $x_c+x_e=1+2x_e$, 
from which the corner exponent $x_c(\theta )=1+x_e
(\theta )$ follows. If $\theta=\pi/2$, one obtains
a cubic corner with three right angles and $x_c=7/2$.
The exponent of the order parameter is calculated as above
$$
\beta_c(\theta )=1+{\pi\over 2\theta}.\eqno(3.8)
$$
For small $\theta$ this is the same law as in (3.6)
but in general $\beta_c>\beta_e$ so that the order
near the corner vanishes faster with temperature
than near the edge. In particular for the cubic
corner one has $\beta_c=2$, i.e. a quadratic
behaviour.
 
The general three-dimensional corner, for which the
exponents depend on three angles, has not been treated.
However, results can be given for a cylindrical
cone of opening angle $\theta$ with respect to the
axis. The Green function of this problem has been
determined in other contexts [28, 29]. For ${\bi r}'$
on the axis and $r>r'$ it has the form
$$
G(\bi r,\bi r')=\sum_\mu A_\mu\left({r'\over
r}\right)^{\mu+1}\P_\mu(\cos\alpha)\eqno(3.9)
$$
where again $0\leq\alpha\leq\theta$. The Legendre
function $\P_\mu(\cos\alpha)$ has to vanish on the
boundary $(\alpha=\theta)$ which determines the allowed
values $\mu=\mu_m$, $m=1,2,\cdots$. The asymptotic 
decay is again a power law with the exponent given by 
the smallest $\mu_1=\mu_1(\theta)$. For the order parameter 
near the apex
of the cone one obtains, in the same way as for the wedge
$$
\beta_a(\theta )=\demi [1+\mu_1(\theta )].\eqno(3.10)
$$
The function
$\mu_1 (\theta )$ is shown in reference [30]. For 
$\theta\rightarrow 0$ it varies as $1/\theta$ and the 
system has the same
features as a narrow wedge. For $\theta=\pi/2$ one has
$\mu_1=1$ and the result $\beta_a=\beta_s=1$ for the
planar surface is recovered.For
$\theta\rightarrow\pi$ the quantity $\mu_1$ vanishes
logarithmically. Thus one obtains the bulk result
$\beta_a=\beta=1/2$, although one is dealing with a
system from which an half-infinite needle is cut out.
\Subsection{Ising models}
A number of results exist for corners in two-dimensional Ising
lattices with certain discrete values of the angle. The corner 
magnetization $m_c$ has been
calculated for triangular and square lattices
in three different ways :
from star-triangle recursion equations [31], from
the corner-corner correlation function [31--33] and
from Baxter's corner transfer matrix [34]. 

In the
recursion method, one starts from a finite lattice,
e.g. in the shape of a triangle with fixed spins along
one edge, and reduces it successively to smaller sizes.
In this way the corner magnetization can be evaluated
exactly, but the procedure has to be done numerically.
In the second method one considers the correlation between 
spins at adjacent corners of
a lattice in the form of a square. Asymptotically, it
factors into the product of the two corner magnetizations
for which an expression
$$
m_c\sim {< 1|\sigma_1^x|B >\over< 0|B >}\eqno(3.11)
$$
is found. Here the operator $\sigma_1^x$ refers to the corner
spin, the
states $|0 >$ and $|1 >$ are the eigenstates of the row 
transfer matrix (Appendix B.1) with largest and next-largest
eigenvalue, and $|B >$ describes the state of free spins in
the upper and lower boundary row. The quantity $m_c$ can then
determined from the solution of a matrix or integral equation.
The corner transfer matrix, finally, is already the partition
function for a whole angular segment (see Appendix B.2). Thus
it is, from a
geometrical point of view, the ideal tool for treating 
the corner geometry. 
However, due to the edges of free spins one has to take a
matrix element $< B|{\cal T}|B >$ and the calculation becomes
relatively difficult. One has to solve a matrix equation
also in this case.
{\par\begingroup\parindent=0pt\leftskip=1cm
\rightskip=1cm\parindent=0pt
\baselineskip=12truept\sips{13.6truecm}{9.7truecm}
\midinsert
\centerline{\epsfbox{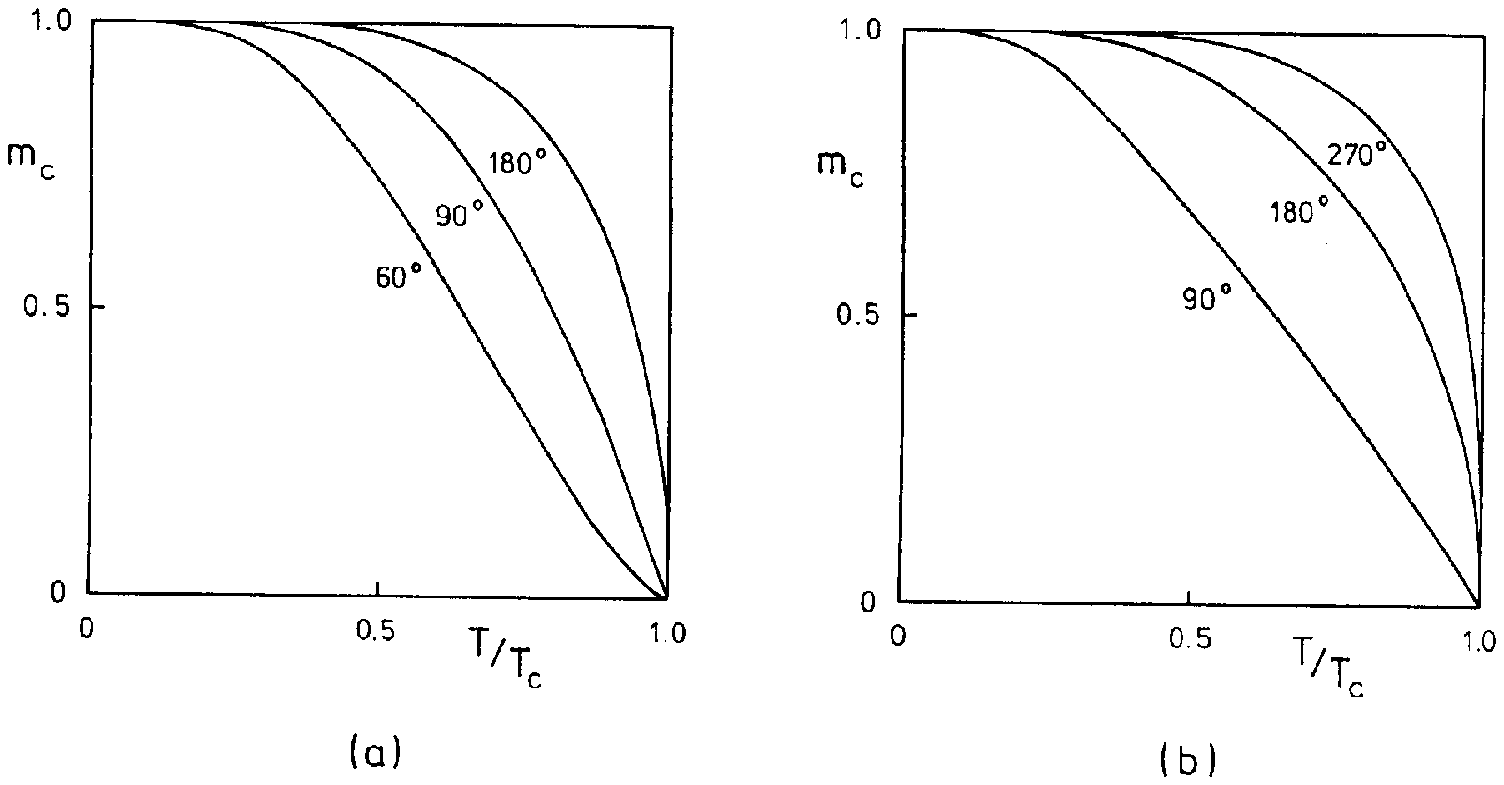}}
\bigskip
\Figcaption{3.1}{\srm Corner magnetization vs. temperature for
isotropic Ising lattices with various opening angles.
Except for the $\sst 60^\circ$ case all results refer to
square lattices. (a) edges along the bonds (b) edges along
the diagonals.}
\par
\endinsert
\endgroup
\par}

In Figure 3.1 results are shown for isotropic systems
with various opening angles at the corner.
The basic feature is that $m_c$ decreases with
decreasing $\theta$ for all non-zero temperatures.
There is no general explicit result for $m_c(t,\theta
)$ because even the transfer matrix calculations
involve some numerics in the end. Based on the
numerical results there is, however, a conjecture for a
square lattice with a $90^\circ$ corner and edges along
the bonds [33] $$
m_c=1-\demi (\coth K_1-1)(\coth K_2-1)\eqno(3.12)
$$
where $K_1$, $K_2$ are the couplings in the two
directions. The orientation of the edges and the
precise location of the chosen spin both affect $m_c$
[32] but not the critical exponent for which the formula
$$
\beta_c={\pi\over 2\theta}\eqno(3.13)
$$
was inferred from the data for isotropic systems with 
discrete angles [31].
Since $\nu=1$ here, this is also the result for $x_c$.
Thus one finds a simple $1/\theta$-dependence on the
angle as in the mean field treatment. If the system is 
anisotropic equation (3.13) is not immediately valid. 
However, by a rescaling as in Appendix B.1, one can 
make the system effectively isotropic. The opening
angle $\theta$ thereby changes to an effective angle 
$\theta_{eff}$, as illustrated in Figure B2, which 
need not be a simple fraction of $\pi$. With the new
angle, equation (3.13) is again satisfied. It was shown
later that this $\theta$-dependence is a consequence of 
conformal invariance (see Section 3.4).
 
Ising models covering the surface of a pyramid have
also been treated [31]. They are obtained by putting
together $k$ lattice segments, each with opening
angle $\alpha$, so that $\theta=k\alpha\neq 2\pi$.
The calculation in this case is much simpler since
the system is closed upon itself and has no edges of
free spins. One can then use corner transfer matrices
and derive the apex magnetization $m_a$ from a simple
trace formula, see (B18--19). The result takes the
form
$$
m_a(t,\theta )=\prod_{l=1}^\infty\tanh\left[\left(
2l-1\right)\varepsilon
(t){\theta\over 2\pi}\right]\eqno(3.14)
$$
where $\varepsilon (t)\sim1/\ln (1/t)$ for $t
\rightarrow 0$. Curves for $m_a(t)$ look qualitatively
similar to those shown in Figure 3.1. The critical behaviour
can be found by converting the product in (3.14) into the
exponential of an integral. The logarithmic dependence of
$\varepsilon$ on $t$ then leads to a power-law behaviour
of $m_a$. From the way the angle $\theta$ enters
it follows that the apex exponent
$\beta_a=x_a$ is related to the bulk value via
$$
\beta_a={2\pi\over\theta}\beta.\eqno(3.15)
$$
This result also follows from conformal invariance. For
a finite system at the bulk critical point, the
variable $t^{-1}$ in $\varepsilon$, which is proportional 
to the correlation length, is~replaced by the
size $L$ [35] so that $m_a$ varies as $L^{-x_a}$ as in
(A8). 
\Subsection{Other systems}
The corner geometry has also been investigated for
the $O(N)$-model in the limit $N\rightarrow 0$. In
this case the spin correlation function is related
to self-avoiding walks (SAW's) on the underlying
lattice via [36]
$$
G(\bi r, \bi r')=\sum_{N=0}^\infty K^NW_N(\bi r,
\bi r')\eqno(3.16)
$$
where $K$ is the coupling constant of the vector
model and $W_N(\bi r, \bi r')$ denotes the number of
SAW's with $N$ steps, going from $\bi r$ to $\bi r'$.
By counting walks up to $N\sim 25$ for several angles,
Guttmann and Torrie [37] found in two dimensions
$$
x_c={5\pi\over 8\theta}.\eqno(3.17)
$$
For wedges in three dimensions they found the edge index
$$
x_e=0.5+0.847{\pi\over\theta}.\eqno(3.18)
$$
So far this is the only result in $d=3$ beyond mean field
theory.
 
One can also go the other way and derive properties of
$W_N$ from those of $G$, invoking conformal invariance
[38, 39]. This was also done for polymers in
corners [40]. For the walks one obtains, for example,
the spatial moments of $W_N$. These were studied for
systems in $d=2$, $3$ with excluded half-infinite lines,
which are special cases of corners and cones,
respectively [41, 39].
\Subsection{Conformal results}
In two dimensions the critical behaviour of a wedge can
be related to that of a half-plane if the system shows
conformal invariance. This is done through the conformal
transformation $w=z^{\theta/\pi}$ [18, 31]. 

The critical correlation function $G(w,w')$ in the wedge 
is found by transforming (2.34) according to (A14). For 
$w'$ near the corner and $w$ far in the bulk, it varies as
$$
G(w,w')\sim{1\over\mid w\mid^{x+x_c}}\eqno(3.19)
$$
where $x_c$ is given by
$$
x_c={\pi\over\theta}x_s.\eqno(3.20)
$$
Thus the corner index is related to the surface index
via the geometrical factor $\pi/\theta$ which also
appears in the mapping. The results (3.13), (3.17) are
special cases corresponding to the surface exponents 
$x_s=1/2$ and $x_s=5/8$ in the two models. This confirms 
that conformal invariance
actually holds in these systems. By relating other
exponents to $x_c$ and repeating this for different
scaling operators one can then derive the whole set
of wedge indices.
 
In the same way a conical system is obtained by mapping
a full plane onto a wedge with periodic boundary
conditions via $w=z^{\theta/2\pi}$. The correlation 
function in this geometry is then obtained from the 
one in the full plane via (A14). This leads to
$$
x_a={2\pi\over\theta}x\eqno(3.21)
$$
relating apex and bulk indices, as already found in the
treatment of the Ising model, cf.~(3.15).
{\par\begingroup\parindent=0pt\leftskip=1cm
\rightskip=1cm\parindent=0pt
\baselineskip=12truept\sips{12.4truecm}{8.9truecm}
\midinsert
\centerline{\epsfbox{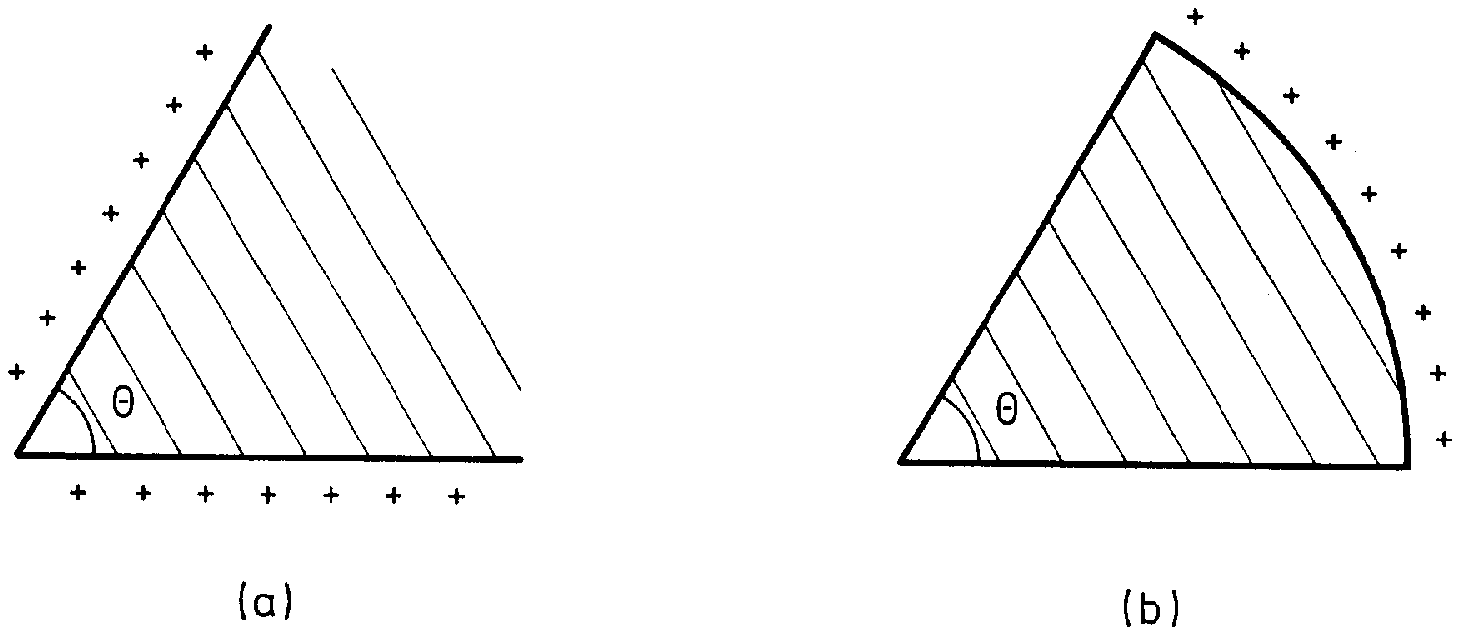}}
\bigskip
\Figcaption{3.2}{\srm Two-dimensional corner geometry with two
types of boundary conditions.}
\par
\endinsert
\endgroup
\par}
 
The mapping from the half-plane can also be used to
determine the universal critical profiles, in the wedge
geometry, which arise by fixing the variables
along certain boundaries. Two situations are shown in
Figure 3.2. For identical edges (Figure 3.2a) one finds
[22]
$$
<\phi (\rho,\alpha)>\,\sim\left[\rho\sin \left({\pi\alpha
\over\theta}\right)\right]^{-x}
\qquad\qquad  0\leq\alpha\leq\theta.\eqno(3.22)
$$
If one of the edges is free, the surface index $x_s$
also enters. In the half-plane this corresponds to a
situation where one has free variables along half of
the real axis and fixed ones along the other half.
The profiles for such a case have so far only been
determined for Ising and Potts
models [42--44]. Taking these results one can 
also find the profiles for the case shown in
Figure 3.2b. The necessary mapping is $w=\zeta^{\theta/\pi}$
with $\zeta=a(\sqrt{z}-1)/(\sqrt{z}+1)$. This transforms the
upper $z$-plane onto a half-disc in the $\zeta$-plane and
onto a wedge with circular boundary of radius R in the $w$-plane.
Near the corner, the result for the profile is
$$
<\phi (\rho,\theta/2) >\,\sim {\rho^{x_c-x\pi/\theta }\over
R^{x_c}}\qquad\qquad \rho\ll R.\eqno(3.23)
$$
For the analogous conical system the profile only
depends on $\rho$ and follows from the result for a disc
[22]. The size dependence is then given by $R^{-x_a}$.
 
With complex mapping one can also study rounded corners.
In agreement with general predictions [5] one finds that
the rounding does not affect the corner exponents or
the asymptotic form of $G$ in (3.19).
 
Conformal invariance also predicts universal
contributions to the critical free energy which are
connected with a corner or an apex [45]. For a finite
two-dimensional system of characteristic size $L$,
the free energy has the form
$$
F=f_bL^2+f_sL+\cdots\eqno(3.24)
$$
where the terms written are the bulk and surface
contributions, respectively. For systems with Euler
number $\chi=0$ like cylinders or tori, the next term
is a universal constant, related to the Casimir effect
[46, 47]. For $\chi\neq 0$, however, as is the case
for simply connected domains, the next term is a
logarithm in $L$. Each corner leads to a contribution
$$
\Delta F_c={c\theta\over 24\pi}\left[ 1-\left({\pi
\over\theta}\right)^2\right]\ln L\eqno(3.25)
$$
where $c$ is the conformal anomaly characterizing the
universality class of the system. For the Gaussian
model ($c=1$) there is a close connection [48]
between this term and the contribution of a corner
to the eigenvalue spectrum of the Laplace operator as
studied by Kac and others [49--51]. Thus the same
geometrical factor appears in both cases. An apex of
a cone gives an analogous contribution as (3.25),
with the substitution $\pi/\theta
\rightarrow 2\pi/\theta$ in the bracket.
 
Adding the contributions (3.25) for a system in the
form of a rectangle gives $\Delta F=-(c/4)\ln L$.
For the Gaussian model with fixed boundary variables
this can be checked by a direct calculation [52, 53].
This result has been used to discuss the shape
dependence of the critical free energy [54].
For a polygon with a large number of edges, $\Delta F$
approaches the result
$$
\Delta F=-{c\chi\over 6}\ln L\eqno(3.26)
$$
valid for smooth boundary curves. Privman has given an
interpretation of the logarithm in terms of an interplay
of singular and non-singular contributions in the free
energy [55]. According to this argument such
terms can also arise from corners in three dimensions.
\si{\vfill\eject}{}
\Section{Parabolic shapes}
A qualitative change in the critical behaviour occurs
for systems which are asymptotically narrower than wedges
or cones. This was observed in a study of two-dimensional
systems bounded by algebraic curves of the form $v=\pm
Cu^\alpha$ [56]. A typical case is the simple parabola
($\alpha =1/2$) and, generally, such shapes will be
called parabolic. The wedge corresponds to the special
case $\alpha =1$. Three typical situations
are shown in Figure 4.1.
{\par\begingroup\parindent=0pt\leftskip=1cm
\rightskip=1cm\parindent=0pt
\baselineskip=12truept\sips{14truecm}{10truecm}
\midinsert
\centerline{\epsfbox{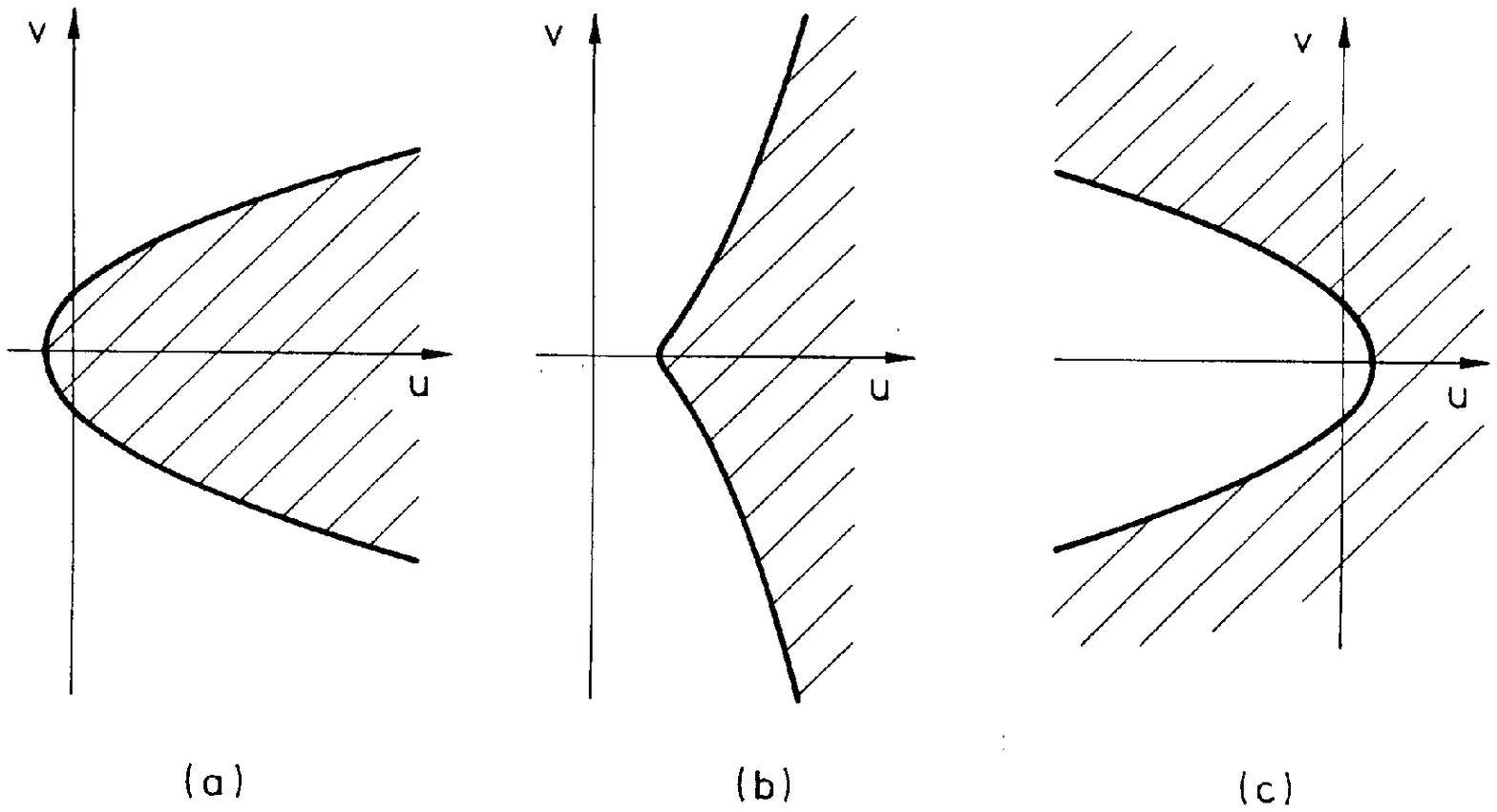}}
\bigskip
\Figcaption{4.1}{\srm Three types of two-dimensional systems
bounded by simple parabolic~curves.}
\par
\endinsert
\endgroup
\par}
 
Using conformal and transfer matrix methods the following
features were found. For $\alpha<1$ the critical
behaviour is no longer characterized by the usual power
laws but by stretched exponentials. Thus thermodyamic
quantities show essential singularities. Furthermore there is
non-universality since the amount of stretching is
determined by the value of $\alpha$. For $\alpha>1$,
on the other hand, one finds normal surface
critical behaviour. The difference between the two
cases can be attributed to the way the dimensional
parameter $C$ behaves under renormalization
for $\alpha<1$ and $\alpha>1$. The wedge geometry,
$\alpha=1$, is thereby seen to be the borderline case.
A corresponding distinction holds in higher dimensions.
For intrinsically anisotropic systems an analogous
classification holds but the marginal case is then
found for $\alpha\neq 1$.
\Subsection{Mean field theory}
The simplest example is a two-dimensional system with
parabolic boundary, $v^2=2pu+p^2$, as in Figure 4.1a. 
Its critical correlation function can be obtained by
direct calculation from (2.12) or using the transformation
$z=\i\cosh\pi\sqrt{w/2p}$ which maps the upper $z$-half-plane
on the interior of a parabola in the $w$-plane.
For points on the axis with $u'$ fixed
and $u$ large, one finds
$$
G(u,u')\sim\exp\left[ -\pi\left({u\over 2p}\right)^{1/2}
\right].\eqno(4.1)
$$
The decay of $G$ is therefore faster than in a wedge,
but slower than in a strip where $G$ varies as $\exp
(-\pi u/L)$, if the width is $L$. The power
$u^{1/2}$ in (4.1) is clearly related to the form
of the boundary. The length scale is set by the
parabola parameter $p$ or, equivalently, by the
quantity $C=\sqrt{2p}$.
 
In three dimensions, a similar result is obtained for
a paraboloid of revolution [57]. Working in parabolic
coordinates, one can express $G$ in terms of Bessel
functions [58]. On the axis it has the form, for $u>u'$
$$
G(u,u')=\sum_{m=1}^\infty A_m(u'){\rm
K}_0\left[\mu_m\left( {u\over
2p}\right)^{1/2}\right]\eqno(4.2) 
$$
where $\mu_m$ is the $m$-th zero of the Bessel function
$\J_0$. Therefore asymptotically
$$
G(u,u')\sim{1\over u^{1/4}}\exp\left[ -\mu_1\left(
{u\over 2p}\right)^{1/2}\right]\eqno(4.3)
$$
with $\mu_1\simeq 2.41$. Here the power in front of 
the exponential is also
related to the shape since it does not appear for a
cylinder. As in Section 3.1, the result may be 
interpreted in electrostatic terms. The induced charge
in the present case is more concentrated near the
source point $u'$ so that the decay of $G$ is faster
than in the conical geometry.
{\par\begingroup\parindent=0pt\leftskip=1cm
\rightskip=1cm\parindent=0pt
\baselineskip=12truept\sips{9.2truecm}{6.6truecm}
\midinsert
\centerline{\epsfbox{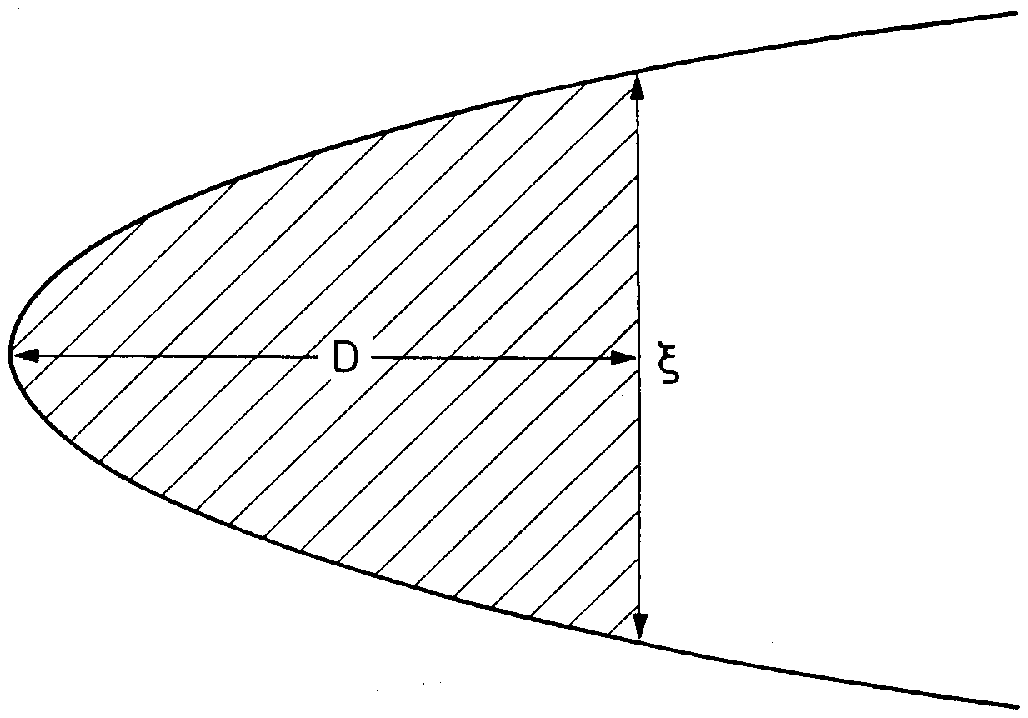}}
\bigskip
\Figcaption{4.2}{\srm Geometry of parabolic system near the tip.
The shaded portion of size $\sst D$ along the axis is the region 
where the shape governs the critical behaviour.}
\par
\endinsert
\endgroup
\par}
 
There exists no exact calculation of the order
parameter $m(\bi r)$ using (2.7). However, near
the critical point one may integrate the linearized
equation from the tip up to $u=D\sim\xi^2/p$, where the
width of the system becomes comparable to $\xi$ and
$m(\bi r)$ reaches its bulk value (Figure 4.2). In this 
region $m$ increases exponentially in $(u/2p)^{1/2}$
and, for fixed $u$
$$
m\sim\exp(-a\xi/p)\eqno(4.4)
$$
where $a$ is a constant. The order therefore vanishes
exponentially in $t^{-1/2}$ at $T_c$.
\Subsection{Conformal and scaling results}
In two dimensions one can go beyond mean field theory
and obtain results for arbitrary  models which are 
conformally invariant. Complex mapping from the upper 
$z$-half-plane allows to obtain the three types of shapes
shown in Figure 4.1. For $\alpha\!<\!1$ (Figure 4.1a) it is 
given by $z\!=\!\i\cosh(w/2p)^\alpha$. The critical correlation 
function is then found as in the case of a wedge.
For points on the axis and large values of $u$ [56]
$$
G(u,u')=A(u'){1\over u^{\alpha x}}\exp\left[ -{\pi x_s
\over 2C(1-\alpha )}u^{1-\alpha}\right].\eqno(4.5)
$$
The functional form of $G$ thus varies continuously
with the parameter $\alpha$. The fall-off is fastest
for $\alpha=0$ which corresponds to a semi-infinite strip
of width $2C$. With increasing $\alpha$ the
exponential becomes more and more stretched and approaches a
power law for $\alpha\rightarrow 1$, the wedge limit.
For $\alpha=1/2$, the mean field (Gaussian)
result (4.1) is reobtained by inserting the bulk and 
surface exponents $x=0$, $x_s=1$ (2.21). One also notes that the
three-dimensional result (4.3) has the same structure
as (4.5) and the power $1/4$ in (4.3) can be interpreted
as $\alpha x$, since $x=\demi$ is the mean field 
exponent in three dimensions.
 
By mapping the full $z$-plane one obtains systems with
identical upper and lower boundaries ("paraboloids").
As in the passage from the wedge to the cone in Section 3.4,
one then has to replace $\pi x_s$ by $2\pi x$ in (4.5).
For the Ising and Potts models one may also find, as in Section
3.4, the complete order parameter profile at the critical
point if the variables are fixed at the right end of the system. 
The size dependence of the order parameter near the tip is then
given, for both geometries, by the same exponential which
appears in $G$, with $u$ replaced by the size $L$.
 
For $\alpha >1$ (Figure 4.1b) the mapping has to be
changed into $z=\i(w^s-p^s)^{1/s}$ where $s=(\alpha -1)
/\alpha$. Then on the axis, asymptotically,
$$
G(u,u')\sim{1\over u^{x+x_s}}\eqno(4.6)
$$
which is the result of the half-plane. Finally, if a
parabolic piece with $\alpha<1$ is cut out of an infinite system
(Figure 4.1c) one obtains
$$
G(u,u')\sim{1\over u^{x+x_c}}\eqno(4.7)
$$
with the corner exponent $x_c(2\pi )=x_s/2$ for a cut (see (3.16)).
 
These results can be understood if one considers the
behaviour of the boundary curves under a rescaling
$\bi r\rightarrow\bi r'=\bi r/b$. Their functional form,
being a power law, is invariant but the parameter $C$
changes according to
$$
C'=b^{\alpha -1}C.\eqno(4.8)
$$
For $\alpha>1$, $C$ grows under renormalization and
the boundary approaches a straight~line. For $\alpha=1$,
$C$ is invariant and thus a marginal variable (the angle
of~Section~3). For $\alpha<1$, $C$ decreases and 
the system approaches either a cut geometry or becomes
locally one-dimensional. In the latter case there is a
destruction of long-range order which leads to the
particular features found above. The same considerations
hold for paraboloids in three dimensions. A parabolic
cylinder, on the other hand, renormalizes to a half-plane
in the latter case which has long-range order. This order,
however, appears only at a lower temperature and stretched
exponentials at the original transition are still possible.
 
The variable $1/C$ may be considered as a scaling field
which plays the same r\^ole as the variable $1/L$ in 
finite-size scaling (Appendix A.1). It  has dimension 
$1-\alpha$ and vanishes at the fixed point
of the half-plane geometry. It appears in
scaling relations like the one for the  magnetization
on the axis (cf (A31))
$$
m\left( t,u,{1\over C}\right) =b^{-x}m\left( b^{1/\nu}t,
{u\over b},{b^{1-\alpha}\over C}\right)\eqno(4.9)
$$
leading to the functional form
$$
m\left( t,u,{1\over C}\right) =t^{\beta}g\left({u\over
t^{-\nu}},{t^{-\nu (1-\alpha )}\over C}\right).\eqno(4.10)
$$

Assuming that, as in the mean field case, the tip 
magnetization is proportional to the critical correlation 
function between $u\!=\!0$ and $u\!=\!D~\sim~(\xi/C)^{1/\alpha}$ 
(cf Figure 4.2), one obtains for $\alpha\!<\!1$ from (4.5) 
$$
m(t)\sim\exp\left[ -{a\over(1-\alpha)}\left(
{t^{-\nu (1-\alpha )}\over C}\right)^{1/\alpha}
\right].\eqno(4.11)
$$
An essential singularity as encountered here has
also been found in the six-vertex model [6] related to
two-dimensional ice problems and to the roughening
transition [59]. In this homogeneous system, however,
all relevant quantities (free energy, correlation
length and order parameter) vary in this way and the
transition, which has been called of
infinite order, is different from the present one.
 
\Subsection{Ising model}
Calculations have been done to confirm in particular
the behaviour (4.11) of the magnetization [56, 60]. For
this the corner transfer matrix method was adapted to
the parabolic geometry. Thus one works with the transfer
matrix relating the spins along the upper and the lower
boundary of the system. In the Hamiltonian limit this
leads  to an inhomogeneous quantum spin chain (Appendix B.2).
If one deals with a lattice, the boundaries
actually are step functions and the same then holds for
the coefficients in the chain Hamiltonian (B3). Within a
continuum interpolation they take the forms 
$$
h_n=(2n+\mu )^\alpha\qquad\qquad\lambda_n=\lambda
(2n+\nu )^\alpha\eqno(4.12)
$$
so that $h_n$ and $\lambda_n$ vary as $n^\alpha$,
reflecting the shape of the system. The parameters
$\mu$, $\nu$ allow for differences in the total number
of vertical and horizontal bonds at position $n$. For
the case $\alpha =1/2$, $\mu =0$, $\nu=2$, the chain
Hamiltonian can be diagonalized with the help of special
polynomials [56]. For other cases the problem was studied
numerically and via a continuum approximation [60]. It was
found that, for $\alpha <1$, the single particle eigenvalues 
$\omega_l$ vary as $(2l-1)/L^{1-\alpha}$ for a finite 
critical system and as $l^\alpha /\xi^{1-\alpha }$ near 
criticality, reflecting again the geometry. For a 
"paraboloid" the formula (B19) can then be used to obtain 
the magnetization at the tip. It reproduces the conformal 
prediction at criticality and gives
$$ 
m\sim t^{-(1-\alpha)/2}\exp\left[
-at^{-(1-\alpha )/\alpha}\right] \eqno(4.13)
$$
near the critical point. The exponential factor
coincides with the expression (4.11) for $\nu=1$.
Formally, the exponential dependence appears, because
the eigenvalues $\omega_l$ scale with a power of $t$
in contrast to the case of a wedge where they vary
logarithmically with $t$.

Calculations for the case of free boundaries have not yet
been done with this method. However, the
problem was studied via Monte Carlo simulations
and the same typical behaviour of $m$ was found [57]. 
One expects a different prefactor in this case
but the data did not allow its determination.
Physically, the exponential vanishing of $m$ can be
understood as follows: a system with $\alpha<1$ is
asymptotically narrower than any wedge and so the magnetization
close to the critical point should
lie below an arbitrary power law. 
\Subsection{Other systems} 
The previous considerations may be extended to systems
displaying anisotropic critical behaviour. In this case
the correlation lengths in two perpendicular directions
diverge with different exponents $\nu_\parallel$ and
$\nu_\perp$. Then rescalings have to
be performed with different factors $b_\perp =b$ and
$b_\parallel =b^z$, where $z=\nu_\parallel/\nu_\perp$
[61]. Choosing the symmetry axis of the figure along the
direction with $\nu_\parallel$, the parameter $C$ 
changes according to 
$$
C'=b^{z\alpha -1}C\eqno(4.14) 
$$ 
and marginal behaviour
occurs for $\alpha =1/z$. Examples are provided by systems
with uniaxial Lifschitz points [62], directed walks or
directed polymers [63]. In the last two cases $z=2$ and
the borderline geometry is the normal parabola or, in
three dimensions, the paraboloid. This situation was
studied in [64, 65].

{\par\begingroup\parindent=0pt\leftskip=1cm
\rightskip=1cm\parindent=0pt
\baselineskip=12truept\sips{9.2truecm}{6.6truecm}
\midinsert
\centerline{\epsfbox{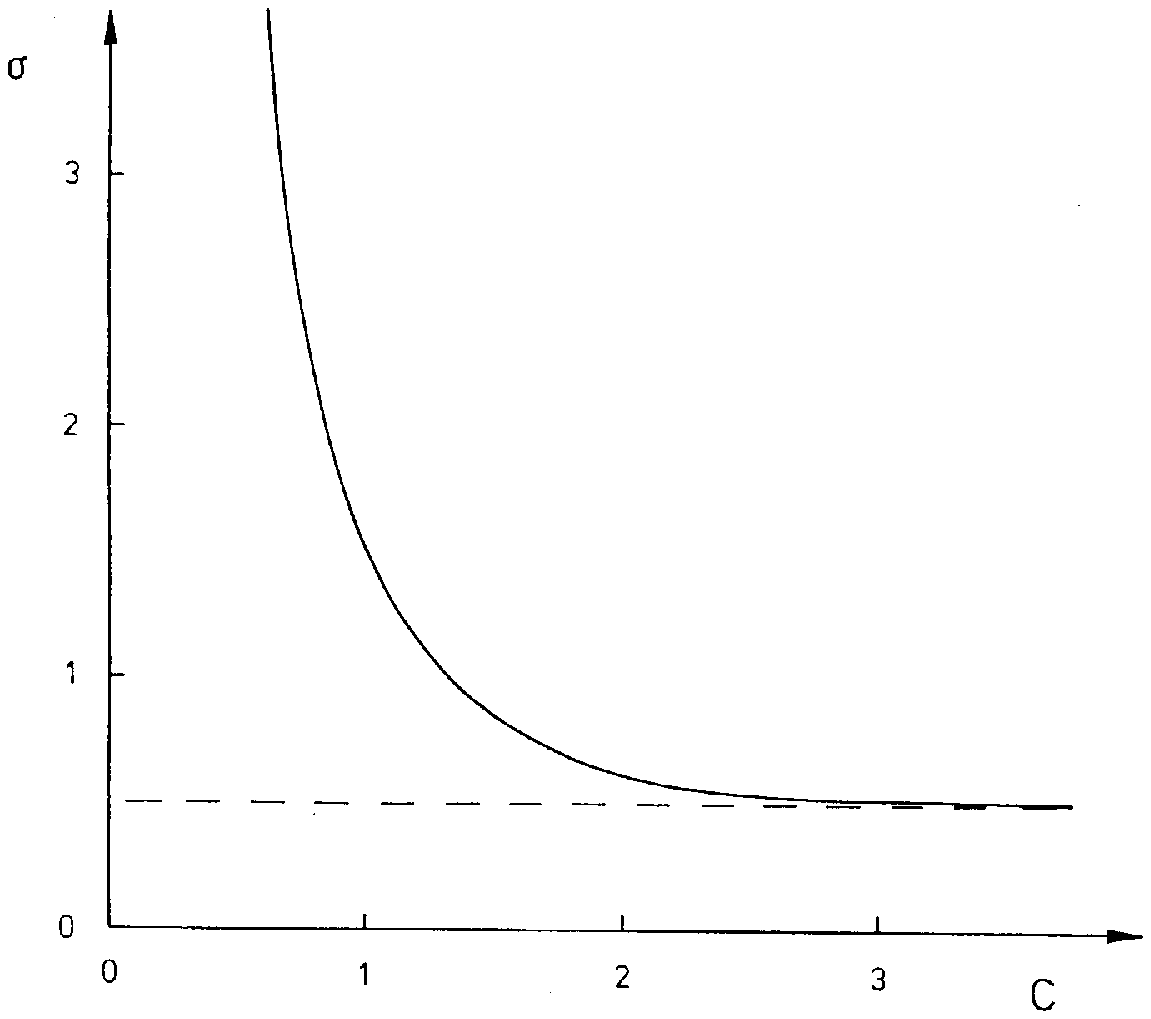}}
\bigskip
\Figcaption{4.3}{\srm Exponent $\sst\sigma$ for the directed
self-avoiding walk as a function
of the parabola parameter $\sst C$.}
\par
\endinsert
\endgroup
\par}
 
The simplest case is the directed self-avoiding walk
in two dimensions. It is obtained by considering a usual
one-dimensional random walk $x=x(t)$ as a path in the 
($x,t$)-plane. In the continuum limit, the relative number 
$P$ of walks between two points $(x,t)$ and $(x',t')$ 
follows from the diffusion equation via
$$
\left({\partial\over\partial t}-\demi{\partial^2\over
\partial x^2}\right) P(x,t\mid x',t')=\delta (x-x')
\delta (t-t').\eqno(4.15)
$$
This is the analogue of the Laplace equation for the
critical mean field correlation functions considered
before. At $x\!=\!\pm\! Ct^\alpha$ a time-dependent boundary
condition is posed. In particular, if paths reaching the 
surface are terminated there, a necessary condition to obtain
equal weights for all the walks with $N$ steps in the 
original problem, one comes back to the Dirichlet boundary 
condition $(P=0)$ considered before. 

In the marginal case $\alpha=1/2$, if the paths start at the
tip, the solution for large~$t$~is~[64] 
$$
P(x,t\mid 0,0)\sim{1\over t^\sigma}\F\left(\sigma ,\demi ;
-{x^2\over 2t}\right).\eqno(4.16)
$$
The confluent hypergeometric function F has to vanish
at the boundary, from which the exponent $\sigma$
follows. It is shown in Figure 4.3. For $C\rightarrow
\infty$ the usual Gaussian distribution with $\sigma
=1/2$ is reobtained. As $C$ is lowered, the exponent
$\sigma$ increases monotonously, thus showing the
expected non-universal behaviour. The decay of $P$ with $t$
thereby becomes faster in narrow systems, as found before 
for $z=1$ in wedges and cones. Other exponents can be 
defined as usual. For example the survival probability, 
i.e. the relative number of walks reaching the time
$t$ (or having $N$ steps in the discrete case) varies
as $t^{\gamma_p-1}$ with a susceptibility exponent
$\gamma_p=3/2-\sigma$.
 
For the case $\alpha<1/2$, the effect of the geometry is 
relevant and the form
$$
P(x,t\mid 0,0)\sim{1\over t^\alpha}\exp\left( -{\pi^2
\over 8C^2}{t^{1-2\alpha}\over 1-2\alpha}\right)
\eqno(4.17)
$$
was deduced. It shows the same stretched exponential
behaviour as the correlation functions in the previous
sections. For $\alpha>1/2$, the boundary is in a region
where $P$ is exponentially small anyway and therefore
does not affect the behaviour in agreement with the scaling
contained in (4.14). The results for the directed polymer
problem, obtained on a discrete lattice via transfer
matrix methods, are very similar [65].
{\par\begingroup\parindent=0pt\leftskip=1cm
\rightskip=1cm\parindent=0pt
\baselineskip=12truept\sips{10truecm}{7.1truecm}
\midinsert
\centerline{\epsfbox{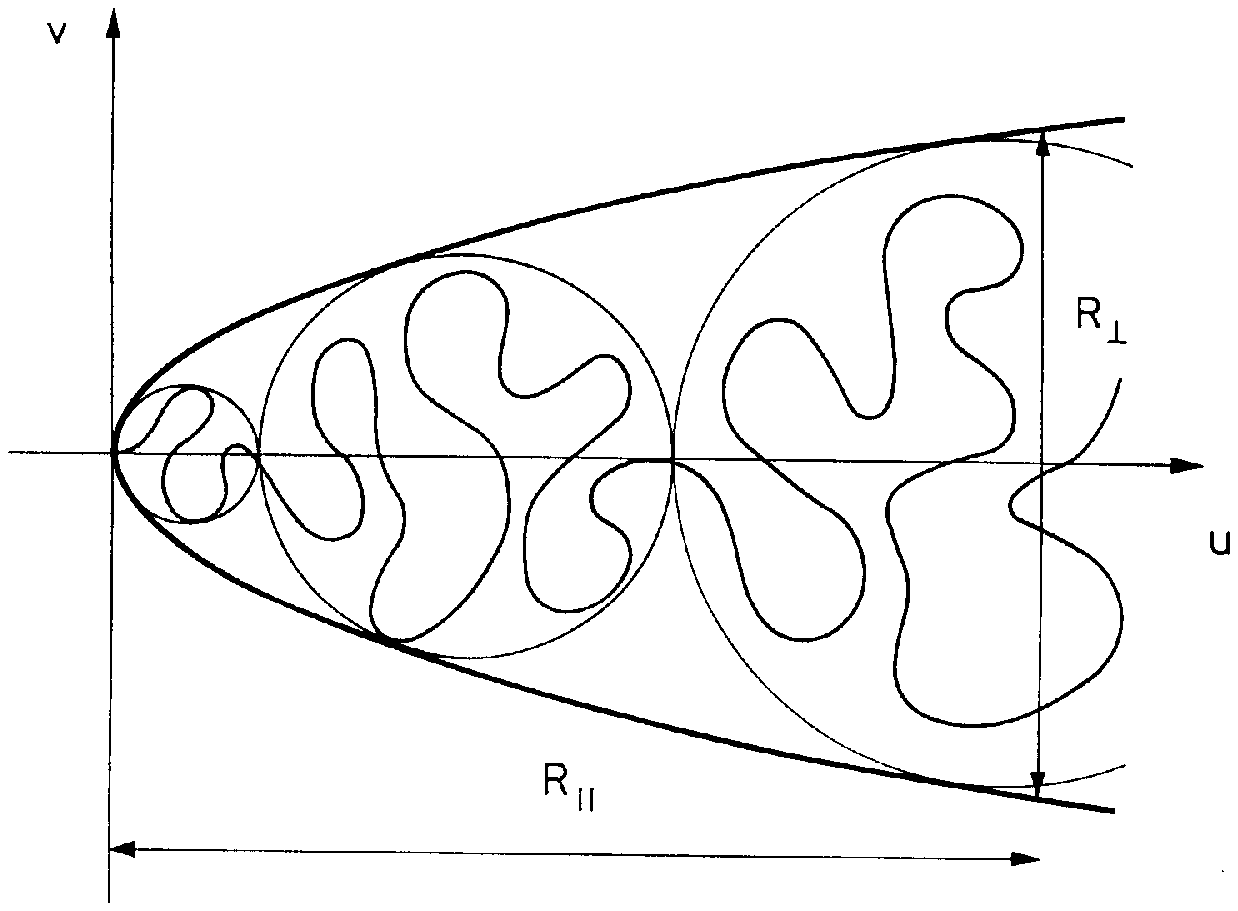}}
\bigskip
\Figcaption{4.4}{\srm Blob picture for a polymer confined inside
a parabola: the chain configuration results from the piling
of self-avoiding fractal blobs (thin circles).}
\par
\endinsert
\endgroup
\par}
{\par\begingroup\parindent=0pt\leftskip=1cm
\rightskip=1cm\parindent=0pt
\baselineskip=12truept\sips{10.8truecm}{7.7truecm}
\midinsert
\centerline{\epsfbox{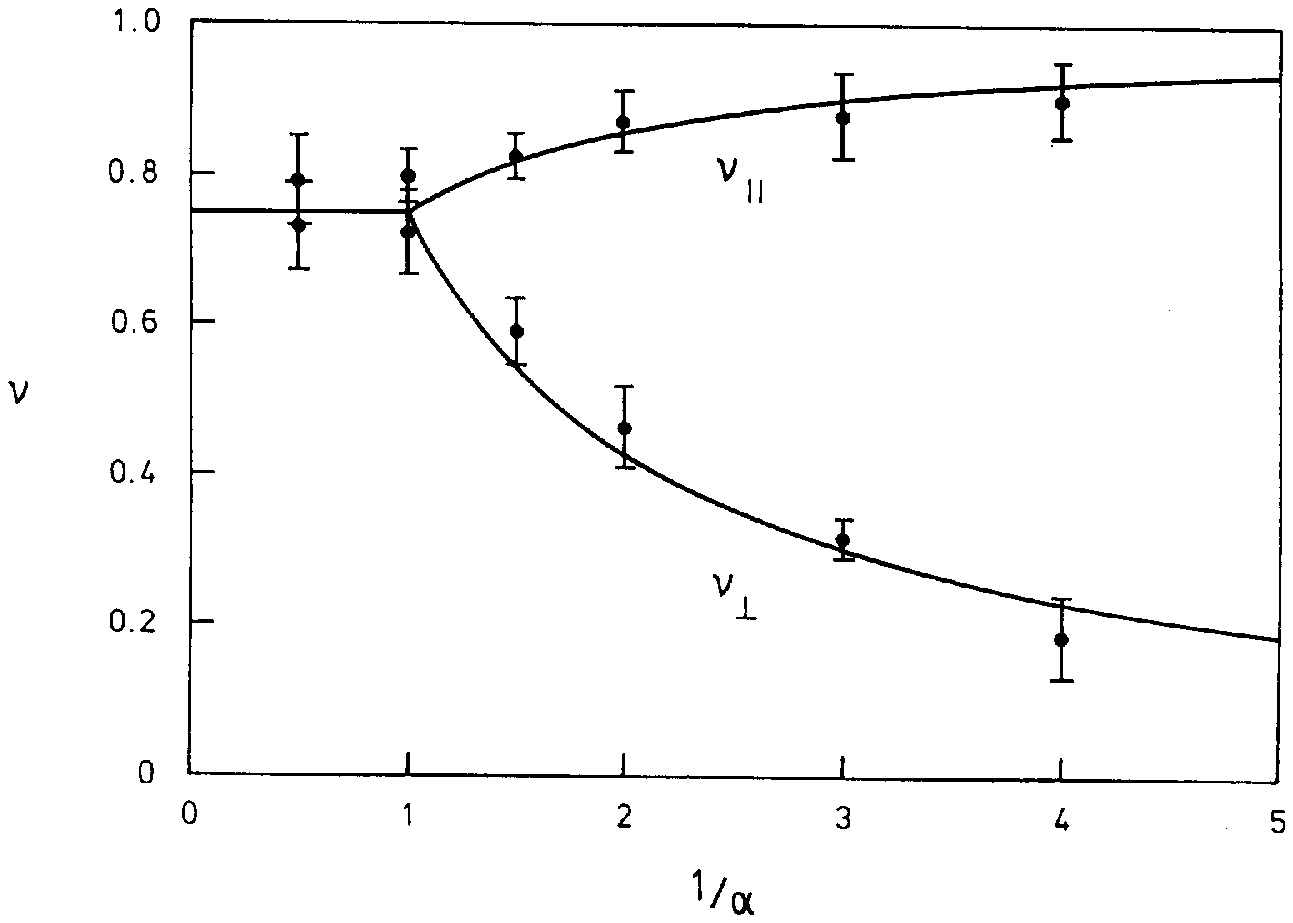}}
\bigskip
\Figcaption{4.5}{\srm Variation with $\sst\alpha^{-1}$ of the SAW 
exponents $\sst\nu_\parallel$ and $\sst\nu_\perp$ in two dimensions
for a chain confined inside a generalized parabola. The 
points are the results of Monte-Carlo simulations on a square 
lattice. The surface geometry is relevant and induces an exponent
anisotropy when $\sst\alpha<1$.}
\par
\endinsert
\endgroup
\par}
 
Finally, normal SAW's with parabolic boundaries have been
studied [66]. It was found that, for $\alpha<1$, this
originally isotropic system becomes anisotropic. A chain
with $N$ monomers then involves two radii of gyration:
$R_\parallel\sim N^{\nu_\parallel}$ along the axis of 
the system and $R_\perp\sim N^{\nu_\perp}$ in the transverse 
direction. The chain configuration can be described using
a blob picture [36] as shown in Figure 4.4. It results 
from the piling up of self-avoiding blobs inside the parabola.
Within each blob the correlations are the same as in an 
unconfined chain. The radius $R_\perp$ is given by the
transverse size of the system at a distance $R_\parallel$
from the tip so that $\nu_\perp=\alpha\nu_\parallel$. This 
implies a new anisotropic fixed point for which, according 
to (4.14), the geometry remains invariant under renormalization
since $z=1/\alpha$. The evolution of the blob size along the
axis of the system simply follows from the geometry and 
$R_\parallel$ is easily obtained as a function of $N$ 
leading to
$$
\nu_\parallel ={\nu\over\alpha+\nu (1-\alpha )}\eqno(4.18)
$$
where $\nu$ is the bulk exponent ($\nu =3/4$ in $d=2$).
Monte-Carlo simulations on a square lattice (Figure 4.5) 
confirmed this picture and gave results
compatible with (4.18). An accurate calculation of the
correlation function like (4.16) has not yet been done.

\Section{Extended defects at surfaces}
In the last sections, systems were treated where the
geometry modifies the critical behaviour. We now turn to
problems where a smoothly varying inhomogeneity in the
couplings, decreasing from the surface, may influence the
local critical properties. In real systems such an
inhomogeneity could result from surface induced elastic
deformations of the lattice. Thus one studies
couplings which deviate from the bulk ones by an amount
$$
\Delta K(y)={A\over y^{\omega}} \eqno(5.1)
$$
$y$ being the distance from the surface.

Under a scale transformation $\bi r'\!=\!\bi r/b$ the coupling
inhomogeneity, being an energy-like perturbation, transforms as
$\Delta K'(y')\!=\!b^{1/\nu}\Delta K(y)$, therefore~[67,~68]
$$
A'=b^{{1/\nu}-\omega}A \eqno(5.2)$$
where $\nu$ is the critical exponent of the correlation length.
Thus, for $\omega>1/\nu$, $A$~decreases and one expects the same
local critical behaviour as at a free surface. On the other
hand, in the marginal case ($\omega=1/\nu$) and in the relevant
case ($\omega<1/\nu)$ scaling theory predicts local
properties different from those of a simple free surface.

\Subsection{Ising model, general results}
An extended defect of the form given in (5.1) was first
studied by Hilhorst and van Leeuwen [69] in the
two-dimensional Ising model. They considered the problem
on a triangular lattice at the critical point. The nearest
neighbour couplings $K_1(y)$ parallel to the surface and 
the diagonal ones $K_2(y)$ were different from their bulk 
values as in (5.1) with amplitudes such that
$\Delta K_1(y)/\Delta K_2(y)=2\sinh (2K_1)/\cosh (2K_2)$.

The calculation of surface quantities was based on the
repeated application of the star-triangle transformation [70]
used also in calculations of the corner magnetization 
(Section 3.2). From the evolution of the couplings one
can deduce the surface magnetization $m_s$ and the surface
spin correlation function $G_\parallel$ for any type of
inhomogeneity in a numerical, iterative way. For smoothly
varying perturbations as in (5.1), the asymptotic
behaviour of the couplings after a large number of
iterations and consequently the surface critical properties
of the model can be determined exactly.

The results for the  critical behaviour [69, 71, 72] are in 
complete agreement with the scaling arguments in the previous 
section. In detail they are:

   i) For $\omega\!>\!1/\nu\!=\! 1$, the perturbation is
irrelevant, $m_s(t)\!\sim\! t^{1/2}$ and $G_{\parallel}(r)\sim r^{-1}$
as in the homogeneous, semi-infinite system.

   ii) In the marginal case, $\omega\!=\!1$, when $A$ is smaller than 
a critical value $A_c>0$, the decay of $G_{\parallel}(r)$
is non-universal with $2x_s\!=\! 1\!-\! A/A_c$, while at $A\!=\!
A_c$ where $x_s\!=\! 0$ the decay is logarithmic,
$G_{\parallel}(r)\sim (\log r)^{-1}$. For strong enough enhancement 
of the couplings, $A>A_c$, there is a spontaneous surface 
magnetization at the bulk critical temperature. As $A$ approaches 
$A_c$ from above, it vanishes as $(A-A_c)^{1/2}$. The spin 
correlations in the surface approach their limiting value $m_s^2$
according to a power law with the non-universal exponent
$2x_s=A/A_c-1$.

   iii) For $\omega\!<\! 1$ the perturbation is relevant. For any
$A\!>\! 0$, there is a spontaneous surface magnetization at the
bulk critical point which vanishes as $A^{1/[2(1-\omega)]}$
as $A$ approaches zero. The correlation function $G_\parallel$
has a stretched exponential form for any sign of $A$
$$
G_{\parallel}(r)\sim \exp\left[-(r/\xi_{\parallel})^{1-
\omega}\right]\qquad\qquad\xi_{\parallel}\sim |A|^{-1/
(1-\omega)}. \eqno(5.3)
$$

The problem was also studied on
the square lattice using Pfaffian methods [73,~74]. In this
case the couplings parallel to the surface were chosen
to be constant $K_1(y)=K_1$, while the perpendicular ones
were modified as in (5.1), writing
$$
\Delta K_2(y)={\overline{A}\over y^\omega},\qquad\qquad
\overline{A}={A\over 4}\sinh (2K_2).\eqno(5.4)
$$

For the marginal problem ($\omega\!=\! 1$) the complete set
of surface exponents, both those defined at the critical
point and those associated with the approach to criticality,
was determined. Furthermore, including a surface
magnetic field $h_s$, the susceptibility exponents were
also calculated. All exponents vary continuously with $A$.
At the critical point the results are analogous to those
obtained on the triangular lattice. The critical exponents 
have the same dependence on $A/A_c$ in both cases. This is 
a kind of "weak universality" as observed also in other 
models containing a marginal operator [75].

The critical behaviour is anomalous in the regime of spontaneous
surface order $A>A_c$. Then the scaling is anisotropic for $T>T_c$ 
and the local exponents are asymmetric, i.e. different for $T<T_c$ 
and $T>T_c$, respectively. This unusual behaviour can be
explained [76] assuming that the local energy density
is a "dangerous irrelevant variable" [77] for $T>T_c$,
whereas it is harmless in the region $T\le T_c$.

\Subsection{Ising model boundary magnetization}
For the Ising model on the square lattice, the surface
magnetization at zero surface field is easily obtained 
using the transfer matrix method (Appendix B.1) [78, 79, 33]. 
It has been calculated first for the homogeneous system and 
later for inhomogeneous ones. The quantity $m_s$ is deduced
from the asymptotic value of the correlation function $G_\parallel$.
In the anisotropic limit from (B11) it is found to be equal 
to the matrix-element of the operator $\sigma_1^x$ between
the ground state $|0>$ and the first excited state $|1>$ of the
Hamiltonian in (B3)
$$
m_s=<1|\sigma_1^x|0>\eqno(5.5)
$$
In the general case, this has to be multiplied by $C_1\!=\!\cosh K_1^*$ 
where $K_1^*$ is the dual coupling defined
below (B2). This expression holds in the low-temperature phase
where the state $|1>$ is degenerate with the ground state in the
thermodynamic limit.

Using the fermion techniques described in Appendix B.1 one can
express the matrix element
by the surface component of the eigenvector ${\bf \Phi}_s$ corresponding
to the smallest excitation energy in (B8)
$$
<1|\sigma_1^x|0>=\Phi_s(1).\eqno(5.6)
$$
Since the vector ${\bf \Phi}_s$ must be normalized, its
component $\Phi_s(1)$ is only non-vanishing in the thermodynamic
limit if the eigenstate is localized near the surface. Such a state,
with a localization length diverging at the critical point, is known 
to exist below $T_c$ [80, 78, 79].

Using (B8) and (B9) with $h_n\!=\!1$, the surface magnetization of
a very anisotropic, semi-infinite system can be expressed in closed form
$$
m_s=\left[ 1+\sum_{l=1}^{\infty} \prod_{n=1}^l
\lambda_n^{-2}\right]^{-1/2} \eqno(5.7)
$$
Including a factor $C_1$ and putting 
$$
\lambda_n={\tanh K_2(n)\over\tanh K_1^*}. \eqno(5.8)
$$
the expression also holds outside the anisotropic limit.
Close to the critical point, $\Phi_s(n)$ extends a long way 
into the bulk, therefore its normalization and thus $m_s$ are 
determined by its asymptotic behaviour for $n\gg 1$. For the 
couplings (5.4), in the extreme anisotropic limit, this gives
$$
\lambda_n \simeq \lambda \left(1+{A \over 2}n^{-\omega}
\right) \eqno(5.9)
$$
where $\lambda=1$ at the bulk critical point while
$\lambda>1$ and $\lambda<1$ correspond to $T<T_c$ and
$T>T_c$, respectively.

For the homogeneous model ($A=0$) the sum in (5.7) is
a geometric series. This gives $m_s\!=\! C_1(1-\lambda^{-2})^{1/2}$ 
which is the result first found by McCoy and Wu [3]. Near the
critical point $m_s\sim t^{1/2}$ so that the magnetic surface
exponent is $\beta_s\!=\! 1/2$. For $\omega>1$ the products in  
(5.7) are convergent, thus for large $n$ the behaviour
of $\Phi_s (n)$ is the same as for $A=0$. As a
consequence $\beta_s$ remains unchanged and the perturbation
is irrelevant.
When $\omega\!<\!1$, $\Phi_s(n)$ varies asymptotically as
$\exp \left[-(A/2)n^{1-\omega}\right]$ corresponding to a
localized state for $A>0$. Then there is a spontaneous
surface magnetization at the critical point which 
vanishes as $A^{1/[2(1-\omega)]}$ when $A$ goes to zero,
in accordance with results on the triangular lattice. For
$A<0$ on the other hand, there is no surface order at the 
critical point and $m_s$ vanishes with an essential singularity
according to
$$
m_s\sim \exp \left[-\alpha(A,\omega)t^{-(1-\omega )/\omega}
\right].\eqno(5.10)
$$
In Figure 5.1 the temperature dependence of $m_s$ is shown 
for an isotropic bulk system and various values of $A$.
{\par\begingroup\parindent=0pt\leftskip=1cm
\rightskip=1cm\parindent=0pt
\baselineskip=12truept\sips{11truecm}{7.8truecm}
\midinsert
\centerline{\epsfbox{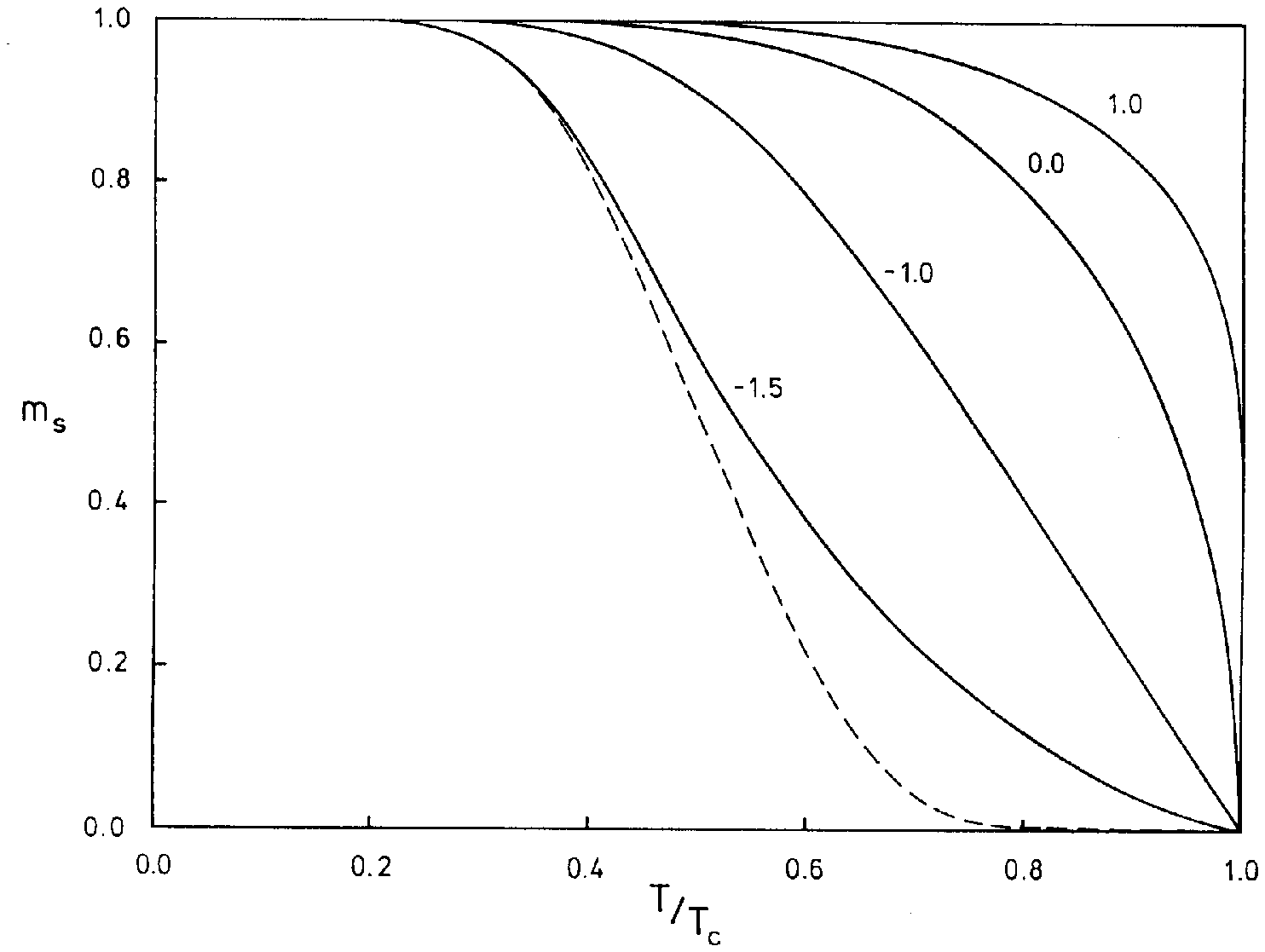}}
\bigskip
\Figcaption{5.1}{\srm Boundary magnetization $\sst m_s$ vs.
temperature in the Hilhorst-van~Leeuwen
model. The full lines correspond to $\sst\omega=1$ and various
values of $\sst A$, the dashed line to $\sst\omega=0.25$, 
$\sst A=-1.5$.}
\par
\endinsert
\endgroup
\par}

In the marginal case $\omega=1$, a closed form for $m_s$
can be obtained if (5.9) is changed into $\lambda_n=\lambda
(1-A/2n)^{-1}$. Then
$$
m_s=\left[ \F\left(1-{A\over 2},1-{A\over 2};1;
\lambda^{-2}\right)\right]^{-1/2} \eqno(5.11)
$$
where F$(a,b;c;z)$ denotes the hypergeometric function
[81].
Expanding (5.11) in powers of $t=1-\lambda^{-2}$ one
obtains 
$$
m_s={\Gamma\left( 1-{A\over 2}\right) \over \sqrt
{\Gamma (1-A)}}t^{(1-A)/2}\qquad\qquad A<1 \eqno(5.12)
$$
thus $\beta_s=(1-A)/2$. Expressions for $A\geq 1$ can 
be obtained in the same way.

\Subsection{Conformal invariance for the marginal case}
In section (2.2) it was shown that the special conformal 
transformation (2.26) which leaves the surface invariant
can be used to obtain information on the critical 
behaviour. For the system considered here the geometry 
is the same and the inhomogeneity (5.1) transforms into 
itself in the marginal case $\omega=1/\nu$ [82]. This suggests
that conformal invariance holds in the same way as for
the homogeneous case. In particular one can then map 
the half-space onto a strip as in (A22) and investigate
the specturm of the transfer matrix in this
finite geometry.
Under this mapping the inhomogeneity is transformed  into 
a sinusoidal form [83]
$$
\Delta K(u)=\overline{A}\left[{L \over \pi}\sin\left(
{\pi u \over L}\right)\right]^{-\omega} \eqno(5.13)
$$
where $L$ is the width of the strip and $u$, with $0<u\le L$,
is the transverse coordinate. $\Delta K$ does not depend
on the position along the strip. 

For the Ising model, a strip with couplings given by (5.13)
was studied at the bulk critical point by two different methods.
The case of a triangular lattice was treated in [83] by a 
numerical method based on the star-triangle transformation
[84]. For the square lattice, the spectrum of the transfer
matrix was obtained analytically in the extreme anisotropic
limit [83]. Thus the Hamiltonian $H$ as in Appendix B.1 was 
investigated for couplings given by $h_n=1$, $\lambda_n=
1+\Delta K(n)/2$. The matrix equation (B8, B9) was transformed
into a differential equation which is solved in terms of 
hypergeometric functions. In this way the lowest gaps, which 
are of order $1/L$, were determined in a continuum approximation 
as in [85].
{\par\begingroup\parindent=0pt\leftskip=1cm
\rightskip=1cm\parindent=0pt
\baselineskip=12truept\sips{11.2truecm}{8truecm}
\midinsert
\centerline{\epsfbox{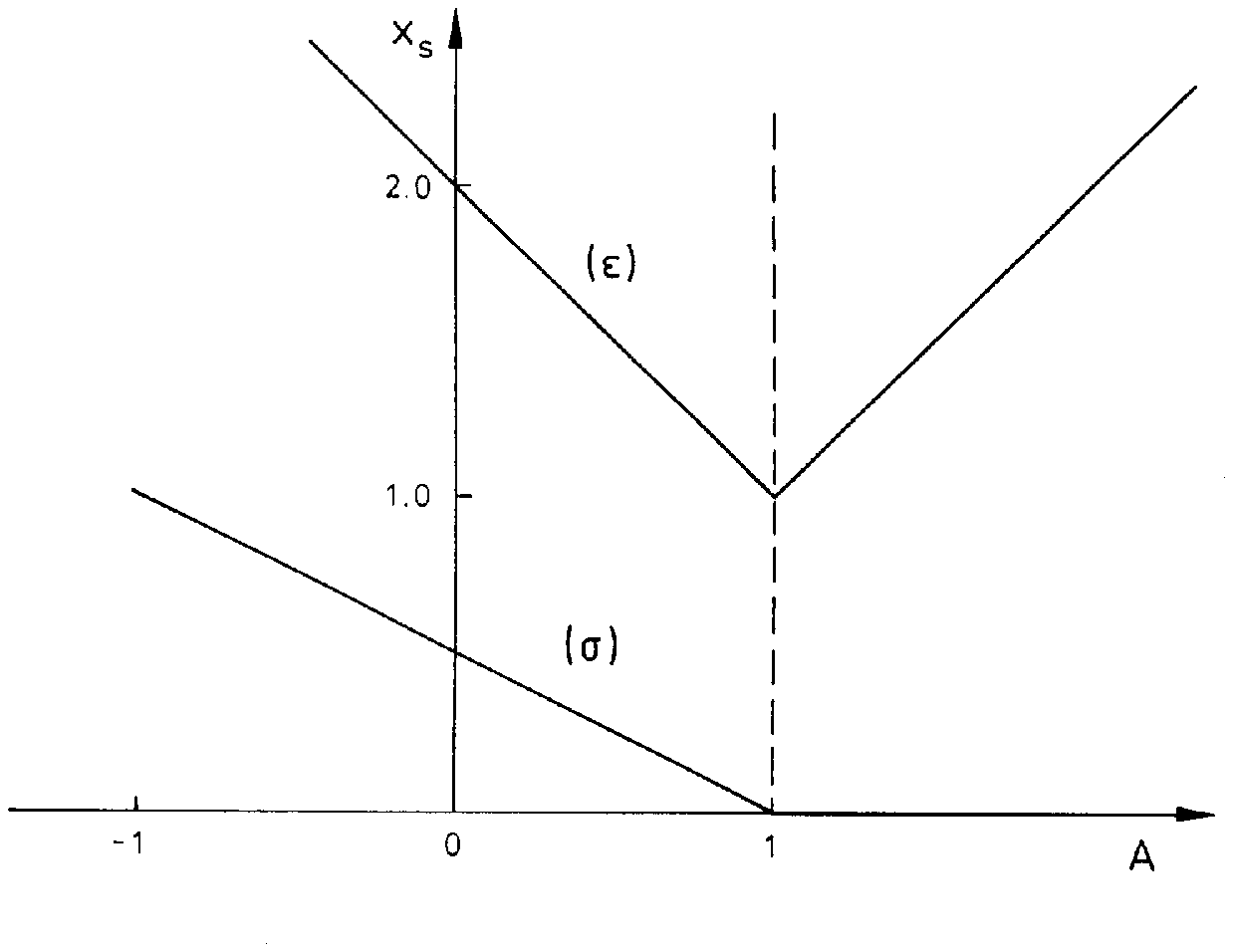}}
\bigskip
\Figcaption{5.2}{\srm Surface scaling dimensions for spin
($\sst\sigma$) and energy density ($\sst\varepsilon$) in the
Hilhorst-van Leeuwen model for $\sst\omega=1$ as a function
of the strength parameter $\sst A$.}
\par
\endinsert
\endgroup
\par}

The spectrum found in the second approach displays the tower-like 
structure typical of conformally invariant systems.{}From the 
lowest gaps surface exponents are deduced, using (A23), 
which are in agreement with those found from the calculations 
for the half-space [73, 74, 86]. They are shown in Figure 5.2 
as functions of the parameter $A$. For $A>1$ the
lowest gap vanishes like $L^{-A}$, i.e. faster than
$1/L$, in the large-$L$ limit. Hence the smallest
magnetic dimension vanishes. This is related to the fact
that for $A>1$ there is magnetic surface order at the bulk
critical point. These scaling dimensions are also in agreement
with the results of the first-order perturbation
expansion (Table C2 in Appendix C), where, according
to (5.4) and Appendix B.3, the defect amplitude becomes
$g=A/4$ in the continuum limit. Similar conclusions are reached 
from the calculations on triangular lattices. These results 
strongly support the validity of conformal invariance for
this particular inhomogeneous system. The correlation functions
in the half-space therefore will have the same structure (2.34)
as for a homogeneous system, but with a different scaling function.

\Subsection{Related problems}
The Hilhorst-van Leeuwen model has been generalized
recently by adding another marginal contribution with angular
dependence to the inhomogeneity in (5.1). The Hamiltonian
spectrum in the associated strip was determined exactly, noticing 
the supersymmetric aspects of the eigenvalue problem [87].

For the most general marginal extended perturbation of the Ising 
model with the form $Af(\theta)/\rho^{1/\nu}$ in polar
coordinates, the gap-exponent relation was shown to remain valid up to 
first order in the perturbation amplitude $A$ [82].  

In addition to the Ising model, the Gaussian model with
a defect of the form (5.1) has also been studied [88]. This
can be viewed as a mean field calculation which, however,
is as complicated as the one for the Ising model.
Using again the star-triangle transformation, the surface
correlation function was determined as in Section 5.1.
In the marginal case, which here corresponds to $\omega\!=\! 2$,
it decays with a non-universal exponent.

For systems with anisotropic scaling as described in Section
4.4, the relation (5.2) has to be used with the exponent $\nu$ in 
the direction of the inhomogeneity. For example, for
a polymer directed along a surface and interacting with it, 
the monomer fugacity can be modified as in (5.1). Then one has 
$\nu_\perp\!=\! 1/2$ in the transverse direction and the marginal 
case corresponds to $\omega\!=\! 2$. For weak enough surface 
interaction ($A<A_c$), the polymer is free and can be characterized
by critical exponents which depend~on~$A$. For stronger
surface interaction, it becomes localized near the
surface and the transition at $A\!=\! A_c$ is of infinite
order [89].

The defects considered so far extended from the surface
into the bulk. One may also consider perturbations which
are confined to the surface but vary along it. For the
Ising model the effect of an inhomogeneous field
$$
h(x,y)={A \over |x|^{\omega}}\delta(y) \eqno(5.14)
$$
was studied in [90]. Under rescaling, the amplitude transforms
as
$$
A'=b^{1-x_s-\omega}A \eqno(5.15)
$$
where the scaling dimension of $h_s$ in (A6) was used. For the
Ising model, $x_s=1/2$ (Table 2.1), which leads to marginal
behaviour when $\omega=1/2$. This also follows from Appendix C.2
with $d_{eff}=x_s$. Mapping the system onto a strip then leads to a
homogeneous boundary field of strength $\sqrt{\pi/L}A$
which shifts the levels of the Hamiltonian by an amount 
of order $1/L$. The magnetic surface exponent determined via (A23)
varies continuously between $x_l\!=\! x_s\!=\! 1/2$ for $A\!=\! 0$ 
and $x_l\!=\! 2$ for $A\!\to\! \infty$. The latter case corresponds 
to fixed boundary conditions and is also realized when the 
perturbation is relevant. The correlation function decays towards 
a finite value in the strip and the exponent refers to the
connected part.

A special form of inhomogeneity is present in  quasiperiodic
systems which interpolate between periodic and random systems and
where the perturbation is spread over  the whole volume. 
Results have recently been obtained for two-dimensional layered
systems with quasiperiodic or aperiodic
modulations. The relevance or irrelevance of the aperiodicity is
connected to the strength of the fluctuations in the couplings~
[91] and the corresponding criterion is obtained~[92] through a 
generalization of Harris's argumentation for random systems~[93].
Exact results have been derived for the specific heat~[91] and the
surface magnetization~[94, 92] for different aperiodic
two-dimensional Ising models and their related quantum chains, as
well as for directed polymers~[92]. It was found that irrelevant
aperiodicities indeed do not modify the critical behaviour,
while for relevant perturbations the thermodynamic quantities show
essential singularities at the critical point like in (5.10). As
expected varying exponents are found in the marginal case.

\Subsection{Scaling considerations for relevant
inhomogeneities}
In the previous sections we have seen that the critical behaviour 
of different systems in the presence of relevant perturbations
($\omega<1/\nu$) is very similar: critical correlations decay  in
a stretched exponential form, while the thermodynamic quantities
have an essential singularity at the critical point. In the
following we show how a general scaling theory together with
plausible assumptions can explain these observations.

We start by considering the scaling behaviour of the surface
magnetization
$$
m_s(t,A)=b^{-x_s} m_s(b^{1/\nu}t,b^{{1/\nu}-\omega}A)
\qquad\qquad A>0 \eqno(5.16)
$$
where we included the relevant scaling field $A$, with the transformation
law in (5.2). Putting $b=\xi$ in (5.16) one obtains $m_s$ in the form
$$
m_s(t,A)=t^{\beta_s}f\left({l\over\xi}\right)\qquad\qquad 
l=\vert A\vert^{-\nu/[1-\nu \omega]}\eqno(5.17)
$$
where $l$ is a characteristic length introduced by the inhomogeneity
which stays finite at the critical point. If $m_s$ is to stay finite
at bulk criticality the temperature dependence has to cancel.
Then
$$
m_s\sim A^{\nu x_s/[1-\nu \omega]}\qquad\qquad A>0.\eqno(5.18)
$$
For the Ising model this is the relation cited before in
Section 5.1.

The scaling behaviour of the parallel correlation function
is given by
$$
G_{\parallel}(t,A,r)=b^{-2x_s}G_{\parallel}(b^{1/\nu}t,
b^{{1/\nu}-\omega}A,{r\over b}).\eqno(5.19)
$$
Taking $b\!=\! r$,
$$
G_{\parallel}(t,A,r)=r^{-2x_s}g_{\parallel}(r^{1/\nu}t,
r^{{1/\nu}-\omega}A) \eqno(5.20)
$$
where the scaling function in the homogeneous case ($A\!=\! 0$)
behaves like $g_{\parallel}(u,0)\sim \exp(-u^{\nu})$.
To obtain information on the case $t=0$, $A\neq 0$ consider the 
choice $\omega=0$. Then one is dealing with an off-critical system
where now $A$ is playing the r\^ole of $t$. Therefore $g_{\parallel}
(0,v)\sim \exp(-v^{\nu})$ in this case. Assuming that this also
holds for $\omega\neq 0$ one arrives at 
$$
G_{\parallel}(r)\sim r^{-2x_s} \exp\left[-(r/l)^{1-\omega
\nu} \right].\eqno(5.21)
$$
for the critical correlations. This is the behaviour which was 
found in (5.3) for the Ising model [72].
A similar argument leads to the scaling form of the
perpendicular correlation function at $t=0$
$$
G_{\perp}(r)\sim r^{-(x+x_s)} \exp\left[-(r/l)^{1-\omega
\nu} \right].\eqno(5.22)
$$

For reduced surface couplings ($A<0$) the surface order,
below the bulk critical temperature, is induced by the
bulk magnetization at a distance $D$ from the surface.
The size of this surface region can be estimated by
equating the thermal and inhomogeneity energy contributions,
$t\sim \vert A\vert/D^{\omega}$. One may then argue as in
Section 4.2 and assume that the order parameter near the surface
is proportional to the correlation function $G_\parallel (r)$
with $r=D\sim (\vert A\vert/t)^{1/\omega}$ so that
$$
m_s(t,A)\sim \exp\left[-a\vert A\vert^{1/\omega}t^{\nu-1/
\omega}\right]\qquad\qquad A<0.\eqno(5.23)
$$
This is consistent with the Ising model result (5.10).

The results (5.21--23) have the same functional form
as (4.5) and (4.11) for the parabolic geometry with the
correspondence $\alpha \leftrightarrow\omega \nu$, $1/C
\leftrightarrow \vert A\vert^{\nu}$. This can be understood
as follows: in both problems, a position-dependent,
smoothly varying local length-scale can be defined near
$T_c$. For the parabola, it is proportional to the local width
$$
\xi(x)=Cx^{\alpha} \eqno(5.24)
$$
while for an extended defect, since $\Delta K(y)$ is a local 
temperature shift
$$
\xi(y)=\Delta K(y)^{-\nu}.\eqno(5.25)
$$
For the calculation of a correlation function between
boundary and bulk variables, one can imagine the system
divided into sections, so that in each of them there is
a decay with the local length $\xi(x)$ or $\xi(y)$. The
complete correlation function is then obtained in the
form of a product, or equivalently, as
$$
G_\perp (y_1,y_2)\sim\exp\left(-\int_{y_1}^{y_2}{\d y
\over\xi(y)}\right).\eqno(5.26)
$$
Inserting for instance the form (5.25) then gives
$$
G_\perp(y_1,y_2)\sim\exp\left[ -a\left( y_2^{1-\omega\nu}
-y_1^{1-\omega\nu}\right) \right]\eqno(5.27)
$$
which corresponds to (5.22).
\Section{Narrow line defects in the bulk}
In the last section the effects of extended defects at 
surfaces were treated. Narrow defects, where the 
couplings at the surface are modified only in a few 
layers do not change the critical behaviour in two
dimensions. If the couplings are modified
along a line, at a finite distance from the surface,
a singularity in the surface correlation length appears
below the bulk critical temperature [95]. This singular
behaviour, however, occurs for a finite value of the
correlation length and is not linked with any new fixed
point. However, if defect lines are situated in the bulk, 
a rich local critical behaviour can be found.
In this section we consider this situation with an emphasis 
on line and star defects. 

First some mean field and scaling
considerations are presented, which show
how the dimensionalities of system and defect, the bulk
and the surface scaling dimensions may affect the local
critical behaviour. This is followed by a discussion of
line and star defects in the two-dimensional Ising model
for which  exact results have been obtained, either
directly in the plane geometry or, using a conformal
mapping, in the strip geometry.

\Subsection{Mean field theory and scaling}
In mean field theory the problem can be discussed for 
arbitrary dimension~$d$. Assume that the system contains 
a defect of dimension $d^*$ (a point, a line or a plane) 
so that the order parameter depends only on one coordinate
$r$. Then the Ginzburg-Landau equation in the ordered phase
reads in terms of reduced variables $\hat m=m/m_b$, $\hat r=r/\xi$
[96, 97]
$$
\hat r^{-d+d^*+1}{\d\over\d\hat r}\left(\hat r^{d-d^*-1}
{\d\hat m\over\d\hat r}\right)=\hat m^3-\hat m.\eqno(6.1)
$$
The spontaneous
magnetization takes the scaling form
$$
m(r)=m_bf({r\over\xi} )\eqno(6.2)
$$
as in (2.17) for the free surface. Equation (6.1) has to
be supplemented by boundary conditions. If the defect
interactions are weaker than bulk ones, the extrapolation
length discussed below (2.3) is positive and one may look 
for the conditions under which a solution exists which 
satisfies Dirichlet boundary conditions at the defect. Since 
$\hat m\ll 1$ near the defect, the local behaviour can be
deduced from a linearized equation
$$
{\d^2\hat m\over\d \hat r^2}+{d-d^*-1\over\hat r}{\d\hat
m\over\d\hat r}+\hat m=0\eqno(6.3)
$$
which is of the Bessel type. An acceptable solution is
obtained only when $d-d^*<2$. It takes the form $\hat
r^\mu \J_\mu (\hat r)$ with $2\mu =2-d+d^*$ and behaves
like $\hat r^{2\mu}$ for small $\hat r$ so that, near 
the defect
$$
m(r)\sim r^{2-d+d^*}t^{(3-d+d^*)/2}\qquad d-d^*<2\eqno(6.4)
$$
The perturbation is then relevant: the defect changes the
local critical behaviour and leads to a magnetization
exponent $\beta_l=(3-d+d^*)/2$. For $d^*=d-1$ one recovers
the ordinary surface behaviour with $\beta_l=\beta_s=1$.
If $d-d^*>2$ the perturbation should be irrelevant and one
expects bulk critical behaviour. This can be shown using
scaling arguments as follows. The coupling perturbation
which belongs to the defect subspace of dimension
$d^*$  and couples to the bulk energy density $\varepsilon$
with dimension $x=d-1/\nu$ has a scaling dimension
$$
d^*-x=d^*-d+{1\over\nu}\eqno(6.5)
$$
As a result the perturbation is relevant (irrelevant) when
$d-d^*<1/\nu$ ($>1/\nu$). With the mean field value
$\nu=1/2$, one finds the condition cited above.

If $d^*\!=\! d\!-\! 1$, the defect divides the system into
two parts. Then if $\nu\!<\! 1$ and the local
interactions are weaker than in the bulk, according to
(6.5) the defect is expected to behave in the same way as a 
cut. Thus ordinary surface critical behaviour will result [15] 
provided the corresponding surface fixed point remains
stable against a weak coupling between the surfaces of the
two half-spaces. The condition for this can be found in the 
following way: starting from two free surfaces the perturbation 
involves the product of two surface magnetization operators 
with scaling dimension $x_s$ and the dimension of the coupling 
is given by 
$$
d-1-2x_s={\gamma_s\over\nu}\eqno(6.6)
$$
where $\gamma_s$ is the local susceptibility exponent at the
ordinary surface transition~(A9). The system with the defect
behaves in the same way as one with a free surface if this
exponent is negative and $\nu\!<\! 1$ [98]. Non-universal
local critical behaviour is expected when the perturbation
is marginal, i.e. when $\nu\!=\! 1$ and $\gamma_s\!=\! 0$.
These scaling results have been confirmed in a series of
calculations on the $O(N)$ model in the limit
$N\!\rightarrow\!\infty$ and using either $\epsilon$- or
$1/N$-expansions [99--107]. 
\Subsection{Ising model with a defect line} 
According to the scaling arguments of the preceding section 
a defect line in the two-dimensional Ising model, which has 
exponents $\nu\! =\! 1$ and $\gamma_s\!=\! 0$
at the ordinary transition, is expected to lead to
continuously varying local exponents. This problem was
first investigated by Bariev [108] who deduced the local
magnetization, as a function of the perturbation strength
and distance to the defect, from the asymptotic behaviour
of the two-spin correlation function below the critical
temperature. This was followed by a detailed study of the
two-spin correlation function by McCoy and Perk [109].
{\par\begingroup\parindent=0pt\leftskip=1cm
\rightskip=1cm\parindent=0pt
\baselineskip=12truept\sips{14truecm}{10truecm}
\midinsert
\centerline{\epsfbox{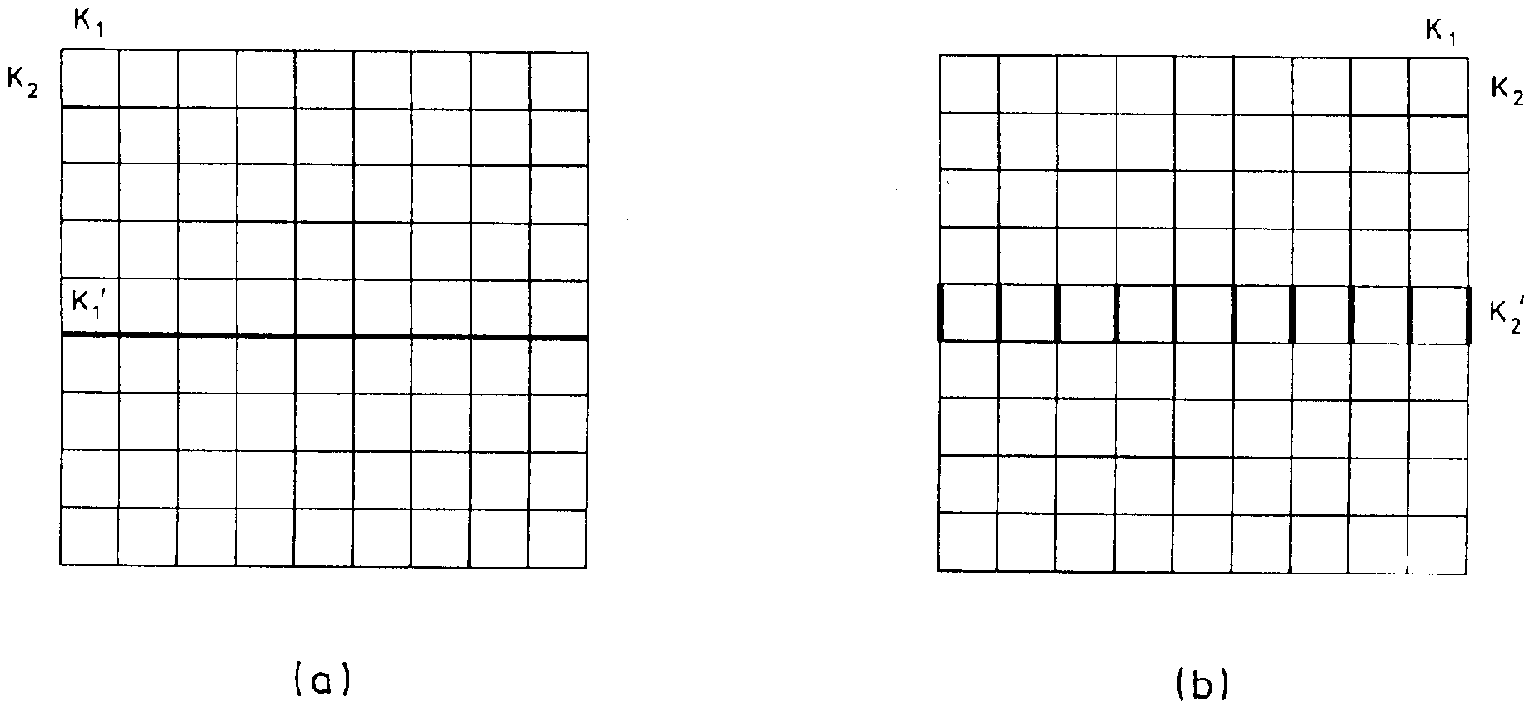}}
\bigskip
\Figcaption{6.1}{\srm Ising square lattice with two kinds of
defect lines (a) chain defect, (b) ladder defect.}
\par
\endinsert
\endgroup
\par}

In the square lattice Ising model one can distinguish two
types of perturbation as shown in Figure 6.1. The chain
perturbation has modified couplings $K'_1$ parallel to the
defect line whereas, for the ladder defect, perturbed
couplings $K'_2$ are in the perpendicular direction. The
local magnetization was found to vary as [108]
$$
<\sigma (y)>\sim t^{\beta_l}y^{\beta_l-\beta}\eqno(6.7)
$$
when the distance $y$ to the defect line is much smaller
than the bulk correlation length $\xi$. In this expression
$\beta =1/8$ is the bulk magnetization exponent and $\beta_l$ 
the continuously varying local magnetization exponent. The 
spatial variation in (6.7) simply follows because $<\sigma 
(y)>$ has the scaling form (6.2) with $\nu =1$.

The local magnetization exponent varies with the type and
strength of the defect according to [108,109]
$$
\Boite{.02\hsbody}{}
\Boite{.3\hsbody}{$\beta_l={2\over\pi^2}\arctan^2\kappa_1$\hfill}
\Boite{.4\hsbody}{$\kappa_1=\e^{-2(K'_1-K_1)}={\tanh {K'}_1^*
\over \tanh K_1^*}$\hfill}
\Boite{.2\hsbody}{${\rm chain}\quad {\rm defect}$\hfill}
\Boite{.08\hsbody}{\hfill(6.8)}
$$
$$
\Boite{.02\hsbody}{}
\Boite{.3\hsbody}{$\beta_l={2\over\pi^2}\arctan^2(\kappa_2^{-1})$\hfill}
\Boite{.4\hsbody}{$\kappa_2={\tanh K'_2\over \tanh K_2}$\hfill}
\Boite{.2\hsbody}{${\rm ladder}\quad {\rm defect}$\hfill}
\Boite{.08\hsbody}{\hfill(6.9)}
$$
where the bulk couplings have their critical values and the
asterix denotes dual variables (see below (B2)). These
exponents are shown in Figure 6.2.
{\par\begingroup\parindent=0pt\leftskip=1cm
\rightskip=1cm\parindent=0pt
\baselineskip=12truept\sips{14truecm}{10truecm}
\midinsert
\centerline{\epsfbox{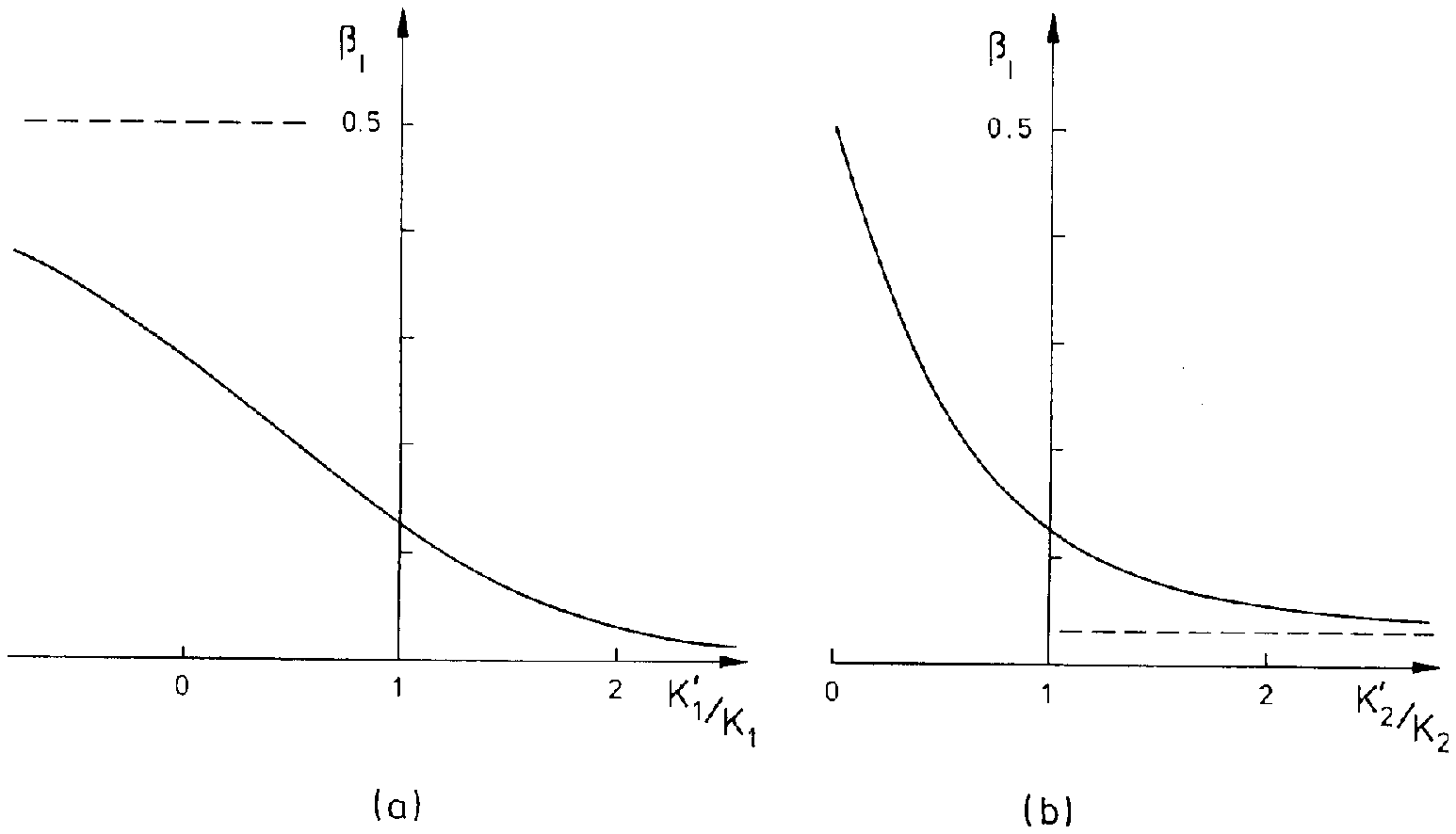}}
\bigskip
\Figcaption{6.2}{\srm Local magnetization exponent $\sst\beta_l$
vs. defect strength in the Bariev model for (a) chain
defect and (b) ladder defect.}
\par
\endinsert
\endgroup
\par}

The chain defect exponent decreases continuously
from $\beta_l=1/2$ when $K'_1\rightarrow -\infty$ to
$\beta_l=0$ when $K'_1\rightarrow +\infty$ (Figure 6.2a).
In the first limit the spins along the defect are frozen
into antiparallel configurations and nearby spins couple
to a vanishing total spin. The local exponent then takes
the free surface value. In the other limit the vanishing
asymptotic value is linked to the onset of local order at
the bulk critical point. When $K'_1=0$ one can sum out the
inner spin and the chain defect effectively becomes a
ladder with strength given by $\tanh K'_2=\tanh^2K_2$.
Then (6.8) and (6.9) give identical results
since $\sinh 2K_1\sinh 2K_2=1$ on the critical line.

For a ladder defect the local exponent is invariant under
the change $K'_2~\!\rightarrow\!~-K'_2$ since the original
defect coupling can be restored through an appropriate
spin reversal in one half of the system. In the positive
sector (Figure 6.2b) $\beta_l$ decreases from the free
surface value when $K'_2\!=\!0$ to a nonvanishing limiting
value when $K'_2\!\rightarrow\!+\infty$. Then the ladder
is also a chain defect with $K'_1\!=\! 2K_1$ and, with the
appropriate values of the perturbed couplings, the two
formulae give identical results.

In their study of the spin-spin correlation function
parallel to the chain defect McCoy and Perk [109] showed
that the correlation length exponent keeps its bulk value
$\nu\!=\! 1$ although the amplitude of $\xi_\parallel$ may 
depend on the local interaction strength. At the critical
point the parallel correlations decay with an exponent
$\eta_\parallel =2x_l$ where $x_l=\beta_l$ as expected from
scaling, compare (A7).

The incremental specific heat introduced by a chain defect
in the square lattice Ising model was obtained by
Fisher and Ferdinand [110]. It has a universal exponent
$\alpha_l=1$ corresponding to the expression (A2) with
$d$ replaced by $d^*$. 

The energy density correlations were
calculated exactly in the scaling region in [111]. Although
the correlation function displays much structure, the decay
exponents are universal, keeping their bulk values. The
same conclusion was reached previously in [112,
113] using the techniques of operator algebra. 
Although the line defect introduces a marginal operator 
which is the energy density, this operator has to keep its
universal scaling dimension $x=1$ in order to
obtain a continuous variation of $\beta_l$ with the defect
strength: Suppose the dimension of the energy density
operator to be changed to $x(\Delta K)$ through the
introduction of a defect $\Delta K$. Then a further change
of the defect strength would no longer constitute a marginal
perturbation since its scaling dimension, given as in
(6.5) by $1\!-x(\Delta K)$, would be nonvanishing. 

Burkhardt and Choi [114] obtained the form of the critical
$n$-point correlation function for the energy density by
summing a perturbation series in the defect strength. The
complete finite-size behaviour of the transfer matrix
spectrum below $T_c$ was investigated in~[115].

The Hamiltonian limit of the line defect problem has been
studied through real-space renormalization [116] and
low-temperature expansion [117]. In this limit one can see
a close connection between the line defect problem and the
$X$-ray problem [118] in which correlation functions
with continuously varying exponents also occur. Using this
analogy [119] one obtains the decay exponents in agreement
with (6.8) and (6.9) where, in the Hamiltonian limit
(Appendix B.1),
$$
\kappa_1={{K'}_1^*\over K_1^*}\qquad\qquad\kappa_2={K'_2\over K_2}
\eqno(6.10)
$$

The size dependence of the lowest gap in the ordered phase
has been investigated through numerical and analytical
calculations [120--123].

A one-dimensional quantum hard-dimer model, belonging to
the Ising universality class, has also been studied [124].
The novel feature in this case is the jump, with increasing
defect strength, from non-universal to ordinary surface
behaviour for a critical value of the defect coupling.

Nigthingale and Bl\"ote [125] used finite-size scaling on
a strip to study line defects in $2d$ models belonging to
the Ising universality class as well as in the $q$-state
Potts model. Varying magnetic exponents are obtained
in the Ising universality whereas the Potts model displays
bulk behaviour for $q=1/2$ (irrelevant perturbation) and
strong crossover effects for $q=3$ (relevant perturbation).

The Potts model with a defect line has also been studied
through exact renormalization on hierarchical lattices~[126]. 
In the relevant case, with strengthened couplings,
the defect remains ordered at the bulk critical point and
the system displays a local first order transition~[76].

\Subsection{Conformal invariance, star-like defects}
For a system with a line defect the symmetries usually
considered to be necessary for conformal invariance 
(Appendix A.2) are partially broken. This is a similar 
situation as for a surface. Nevertheless
the gap-exponent relations and other spectral properties
typical of conformally invariant systems still survive in
the marginal Ising case. This was first conjectured on the
basis of a numerical study of the classical system mapped
onto a strip~[127]. The Hamiltonian limit was then studied
for larger sizes using fermion techniques~[128]. The form
of the whole spectrum was conjectured
in~[129] and exactly calculated in~[130].
{\par\begingroup\parindent=0pt\leftskip=1cm
\rightskip=1cm\parindent=0pt
\baselineskip=12truept\sips{13.6truecm}{9.7truecm}
\midinsert
\centerline{\epsfbox{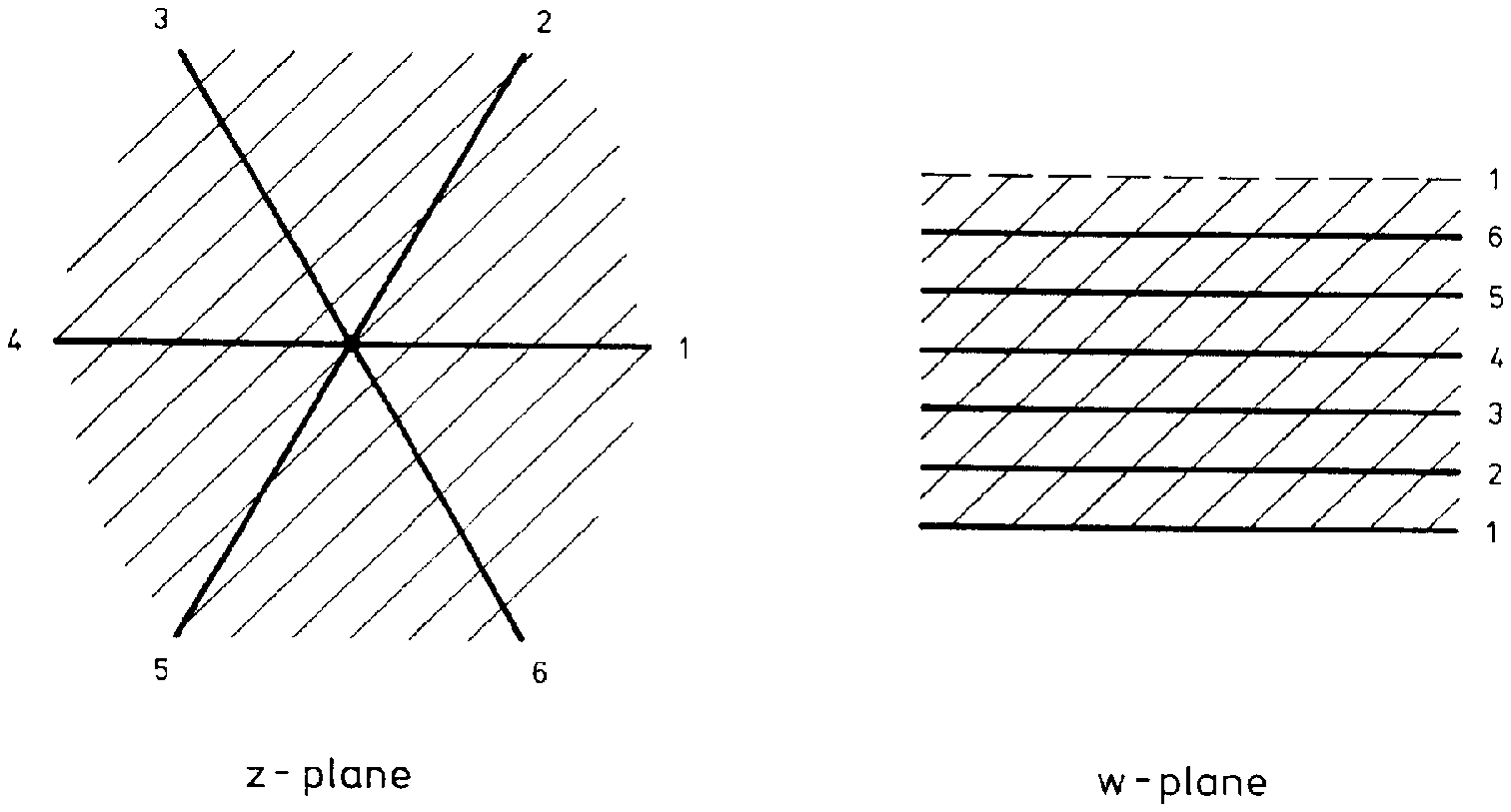}}
\bigskip
\Figcaption{6.3}{\srm A star of defect lines in the $\sst z$-plane
and the corresponding
configuration in the periodic strip of the $\sst w$-plane.}
\par
\endinsert
\endgroup
\par}

Under the logarithmic mapping (A17) a single infinite
defect line in the plane is transformed into a pair of
parallel and equidistant lines along a strip with periodic
boundary conditions [127]. The same mapping can also be used
for star-like defects in the plane [131]. A star with $n_d$
semi-infinite linear branches at the polar angles
$\theta_j=2\pi\delta_j$ ($j=1,\dots n_d$) is then mapped
onto $n_d$ parallel defect lines in a periodic
strip of width $L$ (Figure 6.3). With ladder defects,
in the extreme anisotropic limit, the Hamiltonian
(B3) becomes
$$
{\cal H}=-\demi\sum_{n=1}^L\left[\sigma_n^z+\sigma_n^x
\sigma_{n+1}^x\right]+\demi\sum_{j=1}^{n_d}(1-\kappa_j)
\sigma_{L_j}^x\sigma_{L_j+1}^x\eqno(6.11)
$$
where the defect strength $\kappa_j$ is the ratio of
perturbed to unperturbed couplings like in the second
relation (6.10). The defect positions $L_j$ scale
with the size of the system as $L\delta_j=L_j-L_{j-1}$.

This Hamiltonian has been diagonalized in [130] using
fermion techniques (Appendix B.1). Between the defects the
eigenvectors of the excitation matrix have a plane-wave 
form. The values of the
wave-vectors follow from the $2n_d$ boundary conditions
at the defects. The conformal properties are preserved
only when the $\delta_j$ are commensurate. If $m$ is the
smallest common denominator of the $\delta_j$'s,
either $m=n_d$ for equidistant defects or $m$ is the number
of defects in a corresponding equidistant configuration
which is obtained by adding $m-n_d$ lines with $\kappa_j=1$
(Figure 6.3). Then one gets $\gamma (m)$ fermion
species in the diagonal Hamiltonian. In the plane this
is the number of lines crossing ($m$ even, $\gamma (m)
=m/2$) or meeting ($m$ odd, $\gamma (m)=m$) on the star
defect in the corresponding equidistant configuration.

The $O(L^{-1})$ part of the Hamiltonian can be written in
diagonal form as
$$
\hskip-4mm{\cal H}_p={2\pi\gamma (m)\over L}\sum_{i=1}^{\gamma (m)}
\left( \sum_{r=0}^\infty\left[ (r+\demi -\Delta_i)
\alpha_{ir}^+\alpha_{ir}+(r+\demi +\Delta_i)\beta_{ir}^+
\beta_{ir}\right] -{1\over 12}\left[\demi-6\Delta_i^2
\right]\right)\eqno(6.12)
$$
where $\alpha_{ir}^+$ ($\alpha_{ir}$) and $\beta_{ir}^+$
($\beta_{ir}$) are fermion creation (anihilation)
operators. The shifts $\Delta_i$ depend on the
$\kappa_j$'s and also involve the parity eigenvalues
$p=\pm 1$ due to the periodic boundary conditions (see (B6)).
When $L$ is even the magnetization and energy sectors of
the original Hamiltonian are in correspondence with the
odd states of ${\cal H}_{-1}$ and the even states of
${\cal H}_{+1}$, repectively. Due to the shifts $\Delta_i$
the magnetization sector contains an infinite number of 
conformal towers.

In the case of a single defect line in the plane the two
equidistant lines on the strip have the same strength
$\kappa$. Then $m=2$, $\gamma (m)=1$ and one gets [130]
$$
\Boite{.15\hsbody}{}
\Boite{.5\hsbody}{$\Delta (\kappa )=1-{2\over\pi}\arctan (\kappa^{-1})$
\hfill}
\Boite{.25\hsbody}{$p=-1$\hfill}
\Boite{.1\hsbody}{\hfill(6.13)}
$$
$$
\Boite{.15\hsbody}{}
\Boite{.5\hsbody}{$\Delta (\kappa )=0$\hfill}
\Boite{.25\hsbody}{$p=+1$\hfill}
\Boite{.1\hsbody}{\hfill(6.14)}
$$
The lowest gap gives the dimension of the local
magnetization via (A20) and as~in~(A7)
$$
\beta_l=\demi[\Delta (\kappa )-1]^2={2\over\pi^2}
\arctan^2(\kappa^{-1})\eqno(6.15)
$$
in agreement with the direct calculation in the plane
(6.9). When there are several defects, $\beta_l$ refers
to a star and is something like a generalized corner
exponent depending on all the defect strengths. In
the energy sector the spectrum remains unaffected by
the defects and the scaling dimensions are
universal. According to (6.12) when $m>2$ the gap
normalization differs by a factor $\gamma (m)$ from the
usual one for periodic boundary conditions. This
point is discussed in~[132]. The spectrum-generating 
algebra has been obtained in~[133] for the single line 
problem and in~[131,~134] for general values of $m$. 
Details about this aspect can be found in the review~[135].

In addition to the defect problems described so far, the
following situations also have been studied.
For a 3-state quantum chain with a defect, corresponding 
to a Potts model, the structure of the spectrum was 
studied numerically in~[136]. A defect where the Pauli
matrix $\sigma^x$, in one coupling term of the periodic
quantum Ising chain, was replaced by a general
Hermitian $2\times 2$ matrix was studied in~[137]. Exact 
results could be obtained when the global $Z_2$ symmetry 
of the quantum chain is preserved. Then the critical and conformal
properties are the same as with an ordinary defect.
The same occurs for the Ising quantum chain with staggered
3-spin interactions and the Ashkin-Teller model with
generalized defects [138, 139]. Finally, the Ising quantum 
chain associated with star-like defects
was generalized in [140] introducing modified couplings
over an extended region. In order to keep the system
critical, the same defect strength was taken for the
$\sigma^z$ and $\sigma^x$ parts in the strip Hamiltonian
(B3), which corresponds to a constant change in the
anisotropy inside this region. Conformal invariance is
preserved, as above for stars, for commensurate
configurations. For an arbitrary varying defect amplitude,
the equidistant level structure of the spectrum
is maintained at high energies only.
\Section{Extended line defects in the bulk}
For defects in the bulk, interesting situations do not only
arise if one deals with narrow ones as in Section 6 but also
if they are extended as (5.1). This type of problem was first
introduced and investigated by Bariev [141--144]. For such 
extended defects originating from a core of dimension $d^*$
in a $d$-dimensional system, scaling considerations give two 
separate relevance-irrelevance criteria. One is associated 
with the extended part of the defect and is the same as for 
the surface case. Thus the perturbation is relevant,
irrelevant or marginal for $\omega<1/\nu$, $\omega>1/\nu$
and $\omega=1/\nu$, respectively. The other criterion, related 
to the core of the defect, is the condition on the sign of 
$d^*-d+1/\nu$ found below (6.5). The most interesting situation 
occurs when both conditions predict marginal behaviour.
In this case, although the exponents are varying 
with the defect strength $A$, there are discontinuities at
$A\!=\! 0$ [145]. Connected with that, a perturbation
expansion for the exponents is divergent [141]. This feature is 
outlined in Appendix C. 

\Subsection{Ising model, conformal results}

In the two-dimensional Ising model an extended line defect
with decay exponent $\omega=1$ satisfies both marginality
conditions since $\nu=1$. An exact treatment of this
problem, however, is more difficult than for the extended
surface defect in Section 5, since most of the analytical
methods used there cannot be adapted easily to internal
defects. The only exception is the method based on conformal
mapping.

The problem was treated [145] for a  square lattice as in Figure
6.1 with a defect centered at $y=0$ and couplings $K_2(y)$
varying as
$$
K_2(y)=K_2+{\overline{A}\over |y|} \eqno(7.1)
$$
where $\overline{A}$ is defined in (5.4) and
the bulk couplings $K_1$, $K_2$ have the critical
values. When $K_2(0)\!=\! 0$, the system separates into two
uncoupled semi-infinite ones, with a surface inhomogeneity
as studied in Section 5.

{\par\begingroup\parindent=0pt\leftskip=1cm
\rightskip=1cm\parindent=0pt
\baselineskip=12truept\sips{10truecm}{7.1truecm}
\midinsert
\centerline{\epsfbox{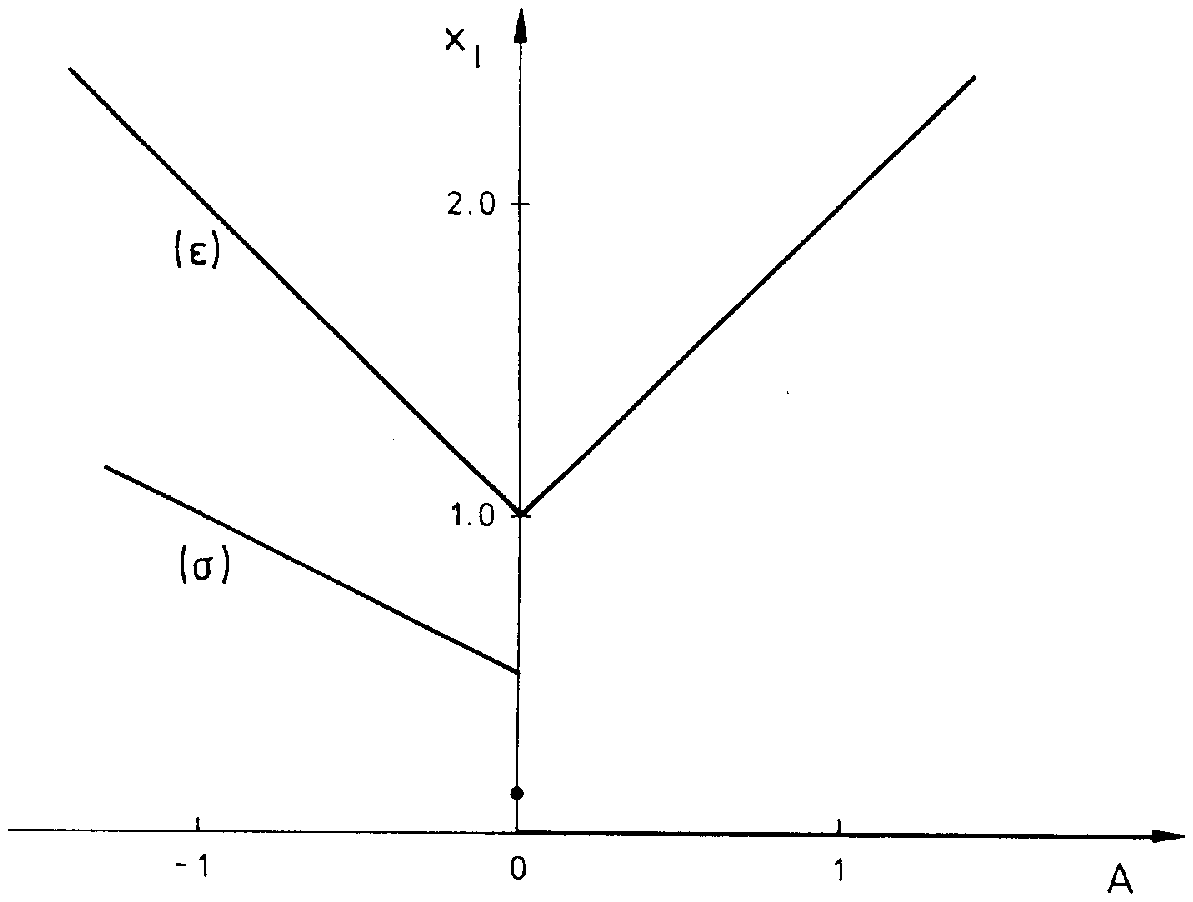}}
\bigskip
\Figcaption{7.1}{\srm Local scaling dimensions for spin
($\sst\sigma$) and energy density ($\sst\varepsilon$) for an
extended defect with $\sst\omega=1$ in the bulk of an
Ising model as a function of~the~strength~$\sst A$.}
\par
\endinsert
\endgroup
\par}
Using the logarithmic transformation (A17) the problem
is mapped onto a strip. The inhomogeneity thereby becomes
periodic with period $L/2$ and is given by
$$
\Delta K(u)=\overline{A}\left\vert{L\over 2\pi}\sin\left(
{2\pi u\over L}\right)\right\vert^{-1} \eqno(7.2)
$$
In each of the intervals, up to a factor of two, this has
the same form as (5.13).

As in Section 5.3 the lowest gaps were determined in a
continuum approximation. Due to the different boundary
conditions, the eigenvalues and hence, the local
critical exponents are different from those in the
surface defect problem. The scaling dimensions
are shown in Figure 7.1 as functions of $A$.

For enhanced local couplings ($A\!>\!0$) the energy of the
lowest excited state, $E_1\!-\! E_0\!\sim\! L^{-1-2A}$, vanishes
faster than $1/L$ so that the scaling dimension of the
magnetization is $x_l\!=\!0$ and the defect region
remains ordered at the bulk critical point. On the other
hand for reduced couplings ($A\!<\! 0$) there is no order at
$T\!=\! T_c$ and the local exponents are linear functions of A.
The magnetic exponent is the same as for two disconnected
semi-infinite systems (see Figure 5.2). At $A\!=\! 0$ it is
discontinuous. As a consequence a finite-order perturbation
expansion for the gaps around the homogeneity point $A\!=\! 0$
is not possible. The expansion for the first gap starts as
[145]
$$
E_1-E_0={2\pi \over L}\left({1\over 8}-{A\over \pi} \log
L+\dots\right) \eqno(7.3)
$$
i.e. the first-order correction term is divergent, but the
singular contributions sum up to a regular term in infinite
order. This observation is in agreement with the results of
the perturbation theory for the system in the original
geometry (Appendix C).

The critical exponent for the energy operator given by
$$
x_l=1+|A|\eqno(7.4)
$$
is continuous at $A=0$ where it reaches its bulk value. It
can be determined independently, working in the plane and
using finite-size scaling for the matrix-element of the
local energy operator as in (A21). The coincidence of the 
two results supports the assumption that the system is
conformally invariant at the critical point.
\Section{Radially extended defects}
The systems studied so far had essentially all a layered
structure. As a final example, we now examine the case where
the inhomogeneity extends radially. The centre may be
located either in the bulk or on a free surface. Only
marginal temperature-like perturbations corresponding to
a change in the interaction strength are considered. After
some scaling considerations, exact results for
the local magnetization in the two-dimensional Ising model
are presented. This is  supplemented by a discussion of
the conformal aspects.
\Subsection{Scaling considerations}
According to the scaling arguments of Section 6.1, a
short-range perturbation introduced by a point defect with
$d^*\!=\! 0$ into a $d$-dimensional system has a scaling dimension
$-d+1/\nu$ which follows from (6.5). Since $1/\nu\!<\!d$ at a 
second order phase transition this type of perturbation
is always irrelevant. This can be verified explicitly for the 
Ising model through a calculation of the local magnetization which may
be expressed in terms of averages for the unperturbed system
[146]. Following the method of Section 2.2 and using the 
rotational symmetry one can also obtain the critical correlation 
functions. They are completely determined and
identical to the unperturbed ones.
 
In the case of an extended defect and a perturbation
amplitude decaying as a power of the distance from the centre,
the relevance-irrelevance criterion is the same as for
perturbations decaying from an interior line defect or a
surface (Sections 5, 7 and Appendix C). For a
temperature-like perturbation, according to Equations (5.2)
and (C7), the amplitude has a scaling dimension $1/\nu-\omega$
where $\omega$ is the decay exponent of the perturbation. In the
following we consider the marginal case where $\omega=1/\nu$.
\Subsection{Ising model}
For the Ising model containing an extended defect with
$\omega=1$ exact results were obtained using the corner 
transfer matrix method [147].

{\par\begingroup\parindent=0pt\leftskip=1cm
\rightskip=1cm\parindent=0pt
\baselineskip=12truept\sips{11.6truecm}{8.3truecm}
\midinsert
\centerline{\epsfbox{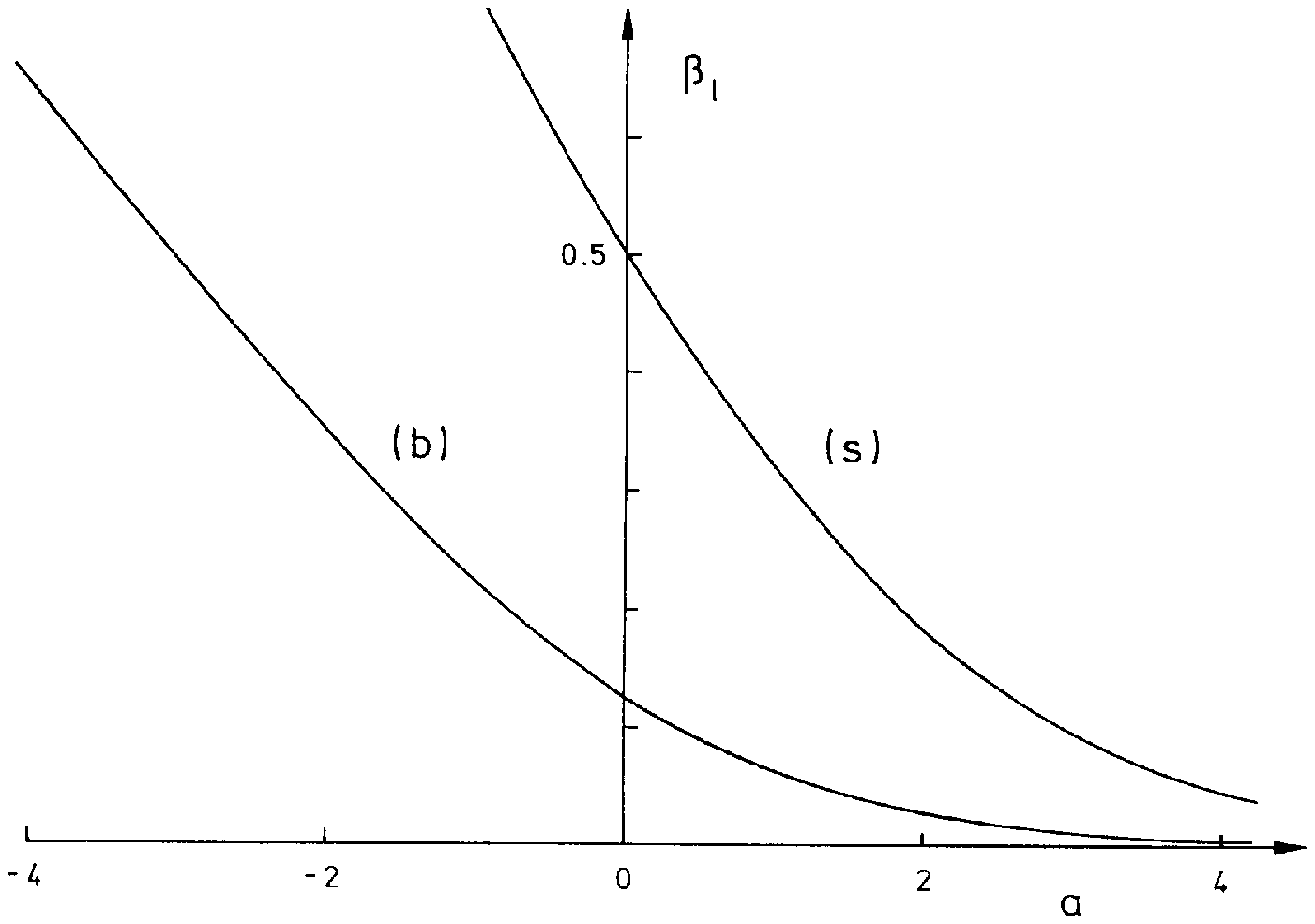}}
\bigskip
\Figcaption{8.1}{\srm Local magnetization exponent $\sst\beta_l$
vs. defect strength for a radially extended defect ($\sst b$)
in the bulk and ($\sst s$) on the surface of
an isotropic Ising model. Here $\sst a$ is equal to $\sst\pi\alpha$.}
\par
\endinsert
\endgroup
\par} 
The section of a square lattice forming the corner transfer 
matrix is shown in Figure B1b. For the present problem one chooses
anisotropic couplings $K_1$ and
$$
K_2(n)=K_2\left(1+{\alpha\over 2n+1}\right)\eqno(8.1)
$$
where $n$ is the row index increasing from the centre.
The Hamiltonian limit leads to a study of an inhomogeneous 
quantum chain (B3) with couplings $h_n=2n$, $\lambda_n=
\lambda (2n+\alpha +1)$. As it stands, the corner transfer 
matrix describes a segment of an anisotropic system with 
opening angle $90^{o}$. However, as explained in Appendix B.3, 
it also describes an isotropic system for which the opening 
angle is infinitesimal. Therefore an isotropic system with
rotational symmetry and opening angle $\theta$ can be 
described by the transfer matrix
$$
{\cal T}(\theta )=\exp\left(-\demi\theta {\cal H}\right)
\eqno(8.2)
$$
 
According to (B19) the magnetization at the
centre involves the single particle excitations $\omega_l$
of the quantum chain. The critical excitation
spectrum can be obtained by taking the continuum limit. Then
the very centre of the system has to be excluded [148].
Taking a cut-off at $r$ and an outer radius $R$, the
critical excitations are given by
$$
\omega_l =\sqrt {\alpha^2+\left({2z_l\over\ln
R/r}\right)^2}\eqno(8.3) 
$$
where the $z_l$ are solutions of $\alpha\ln (R/r)\tan z=-2z$.  
An additional bound-state-like solution occurs when $\alpha <0$. 
The local magnetization exponent $\beta_l$ then
follows from the finite-size behaviour of the centre
magnetization which is analogous to (A8). This gives
$$
\beta_l={\alpha^2\over 2}\int_0^\infty\d u{\sinh^2u\over
\sinh (\pi\mid\alpha\mid\cosh u)}\eqno(8.4)
$$
A bound-state contribution ${\mid\alpha\mid/2}$ must
be added when $\alpha <0$. The variation of the
magnetization exponent with the perturbation strength is
shown in Figure~8.1. The exponent decreases smoothly when the defect
amplitude is increased but, contrary to the case of surface
or extended line defects, it vanishes only asymptotically.
Thus there is no local order at the bulk critical point
for any finite value of the defect strength.
 
\Subsection{Conformal considerations}
The previous Ising results can also be deduced from a
mapping of the perturbed critical system onto a strip
[149, 150]. This mapping can done for an arbitrary 
conformally invariant two-dimensional system containing
an extended perturbation with rotational
symmetry. Then, in the continuum limit,
$$
-\beta {\cal H}=-\beta {\cal H}_c+g\int r^{-\omega}\psi
(r,\theta )r\d r\d\theta\eqno(8.5)
$$
where $\psi (r,\theta )$ is some local field with scaling
dimension
$x_\psi $. In the following, as before, we assume that
the decay exponent takes its marginal value $\omega =2-x_
\psi$. Under the transformation (A17) the dilatation factor 
just compensates the decay of the perturbation in the 
marginal case. Therefore, in the periodic strip of width $L$, 
one obtains an homogeneous $L$-dependent deviation from 
criticality
$$
\Delta =g \left({2\pi\over L}\right)^{2-x_\psi }\eqno(8.6)
$$
for the couplings which are conjugate to $\psi$. One
should note that such a homogeneous deviation from the
bulk critical couplings leads to a massive excitation
spectrum as in (8.3) and, contrary to the case of extended
line defects treated in Sections 5 and 7, the tower-like,
equidistant-level structure of the spectrum is lost here.
 
Let $\phi$, with bulk scaling dimension $x$, be either the
energy or the magnetization density. Its local scaling
dimension at the defect, $x_l$, can be deduced from the
corresponding gaps for the strip through (A20). According
to finite-size scaling, the perturbed gap $G_\phi$ 
transforms as
$$
G_\phi(\Delta ,L)=L^{-1}G_\phi(L^{2-x_\psi}\Delta ,1)=
L^{-1}f_\phi\left[(2\pi)^{2-x_\psi}g\right]\eqno(8.7)
$$
where $f_\phi$ is a universal scaling function [151]. 
This leads to the local exponent
$$
x_l={1\over 2\pi}f_\phi \left[ (2\pi)^{2-x_\psi}g\right]
\eqno(8.8)
$$
It depends on the perturbation amplitude $g$, as expected
for a marginal perturbation, but does it in an universal way
[150]. Different systems belonging to the same class of
universality will show the same dependence on the 
perturbation amplitude.
 
For the Ising model, the perturbed gaps (or gap scaling
functions) can be calculated either for an isotropic strip
or using the Hamiltonian limit [152--155]. In the latter 
case the defect amplitude has to be rescaled as discussed 
in Appendix B.3. On a square lattice with modified 
interactions $K(r)=K_c+A/r$ one recovers the local magnetic 
exponent given in (8.4) with $\alpha=8A$ [149]. For small 
values of $A$, it has the expansion
$$
\beta_l=x_l={1\over 8}-2A+16\ln 2\ A^2+O(A^3)\eqno(8.9)
$$
To first order this is identical to the perturbation result
in Table C2 with $g=2A$. The factor of two in the continuum
parameter $g$ reflects the two bonds per unit cell on the
square lattice. Due to the dual symmetry of the Ising model 
[156] the exponent of the local energy density has an 
expansion involving only even powers of the perturbation 
amplitude and reads [150]
$$
x_l=1+32A^2+O(A^4)\eqno(8.10)
$$
The same procedure applies with minor changes when the
defect centre is located on a free surface. This was also
studied for the Ising model [149]. To first order the result
is again in agreement with the perturbation approach
(Table C2). The corresponding magnetic exponent is also
shown in Figure 8.1.
\Section{Conclusion}
In this review we have presented a number of systems and
situations where the universality which is normally found
at continuous phase transitions can be absent. Superficially,
the examples look quite different, but in each case
the tendency towards order is modified locally in a
particular systematic way. This is the basis for the common
features one observes.
 
It is not difficult to understand the results qualitatively.
For example, if order is favoured this leads to a steeper
magnetization curve and will tend to reduce the magnetic
exponent. This explains the general form and the similarity
of the graphs in the various marginal cases. There are, of
course, some differences in details. The close parallel
between effects from the geometry and from the modification
of interactions inside a system is nevertheless
remarkable. A similar relation is found if one compares
homogeneous integrable systems at and off the critical point
[35, 157, 158, 48]. In this case the corner transfer matrix
can be used as a link. For the present systems the
direct connection is less obvious.
 
In two dimensions conformal mappings play an important
r\^ole in the studies. The results show that they can be
used even beyond their obvious domain. The gap-exponent
relation also holds in cases like in Section 8 where no
tower of equidistant levels exists. This is an aspect which
might deserve further study. The same is true for the situation
in three dimensions where only a few, essentially mean
field results have been obtained so far.
 
The last point is also important since it should be possible
to measure some of the effects. The most obvious experiment
would be to look for the local order in systems with
wedge-like, conical or parabolic shape. This is not easy since
the local order will be smaller than inside the bulk,
and already for planar surfaces such measurements are rare.
Nevertheless they should be feasible with proper local probes
and would add an interesting aspect to the general picture
of critical phenomena.
\ack
This work is the result of a collaboration with mutual
visits of the authors to Nancy, Berlin and Budapest. F. I. is
indebted to the Minist\`ere Fran\c cais des Affaires
Etrang\`eres and the CNRS for research grants. I. P. thanks
the University of Nancy I and the Research Institute for 
Solid State Physics in Budapest for hospitality. L. T. 
acknowledges the support of DFG, Projekt "Zweidimensionale 
kritische Systeme" and together with F. I. the support of the 
CNRS and the Hungarian Academy of Sciences through an
exchange project. The authors benefitted from 
conversations with R. Z. Bariev and B. Berche.
\vfill\eject
\appendice{A}{Scaling and conformal invariance}
Near a second order phase transition the singular parts of 
thermodynamic quantities vary as powers of the parameters
(scaling fields) $t\sim\mid T-T_c\mid$, $h$, etc. measuring
the deviation  from the critical point. This type of
behaviour is linked  to the self-similarity of critical
fluctuations inside the  correlation volume $\xi^d$ where
$\xi\sim t^{-\nu}$ is the correlation length. The system is
then covariant under a global change of the length scale
and singular quantities are homogeneous functions of their
arguments. These properties form the basis of the scaling 
hypothesis [159] which lead to scaling laws relating
critical exponents to a small number of fundamental ones.
The scaling behaviour was also the source of the
renormalization group ideas [1], allowing a calculation
of the fundamental exponents. 

At the critical point, $\xi$ diverges and the system
becomes scale invariant. This, together with more usual
symmetries (rotation, translation), leads to the invariance
under conformal transformation, first used in field
theory [160] and later fully exploited in two-dimensional
critical systems where it allows an exact determination
of the critical exponents and much more [161, 16, 162] .
  
\subappendice{A.}{Scaling and critical exponents}
According to scaling theory, when the lengths are rescaled
by a factor $b>1$, i.e. when $\bi r\to\bi r/b$, the
scaling fields (like $t$ and $h$) are changed by a factor
$b^{d-x}$ where $d$ is the dimension of the system and $x$
the scaling dimension of conjuguate quantities (e.g. the
energy density for $t$, the magnetization density for
$h$). When $d-x>0\ (<0)$ the corresponding scaling field
grows (decreases) under rescaling. Such a field is said
to be relevant (irrelevant) whereas it is marginal when
$x=d$. The system becomes invariant under rescaling only
when the relevant scaling fields vanish which corresponds
to the critical point.
 
Since irrelevant variables finally vanish under
rescaling, only relevant and marginal scaling fields
influence the critical properties, the marginal ones
generally leading to varying exponents. 

Near the critical point, the singular part of the free
energy density is a homogeneous function of the scaling
fields and transforms according to
$$
f_b\left( t,h,{1\over L}\right)=b^{-d}f_b\left( b^{1/\nu}t,
b^{d-x}h,{b\over L}\right).\eqno(A1)
$$
In (A1) we kept only relevant variables and included the
inverse of the linear size of the system as a new
relevant scaling field, since at its bulk critical point a
system is truly critical only in the thermodynamic limit,
i.e. when $1/L=0$. 

The critical behaviour of conjuguate quantities 
and their derivatives can be deduced from (A1) taking
advantage of the arbitrariness of the dilatation factor
$b$. The corresponding exponents are all related to the
basic ones $x$, $\nu$ through scaling laws and involve $d$.

For example, the specific heat exponent $\alpha$ can be
obtained from the second $t$-derivative of both sides of
(A1) at $h=0$, $1/L=0$ and taking
$b=t^{-\nu}$ with the result 
$$
\alpha=2-d\nu \eqno(A2)
$$
The magnetization $m=-{\partial f_b/\partial h}$ transforms as
$$
m\left( t,h,{1\over L}\right)=b^{-x} m\left( b^{1/\nu}t,
b^{d-x}h,{b\over L}\right) \eqno(A3)
$$
and the exponent $\beta$, defined through
$m\sim t^{\beta}$, can be obtained from (A3) taking $h=0$,
$1/L=0$ and $b=t^{-\nu}$ as before, so that
$$
\beta=\nu x.\eqno(A4)
$$

At the bulk critical point $t=0$, $h=0$, critical
singularities are suppressed by a finite $L$ value.
Finite-size scaling exploits the way they develop
when $L\rightarrow\infty$ in order to determine the critical
exponents. For example, from (A1) with $b=L$, the free
energy density behaves as $L^{-d}$ at criticality
whereas the magnetization density in (A3) gives
$$ 
m(L) \sim L^{-x}.\eqno(A5) 
$$ 

When the system is limited in space through a surface,
besides the bulk term $f_b$ in (A1), new local
contributions  (surface, corner, etc.) to the free energy
density appear, with singular parts behaving as above for
the bulk. For example, the contribution of a surface with
dimension $d-1$ transforms according to [4] 
$$ 
f_s\left( t,h_s,{1\over L}\right)=b^{-(d-1)}f_s\left(
b^{1/\nu}t,b^{d-1-x_s}h_s,{b\over L}\right).\eqno(A6) 
$$
where a surface magnetic field $h_s$ has been
included. Repeating the previous argument, the scaling
relations (A4--5) remain valid with the appropriate surface
exponents replacing bulk ones,
$$\eqalignno{ 
\beta_s&=\nu x_s,&(A7)\cr
m_s(L)&\sim L^{-x_s}.&(A8)\cr} 
$$ 
The surface susceptibility exponent $\gamma_s$ defined
through ${\partial^2 f_s /\partial h_s^2}\sim t^{-\gamma_s}$
is obtained as 
$$ 
\gamma_s=\nu(d-1-2x_s).\eqno(A9) 
$$
The bulk two-point correlation function for an operator
$\psi$ is  obtained taking a functional derivative of the
free-energy with respect to the appropriate
position-dependent field  
$$
G(r,t)\equiv<\psi(0) \psi({\bf r})>={\delta^2 F \over
\delta h(0) \delta h({\bf r})}.\eqno(A10)
$$ 
If e.g. $\psi({\bf r})$ stands for the magnetization
operator, then its average $<\psi({\bf r})>=m({\bf r})$
gives the local magnetization. The transformation law of
the two-point function then follows from (A10) and (A1) 
$$
G(r,t)=b^{-2x}G\left({r\over b}, b^{1/\nu}t\right)
\eqno(A11)
$$
where $x$ is the bulk scaling dimension of $\psi$.
At the critical point, $t=0$, taking $b=r$, the power-
law decay of correlations follows
$$
G(r,t=0)={A\over r^{2x}}.\eqno(A12)
$$
Surface correlations can be investigated in the same way.
The  correlation function with  two points at the surface
behaves as 
$$ 
G_{\parallel}(r,t=0) \sim r^{-2x_s}\eqno(A13)
$$
while the perpendicular correlations decay like
$r^{-(x+x_s)}$ (see Section 2.2).
\subappendice{A.}{Conformal invariance}
Covariance under conformal transformations is expected to
hold at the critical point of systems with short range
interactions, which possess translational and rotational
symmetry and are invariant under uniform scaling. When some
of the above symmetries are broken by a marginal
perturbation (e.g. for inhomogeneous systems) some of the
properties associated with conformally invariant systems,
like the gap-exponent relation, can be preserved as
observed in specific examples. 

A conformal transformation $\bi r\to\bi r'(\bi r)$ can be
seen as a generalization of uniform scaling, where the
structure of the lattice is  locally preserved, but the
rescaling factor $b({\bf r})$ becomes a smooth function of
the position. It follows from the Jacobian of the
transformation as $b(\bi r)^{-d}=\det (\partial \bi r' /
\partial \bi r)$.  Since local fields transform as $h({\bf
r}) \to h'( {\bf r}')=b({\bf r})^{d-x}h({\bf r})$
the two-point function in (A10) transforms like
$$
<\psi(\bi r_1)\psi(\bi r_2)>=b(\bi r_1)^{-x}b(\bi r_2)^{-x}
<\psi(\bi r'_1)\psi(\bi r'_2)> \eqno(A14)
$$
under a conformal transformation.
This is a straightforward generalization of (A11) with
$t=0$. Similarly the transformation law for an operator
profile $<\psi(\bi r)>$ is obtained as
$$
<\psi(\bi r)>=b(\bi r)^{-x}<\psi(\bi r')>.\eqno(A15)
$$

The conformal group for dimensions larger than two is
finite-dimensional and contains rotations, uniform dilatations,
translations and inversions. The special conformal
transformation with an arbitrary translation ${\bi
a}$,
$$ 
{{\bi r'}\over {r'}^2}={{\bi
r}\over r^2}+{\bi a},\eqno(A16)
$$ 
is constructed from the two last ones. Invariance under
this transformation fixes the form of three-point
functions [160] like scaling does with the two-point
functions in (A12) and (A13). If ${\bi a}$ in (A16) is
an infinitesimal surface vector, then one can use this
transformation to find restrictions on the
form of surface correlations, as shown in Section 2.2.
 
The method of conformal invariance is especially powerful
in two dimensions where the conformal group, being
isomorphic with the group of complex analytic functions,
becomes infinite-dimensional and strongly restricts the
possible values of critical exponents for a broad class of
systems [161, 163, 16, 162]. 

In two dimensions one may also use a complex mapping $w(z)$ 
to go from one geometry to another [164]. When some
critical correlation is known in the first geometry, it can
be transformed into the second one. The local dilatation
factor is then $b(z)=\vert\d w/ \d z\vert^{-1}$. 

Two basic geometries are connected by the
logarithmic transformation
$$
w={L \over 2\pi} \ln z \eqno(A17)
$$
which maps the infinite $z$-plane onto a periodic
strip of width $L$ in the $w$-plane. Transforming the correlation
function in (A12) according to (A14), one obtains  
$$
<\psi(u_1,v_1)\psi(u_2,v_2)>={(2\pi/L)^{2x} \over \left[2
\cosh{2\pi\over L}(u_1-u_2) - 2\cos{2\pi\over L} (v_1-v_2)
\right]^x}\eqno(A18)
$$
in the strip geometry where $u$ measures the distance along
the strip and $0<v\leq L$ denotes the transverse periodic
coordinate. An expansion of the r.h.s. of (A18) for
$u_1>u_2$ leads to
$$\eqalign{
<\psi(u_1,v_1)\psi(u_2,v_2)>=&({2\pi\over
L})^{2x}\sum_{m,\overline{m}}^\infty
a_ma_{\overline{m}}\exp\left[-{2\pi\over L}
(x+m+\overline{m})(u_1-u_2)\right]\cr
&\times\exp\left[{2\i\pi\over L}
(m-\overline{m}) (v_1-v_2)\right]\cr}\eqno(A19) 
$$
where $a_m\!=\!\Gamma(x+m)/\Gamma(x)m!$. The strip
correlation function can be also determined through a direct
calculation using the transfer matrix method as described
in Appendix B. Then, comparing (A19) with expressions like
(B13), one reaches the following conclusions:

   i) The eigenstates of the Hamiltonian in (B1) are labelled
by pairs of integers ($m,\overline{m}$), so that the spectrum
exhibits a tower-like structure with energies
$E_0~+~(~x~+~m~+~\overline{m}~)~{2~\pi ~/L}$ and ~
momenta given by $(m-\overline{m})\ {2\pi /L}$.

   ii) The gap between the ground state and the lowest excited
state for which the matrix element of $\psi$ is
non-vanishing, is related to the scaling dimension $x$ via
$$
E_1-E_0={2\pi \over L}x.\eqno(A20)
$$
This is the  gap-exponent relation.

   iii) Finally, the matrix element itself is given as
$$
<1|\psi |0>=\left({2\pi \over L}\right)^x \eqno(A21)
$$
For the magnetization, this is in agreement with
the finite-size scaling result (A5).
\bigskip
To study surface correlations, the half-plane ($y>0$)
can be mapped onto a strip with free boundaries through the
conformal transformation 
$$
w={L \over \pi} \ln z.\eqno(A22)
$$
Now the eigenstates of the Hamiltonian in the strip are 
labelled by
an integer $m=0,1,\dots$, the spectrum has a tower-like
structure with energy eigenvalues $E_0+(x_s+m){\pi /L}$ and
the gap-exponent relation reads
$$
E_1-E_0={\pi \over L}x_s.\eqno(A23)
$$
Finally, the matrix element for $\psi$ near the boundary
is given by
$$
<1|\psi_s|0>=\left({\pi \over L}\right)^{x_s}.\eqno(A24)
$$
\vfill\eject
\appendice{B}{Transfer Matrices}
In the transfer matrix method [156, 165, 166] a system of
spins or other classical variables with short-range
interactions is built from identical smaller units. This
technique can also be used for the inhomogeneous systems
considered here. In two dimensions one can choose the basic
units as shown in Figure B1. The corresponding transfer
matrices ${\cal T}$ are the partition functions of either two 
consecutive rows (case a) or
a whole angular segment (case b). Their matrix character
results from their dependence on the two sets
$\{\sigma\},\{\sigma'\}$ of boundary spins (full points).
The partition function Z of the whole system is then obtained
by taking the product of an appropriate number of
${\cal T}$'s. For periodic boundary conditions in the direction
of transfer, Z is given by the trace over this product
and for free boundary conditions by a particular matrix element.
Thermal expectation values can be calculated by
inserting additional operators. In this way, the complete
thermodynamical behaviour is contained in ${\cal T}$.
{\par\begingroup\parindent=0pt\leftskip=1cm
\rightskip=1cm\parindent=0pt
\baselineskip=12truept\sips{13.2truecm}{9.4truecm}
\midinsert
\centerline{\epsfbox{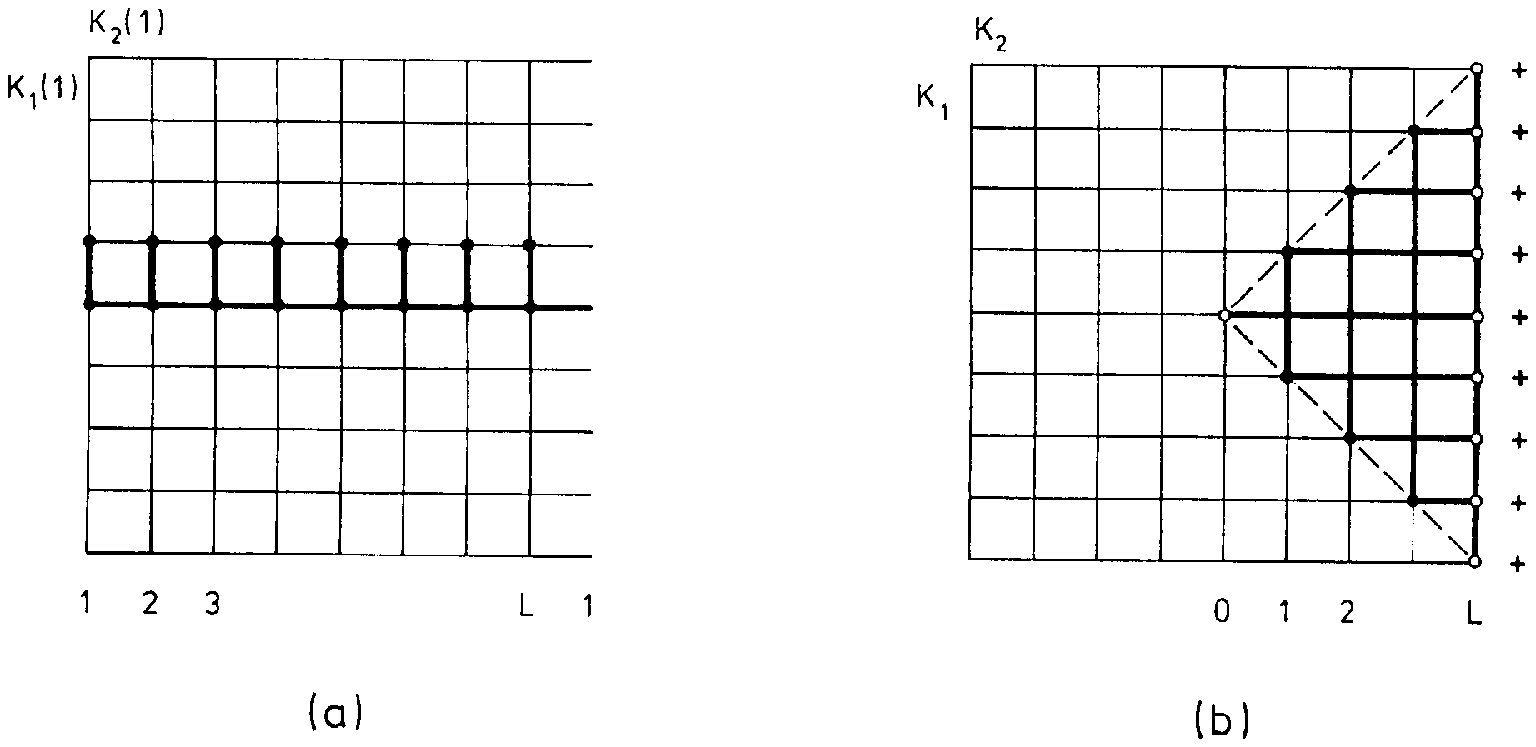}}
\bigskip
\Figcaption{B1}{\srm Sections of a square lattice (thick) which
form (a) the row transfer matrix and (b) the corner
transfer matrix. Open circles correspond to fixed spins.}
\par
\endinsert
\endgroup
\par}

It is customary to write the transfer matrices as
$$
{\cal T}=\exp (-H)\eqno(B1)
$$
in analogy with time evolution operators. For the Ising
model, ${\cal T}$ can be expressed in terms of Pauli matrices
and the corresponding $H$ may be regarded as the
Hamiltonian for a certain spin one-half quantum chain. In the
following this is described in some detail.
\subappendice{B.}{Row transfer matrix} 
This quantity (Figure B1a) is appropriate for a homogeneous
system or a layered one with couplings $K_i(n)$ varying along
the horizontal direction. It is given explicitely by [167] 
$$
{\cal T}=\exp \left[\sum_nK_1^*(n)\sigma_n^z\right]\exp\left[
\sum_nK_2(n)\sigma_n^x\sigma_{n+1}^x\right]\eqno(B2)
$$
where the $K_1^*(n)$ are dual couplings ($\tanh K_1^*=\exp
(-2K_1)$) and periodic boundary conditions are assumed
horizontally, $\sigma_{L+1}^\alpha=\sigma_1^\alpha$. An open
system can be obtained by setting one $K_2(n)$ equal to zero.

The matrix ${\cal T}$, or its symmetrized versions one uses
normally, can in principle be diagonalized for arbitrary
couplings [168, 78]. In this sense, the two-dimensional
Ising model is called a completely integrable system. A
considerable simplification occurs, however, in the
anisotropic (Hamiltonian) limit of strong vertical and weak
horizontal bonds [169, 170, 135]. Then $K_1^*(n)$ and $K_2(n)$ are
both small and one can combine the exponentials to obtain
$H=2K_1^*{\cal H}$, where $K_1^*$ is some reference value and
$$
{\cal H}=-\demi\sum_{n=1}^Lh_n\sigma_n^z-\demi\sum_{n=1}^L
\lambda_n\sigma_n^x\sigma_{n+1}^x\eqno(B3)
$$
This is the Hamiltonian of an Ising chain with couplings
$\lambda_n$ and transverse fields $h_n$. The normalization
is such that for $h_n=\lambda_n=1$ the excitations have
velocity one.
 
The operator ${\cal H}$ can be expressed in terms of Fermi
operators $c_n$, $c_n^+$ via the Jordan-Wigner
transformation [171, 167, 172] 
$$\eqalignno{
\sigma_n^+&=c^+_n\exp\left(\i\pi\sum_{l=1}^{n-1}c^+_lc_l
\right)&(B4)\cr
\sigma_n^z&=2c^+_nc_n-1&(B5)\cr}
$$
This gives
$$
\hskip-4mm{\cal H}=-\sum_{n=1}^Lh_n\left( c^+_nc_n-\demi\right)
-\demi\sum_{n=1}^{L-1}\lambda_n(c^+_n-c_n)(c^+_{n+1}
+c_{n+1})+\demi\lambda_L(c^+_L-c_L)(c^+_1+c_1)
{\cal P}\eqno(B6)
$$
where ${\cal P}=(-1)^L\prod_n\sigma_n^z=\exp\left(\i\pi
\sum_nc_n^+c_n\right)$. This operator with eigenvalues
$p=\pm 1$ commutes with ${\cal H}$ and distinguishes
subspaces with even and odd fermion number. In~each subspace
 ${\cal H}$ is diagonalized by a Bogoljubov transformation
[80, 173]. In terms of new operators $\alpha_q$,
$\alpha_q^+$ it takes the form
$$
{\cal H}=\sum_q\varepsilon_q\left(\alpha_q^+\alpha_q-
\demi\right)\eqno(B7)
$$
The single-fermion eigenvalues $\varepsilon_q$ follow
from a $L\times L$ matrix equation which reads, in the
notation of reference [80]
$$
(\bss{A}-\bss{B})(\bss{A}+\bss{B}){\bf\Phi}_q=
\varepsilon_q^2{\bf\Phi}_q\eqno(B8)
$$
with the matrix given by
$$
\pmatrix
{h_1^2+\lambda_L^2&h_1\lambda_1&{}&{}&-p\ h_L
\lambda_L\cr h_1\lambda_1&h_2^2+\lambda_1^2&h_2
\lambda_2&{}&{}\cr {}&\ddots&\ddots&\ddots&{}\cr
{}&{}&h_{L-2}\lambda_{L-2}&h_{L-1}^2+\lambda_{L-2}^2
&h_{L-1}\lambda_{L-1}\cr -p\ h_L\lambda_L&{}&{}&h_{L-1}
\lambda_{L-1}&h_L^2+\lambda_{L-1}^2\cr}\eqno(B9)
$$
For a homogeneous system the equations are solved by
Fourier transformation. The fermion states are running waves
with momenta $q$ given by $q=2n\pi/L$ $(p=-1)$ or
$q=(2n+1)\pi/L$ $(p=+1)$ and integers $n$ such that $\mid
q\mid\leq\pi$. Putting $h_n=1$, $\lambda_n=\lambda$, the
eigenvalues are
$$ 
\varepsilon_q=(1+\lambda^2-2\lambda\cos q)^{1/2}.\eqno(B10) 
$$
This expression is the Hamiltonian limit of Onsager's general
result [174]
$$
\cosh\omega_q=\cosh 2K_1^*\cosh 2K_2-\sinh 2K_1^*\sinh
2K_2\cos q\eqno(B11)
$$
where $\omega_q=2K_1^*\varepsilon_q$.

If one multiplies a large number of ${\cal T}$'s, their
largest eigenvalues are most important. They are given by the
smallest ones of ${\cal H}$. The two lowest states of ${\cal
H}$ are the ground states in the two subspaces $p=\pm 1$. If
$\lambda>1$, they become degenerate for $L\rightarrow\infty$
and this leads to the appearance of long-range order in the
system.

If the system has free ends, no distinction occurs between
$p\!=\!\pm 1$. The solutions are standing waves with $q$-values
determined by the boundary conditions. For $\lambda\!>\! 1$ one
finds an additional solution with imaginary wave number,
localized near one surface. The corresponding eigenvalue
vanishes exponentially as $L\!\rightarrow\!\infty$ so that this
state is again related to the long-range order. In
particular, it determines the surface magnetization 
(cf.~Section~5.2).

At the critical (self-dual) point $h=\lambda=1$ the
eigenvalues are 
$$
\varepsilon_q=2\mid\sin{q\over 2}\mid.\eqno(B12)
$$
The $q$-values for the homogeneous case are given above,
while for free ends $q\!=\!(2n+1)\pi/(2L+1)$ and $0\!\leq
\! q<\!\pi$. The low-lying part of the spectrum has a linear
energy-momentum relation, $\varepsilon_q\!=\! q$, and the
towers of equidistant eigenvalues are in complete agreement
with the conformal predictions (Appendix A). The left- and
right-moving particles in the homogeneous system correspond to
the two Virasoro algebras present in this case.
 
The layered systems treated in Sections 5-7 are special
cases of the problem formulated in (B3) or (B8--9).
 
For periodic boundary conditions the existence of two
subspaces leads to considerable complications if an operator
like $\sigma^x$ connects the two. Thus a spin correlation
function in the direction of the transfer has the form
$$
<\sigma_l^x\sigma_0^x>=\sum_m\mid <m,-p\mid\sigma^x
\mid 0,p>\mid^2\exp\left[ -l\left( E_m(-p)-E_0(p)\right)
\right]\eqno(B13)
$$
The non-universal behaviour of this function for the case
of a line defect (Section~6) can be related to its
particular structure involving both subspaces~[119].
\subappendice{B.}{Corner transfer matrix}
This quantity is the partition function for a
whole angular segment of a lattice. For the case shown in
Figure B1b it is a $90^\circ$ segment. The inner variables
are supposed to be summed over. The partition function of
the total system is then given by the trace over the product
of four such $90^\circ$ matrices. If the system is invariant
under $90^\circ$ rotations, all matrices are identical. Otherwise 
one has two different types. 
The importance of these
matrices is connected with their use in calculating order
parameters. For this the 
variables along the outer boundary are fixed as shown and the
order at site $0$ is considered in the thermodynamic limit.
That this is a practical and quite efficient procedure was
noted by Baxter who introduced this type of transfer
matrix and studied it for various solvable models [175, 176,
6, 177].

{}From Baxter's results it follows that the operator ${\cal H}$
related to the corner transfer matrix of a homogeneous Ising
model has again the form (B3). In the Hamiltonian limit this
can be seen directly. The coefficients $h_n$, $\lambda_n$
then follow from the geometry and are proportional to the
number of vertical and horizontal bonds at distance $n$ from
the tip, respectively
$$
h_n=2n\qquad\lambda_n=\lambda (2n+1)\qquad n=0,1,\cdots
(L-1)\eqno(B14)
$$
In addition $h_L=0$ due to the fixed boundary spins. The
resulting eigenvalue problem (B8--9) can be solved via
generating functions and special polynomials [178, 179].
Due to the boundary conditions, one eigenvalue is zero.
The other low-lying eigenvalues in the ordered phase ($\lambda>1$)
are, with $q$ replaced by $l$
$$
\varepsilon_l=(2l-1)\varepsilon\qquad l=1,2,\cdots\eqno(B15)
$$
where $\varepsilon =\pi\lambda /2${\bf K}$(1/\lambda )$ and
{\bf K} is the complete elliptic integral of the first kind.
This result also follows by taking the Hamiltonian limit in the
general formula	 for the eigenvalues
$\omega_l=2K_1^*\varepsilon_l$ of the infinite system [6]
$$
\omega_l=(2l-1){\pi u\over 2{\bf K}(k)}\eqno(B16)
$$
where $k=(\sinh 2K_1\sinh 2K_2)^{-1}$ and $u$, measuring the
anisotropy, is defined via the elliptic function $\sn$ by
$\sinh 2K_2=-\i\sn(\i u,k)$. The equidistance of the levels
$\varepsilon_l$, $\omega_l$ which directly reflects the
geometry, is a very remarkable property. Such a level structure
is also found for the corner transfer matrices of other
integrable models. {}From (B16) it follows that near criticality
the $\omega_l$ scale as $1/\ln(1/t)$.Thus the spectrum
collapses at the critical point, but
(B15) still holds for a large finite system.
Then $\varepsilon\cong\pi /\ln L$ in agreement with conformal
predictions [35, 180, 181].
 
For radially inhomogeneous systems as in Section 8.2, the
coefficients $h_n$, $\lambda_n$ are modified compared to
(B14). The same holds if the shape is different from a
wedge, as in Section 4.3.
 
The magnetization $<\!\sigma_0\!>$ in the centre of a system built
from $m$ segments and closed
upon itself, can be written as
$$
<\sigma_0>={{Z_{+}-Z_{-}}\over{Z_{+}+Z_{-}}}\eqno(B17)
$$
where $Z_{+}$ and $Z_{-}$ are the partition functions with 
$\sigma_0$ parallel and antiparallel to the boundary spins, 
respectively. 
In terms of the corner transfer matrix ${\cal T}$
of one segment this becomes a quotient of traces
$$
<\sigma_0>={\Tr\ (\sigma_0^x\sigma_N^x{\cal T}^m)\over\Tr\ 
({\cal T}^m)}\eqno(B18)
$$
The operator $\sigma_0^x\sigma_N^x$ measures the relative 
orientation of the spin at 0 and the outer spins. It can be
expressed in terms of the Fermi operators which diagonalize
${\cal T}$ as
$\exp(\i\pi\sum_l\alpha_l^+\alpha_l)$ where $l\geq 1$.
Thus it distinguishes states with even or odd number of
excited fermions. Inserting (B1) and the diagonal form of $H$, 
the traces can be performed independently over the
single-fermion states and one obtains the magnetization
in the form of a product
$$
<\sigma_0>=\prod_l\tanh\left({m\omega_l\over 2}\right)
\eqno(B19)
$$
This is a central result of the corner transfer matrix
approach.

One may note that 
the formula (B19) is identical to the one for the
corresponding end magnetization of an inhomogeneous Ising chain
with couplings ${m\omega_l/2}$. 
Onsager's result [182] for the spontaneous bulk
magnetization $<\sigma_0>=(1-k^2)^{1/8}$
follows directly from it by setting $m=4$ and using 
(B16) together with
elliptic function identities. As mentioned in Section 3.2, the
situation is more complicated for systems having
free edges [34].
\subappendice{B.}{Rescaling}
{\par\begingroup\parindent=0pt\leftskip=1cm
\rightskip=1cm\parindent=0pt
\baselineskip=12truept\sips{12.4truecm}{8.9truecm}
\midinsert
\centerline{\epsfbox{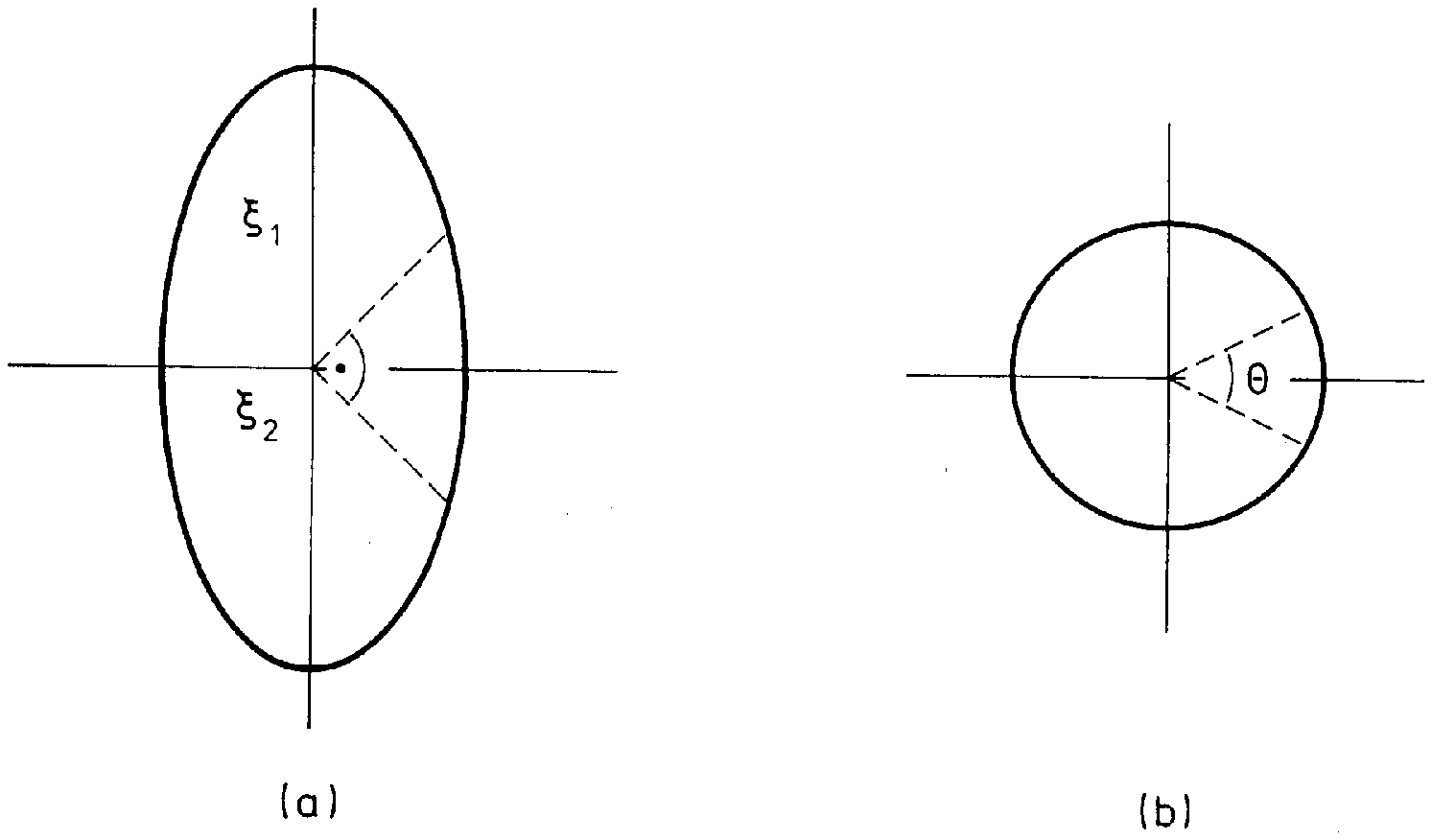}}
\bigskip
\Figcaption{B2}{\srm Rescaling of a system with different
correlation lengths (a) into an isotropic one (b). The
change of an angle is also indicated.}
\par
\endinsert
\endgroup
\par}
The results of the Hamiltonian limit may also be used to
discuss isotropic systems. The problem becomes effectively
isotropic if the correlation lengths in the two directions
are rescaled as shown in Figure B2. For this, one introduces
lattice constants $a_i$ such that $a_1\xi_1=a_2\xi_2$
[26, 31]. For the square lattice near criticality one has
$\xi_2/\xi_1=\cosh 2K_2/\cosh 2K_1\cong 2K_1^*\ll 1$.
Thus $a_2=1$ gives $a_1=2K_1^*$. The operator ${\cal H}$
for row transfer then corresponds to a unit step in the
isotropic system. For the wedge shown in the figure, the
original opening angle of $90^\circ$ is reduced to $\theta
=4K_1^*$. Therefore the operator ${\cal H}$ for corner
transfer corresponds to a step of twice the unit angle in the
isotropic case. There is no change in the angle if the wedge
is bounded by the principal axes.
\vfill\eject
\appendice{C}{Perturbation theory for extended defects}
A perturbation approach to the extended defect critical
behaviour has been proposed by Bariev [141]. It provides
a criterion for marginal behaviour and, when the system
is marginal, allows a determination of local first-order
corrections to the scaling dimensions of either the order
parameter $\sigma (\bi r)$ or the energy density
$\varepsilon (\bi r)$. These corrections are obtained
through a calculation of the two-point correlation functions
to logarithmic accuracy, making use of the operator product
expansion (OPE) [183, 184]. This is an extension to
inhomogeneous sytems of techniques introduced by Kadanoff
and Wegner [185] in statistical physics and Polyakov [186]
in quantum field theory. In the following the method is
presented for a temperature-like extended perturbation and
the correction to the order-parameter correlation function
is determined.
\subappendice{C.}{General}
One considers, in the continuum limit, the following
perturbed $d$-dimensional system
$$
-\beta {\cal H}=-\beta {\cal H}_0+g\int {\cal Z}(\bi r)
\varepsilon (\bi r)\d\bi r\eqno(C1)
$$
where ${\cal H}_0$ is the unperturbed Hamiltonian and $g$
the perturbation amplitude. The shape function ${\cal Z}
(\bi r)$, which gives the form of the inhomogeneity, is
assumed to be homogeneous in $r$ so that it may be
generally written as
$$
{\cal Z}(\bi r)={f(\bi u)\over r^\omega}\eqno(C2)
$$
where $\bi u$ is a unit vector along $\bi r$.
The numerator contains the angular dependence of the
perturbation which decays from the core of the defect
with dimension $d^*$.	For a point source $f(\bi u)=1$
whereas for a line or a plane $f(\bi u)=\mid\cos\theta
\mid^{-\omega}$ where $\theta$ is the angle measured
from a polar axis which is orthogonal to the core subspace.
 
At the critical point of the unperturbed system the OPE is
used, in multipoint correlation functions, to expand the
product of two local operators at ${\bi r}_1$ and ${\bi
r}_2$ on the complete set of operators associated
with the fixed point of ${\cal H}_0$. The distance
$r_{12}$ has to be smaller than the distances for other
pairs appearing in the correlation function. Only the most
singular term in the expansion is retained. The following
reduction equations are needed:
$$
\sigma ({\bi r}_1)\varepsilon ({\bi r}_2)\simeq a\sigma
({\bi r}_1)r_{12}^{-x}\qquad\varepsilon ({\bi r}_1)
\varepsilon ({\bi r}_2)\simeq b\varepsilon
({\bi r}_1)r_{12}^{-x}\eqno(C3)
$$
They are written for a translationally invariant system, $x$
is the bulk scaling dimension of the energy density and
averages are supposed to be subtracted.
The powers in (C3) follow from scaling.
 
In a semi-infinite system translational
invariance is lost in the direction perpendicular to the
surface. When ${\bi r}_1$ belongs to the surface, the
structure constants $a$ and $b$ in (C3) only acquire an
angular dependence. When ${\bi r}_2$ goes to the surface,
i.e. when $\theta\simeq\pi/2-r_{12}^{-1}$, the surface
scaling dimension $d$ (Section 2.3), has to replace the
bulk one in (C3). Then, as for surface scaling functions
[18], one expects quite generally $a(\theta )\sim b
(\theta)\sim (\cos\theta )^{d-x}$.
 
For the $2d$ Ising model with $x=1$, one indeed obtains
[187] 
$$
a(\theta )={2\over\pi}\cos\theta\qquad b(\theta )={8\over\pi}
\cos\theta\sin^2\theta\eqno(C4)
$$
The bulk values are $a=1/2\pi$ and $b=0$. The last one
vanishes due to the dual symmetry ($\varepsilon\rightarrow
-\varepsilon$) of the bulk Ising model in two dimensions
[156]. 
\subappendice{C.}{Relevance-irrelevance criterion}
The order parameter correlation function has the following
expansion in $g$
$$
G(\bi R, g)=\sum_{n=0}^\infty {g^n\over n!}\int_{{\cal E}'}
\ll\sigma (0)\sigma (\bi R)\varepsilon (\bi r_1)\cdots
\varepsilon (\bi r_n)\gg\prod_{i=1}^n{\cal Z}(\bi r_i)
\d\bi r_i\eqno(C5)
$$
where the double brackets denote the irreducible part of
the multi-point correlation function. The integral
extends over the subspace ${\cal E}'$ with dimension $d'$
where the perturbation is nonvanishing. $\bi R$ is assumed
to be orthogonal to the core subspace.
 
The OPE allows an evaluation of the first-order correction
in (C5) giving
$$
\delta G^{(1)}\sim g G(\bi R,0)R^{d_{eff}-x}\eqno(C6)
$$
where $x$ is the bulk scaling dimension of the energy
density and, for a bulk perturbation, the effective
dimension of the defect is defined as
$$
d_{eff}=\max\  [d'-\omega,d^*]\eqno(C7)
$$
The second alternative, $d_{eff}=d^*$, is met when the
angular integral in (C5) becomes divergent at $\theta=
\pi/2$ due to the singularity of ${\cal Z}(\bi r)$ at the
core of the defect. This is cured by introducing a
cut-off at $\theta=\pi/2-r^{-1}$ which modifies the
$R$-dependence in (C6) [141].
 
When $d_{eff}<x$, the leading correction to the long-distance
behaviour is small and the perturbation is irrelevant. In
the opposite case, the correction factor in (C6) grows as
a power of $R$, indicating a relevant perturbation and the
expansion becomes useless. Finally, when $d_{eff}=x$,
the perturbation is marginal and logarithmic corrections
leading to $g$-dependent local exponents are obtained. As
to the value of $d_{eff}$, Table~C1 lists it
for the systems studied in this review.
{\medskip\par\begingroup\parindent=0pt\leftskip=1cm
\rightskip=1cm\parindent=0pt
\midinsert
\hfil\boldrule{.8\hsbody}\hfil\par
\centerline{\Boite{.36\hsbody}{\rm defect type}
\boite{$d^*$}\boite{$d'$}\boite{$d$}\Boite{.16\hsbody}{$d_{eff}$}}
\hfil\medrule{.8\hsbody}\hfil\par
\centerline{\Boite{.36\hsbody}{\rm narrow line defect $(\omega=0)$}
\boite{$1$}\boite{$1$}\boite{$2$}\Boite{.16\hsbody}{$1$}}
\centerline{\Boite{.36\hsbody}{\rm surface extended line defect}
\boite{$1$}\boite{$2$}\boite{$2$}\Boite{.16\hsbody}{$2-\omega$}}
\centerline{\Boite{.36\hsbody}{\rm bulk extended line defect}
\boite{$1$}\boite{$2$}\boite{$2$}\Boite{.16\hsbody}{$\max\  [2-\omega,1]$}}
\centerline{\Boite{.36\hsbody}{\rm radial defect}
\boite{$0$}\boite{$2$}\boite{$2$}\Boite{.16\hsbody}{$2-\omega$}}
\centerline{\Boite{.36\hsbody}{\rm surface field}
\boite{$0$}\boite{$1$}\boite{$2$}\Boite{.16\hsbody}{$1-\omega$}}
\hfil\boldrule{.8\hsbody}\hfil
\baselineskip=12truept
\tabcaption{C1}{\srm Values of the effective dimension
of the defects, $\sst d_{eff}$, for the different systems studied
in this review.}
\endinsert
\endgroup
\par}
 
For slow decay ($\omega$ small), $d_{eff}=d'-\omega$, the
perturbation is long-ranged and marginal behaviour sets
in when $\omega=d'-x$. This is an extension to $d'<d$
of the condition obtained in (5.2) via scaling arguments.
For large enough $\omega$, $d_{eff}=d^*$, the perturbation
is effectively short-ranged  and one recovers the marginality
criterion~(6.5). 
Similar conclusions can be drawn for an order parameter
perturbation as in Section~5.4.
 
With a free surface, the angular integral is modified
through the angular dependence of the structure constants
in (C4). As a result, $d^*$ in (C7) has to be replaced by
$d^*-d+x$ and marginal behaviour can no longer be induced
by an effectively short-range defect since $d^*<d$.
\subappendice{C.}{Marginal behaviour}
In the following, one assumes a bulk temperature-like
perturbation and a defect with $d_{eff}=d'-\omega=x>d^*$.
The $n$-th order term in (C5) can be rewritten so that
the integration extends over $r_1<r_2\cdots<r_n$.
The leading contribution to the multi-point function comes
from pairs of points which are close to each other.
Assuming that the shortest distance is between ${\bi r}_1$
and the origin and using the first reduction
relation in (C3), the integral over ${\bi r}_1$ gives
$$
a\sigma (0)\int_1^{r_2}\d r_1r_1^{d'-1-\omega -x}\int f(\bi
u_1)\d\bi u_1\eqno(C8)
$$
where we wrote $\d\bi r_1=r_1^{d'-1}\d r_1\d\bi u_1$ for
the volume element. Since $d_{eff}>d^*$, the angular
integral is regular and contributes a
geometrical factor $S_{d'd^*}(\omega )$ leading to
$$
\hskip-3mm\delta G^{(n)}= g^naS_{d'd^*}(\omega )\int_{{\cal E}'
(r_2<r_3\cdots<r_n)}\ln r_2\ll\sigma (0)
\sigma (\bi R)\varepsilon (\bi r_2)\cdots\varepsilon
(\bi r_n)\gg\prod_{i=2}^n{\cal Z}(\bi r_i)\d\bi r_i
\eqno(C9)
$$
The same process can be iterated $n$ times giving
$$
\delta G^{(n)}={1\over n!}\left[gaS_{d'd^*}(\omega )\ln
R\right]^n\ll\sigma (0)\sigma (\bi R)\gg\eqno(C10)
$$
It may be shown [141] that when other pairs like
$\varepsilon (\bi r_1)\varepsilon (\bi r_2)$ are first
contracted, the result is logarithmically smaller. Thus,
to the leading logarithmic order, the perturbed correlation
function is given by
$$
G(\bi R,g)=G(\bi R,0)
R^{ gaS_{d'd^*}(\omega )}\eqno(C11)
$$
{}From it, one deduces the first-order local change of the
bulk order parameter scaling dimension $x$,
$$
x_l=x-gaS_{d'd^*}(\omega)+O(g^2)\eqno(C12)
$$
In the same way, for the local scaling dimension of the
energy density, one obtains
$$
x_l=x-gbS_{d'd^*}(\omega )+O(g^2)\eqno(C13)
$$
where $x$ is the unperturbed bulk value [141].
 
These results are valid provided $\mid\! g\!\mid\ll 1$ and
$\mid\! g\!\mid\ln^2R\gg 1$ and can be used in the scaling
region. When the two alternative conditions for marginal
behaviour are simultaneously fulfilled, i.e. when $d'-\omega=
d^*=x$, the perturbation results are only valid in a
crossover regime with $\mid g\mid\ln R\ll 1$ and $\mid
g\mid\ln^3R\gg 1$ [187]. Then, due to the onset of local
order at the bulk critical point when $g>0$, the local
dimensions are singular in $g$ at $g=0$ (see Section~7).
\si{\vfill\eject}{}
{\medskip\par\begingroup\parindent=0pt\leftskip=1cm
\rightskip=1cm\parindent=0pt
\midinsert
\hfil\boldrule{0.7\hsbody}\hfil\par
\centerline{
\Boite{0.36\hsbody}{}
\Boite{0.17\hsbody}{\rm point source}
\Boite{0.17\hsbody}{\rm line source}}
\hfil\medrule{0.7\hsbody}\hfil\par
\centerline{
\Boite{0.36\hsbody}{\rm bulk order parameter}
\Boite{0.17\hsbody}{${1\over 8}-g$}
\Boite{0.17\hsbody}{\medrule{0.05\hsbody}}}
\centerline{
\Boite{0.36\hsbody}{\rm surface order parameter}
\Boite{0.17\hsbody}{$\demi-{4\over\pi}g$}
\Boite{0.17\hsbody}{$\demi-2g$}}
\centerline{
\Boite{0.36\hsbody}{\rm surface energy density}
\Boite{0.17\hsbody}{$2-{16\over 3\pi}g$}
\Boite{0.17\hsbody}{$2-4g$}}
\hfil\boldrule{0.7\hsbody}\hfil\par
\baselineskip=12truept
\tabcaption{C2}{\srm Local scaling dimensions for extended,
temperature-like,
marginal defects in the $\sst 2d$ Ising model, up to first order
in the perturbation
amplitude $\sst g$.}
\endinsert
\endgroup
\par}
 
In the semi-infinite geometry, with $d_{eff}=d'-\omega=x$,
the calculation proceeds as in the bulk. The only change
is introduced by the angular dependence of the structure
constants [187]. Explicit results for marginal, bulk and
surface extended perturbations in the two-dimensional
Ising model are given in Table~C2.
\references
\numrefbk{[1]}{see for example 1976}{Phase Transitions and
Critical Phenomena}{vol 6 ed. C~Domb and M S Green (London:
Academic)}
\numrefjl{[2]}{McCoy B M and Wu T T 1967}{\PR}{162}{436}
\numrefbk{[3]}{McCoy B M and Wu T T 1973}{The
Two-Dimensional Ising Model}{(Cambridge: Harvard University
Press)} 
\numrefbk{[4]}{Binder K 1983}{Phase Transitions and Critical
Phenomena}{vol 8 ed. C Domb and J L Lebowitz (London:
Academic) p 1}
\numrefbk{[5]}{Diehl H W 1986}{Phase Transitions and Critical
Phenomena}{vol 10 ed. C Domb and J L Lebowitz (London:
Academic) p 75}
\numrefbk{[6]}{Baxter R J 1982}{Exactly Solved Models in
Statistical Mechanics}{(London: Academic)}
\numrefbk{[7]}{Stinchcombe R B 1983}{Phase Transitions and
Critical Phenomena}{vol 7 ed. C~Domb and J L Lebowitz (London:
Academic) p 151}
\numrefjl{[8]}{Wang J S, Selke W, Dotsenko V and Andreichenko
V B 1990}{Physica}{164}{221}
\numrefbk{[9]}{Abraham D B 1986}{Phase Transitions and Critical
Phenomena}{vol 10 ed. C~Domb and J L Lebowitz (London:
Academic) p 1}
\numrefbk{[10]}{Jasnow D 1986}{Phase Transitions and Critical
Phenomena}{vol 10 ed. C Domb and J L Lebowitz (London:
Academic) p 269}
\numrefbk{[11]}{Forg\'acs G, Lipowsky R and Nieuwenhuizen
Th M 1991}{Phase Transitions and Critical Phenomena}{vol 14
ed. C Domb and J L Lebowitz (London: Academic) p 135}
\numrefjl{[12]}{Hoever P, Wolff W F and Zittartz J 1981}
{Z. Phys. B}{41}{43}
\numrefjl{[13]}{\dash\  1981}{Z. Phys. B}{42}{259}
\numrefjl{[14]}{\dash\  1981}{Z. Phys. B}{44}{129}
\numrefjl{[15]}{Bray A J and Moore M A 1977}{\JPA}{10}{1927}
\numrefbk{[16]}{Cardy J L 1987}{Phase Transitions and Critical
Phenomena}{vol 11 ed. C Domb and J L Lebowitz (London:
Academic) p 55}
\numrefjl{[17]}{Stella A L and Vanderzande C 1990}
{Int. J. Mod. Phys. B}{4}{1437}
\numrefjl{[18]}{Cardy J L 1984}{Nucl. Phys. B}{240}{[FS12] 514}
\numrefjl{[19]}{Burkhardt T W and Guim I 1991}{\JPA}{24}{1557}
\numrefjl{[20]}{Burkhardt T W, Eisenriegler E and Guim I 
1989}{Nucl. Phys. B}{316}{559}
\numrefjl{[21]}{Privman V 1985}{Phys. Rev. B}{32}{6089}
\numrefjl{[22]}{Burkhardt T W and Eisenriegler E 
1985}{\JPA}{18}{L83}
\numrefjl{[23]}{Fisher M E and de Gennes P G 
1978}{C. R. Acad. Sci., Paris B}{287}{207}
\numrefjl{[24]}{Burkhardt T W and Cardy J L 1987}{\JPA}{20}{L233} 
\numrefjl{[25]}{Dietrich S and Diehl HW 1981}{Z. Phys.
B}{43}{315}  
\numrefjl{[26]}{Cardy J L 1983}{\JPA}{16}{3617} 
\numrefjl{[27]}{Duplantier B 1991}{\PRL}{66}{1555} 
\numrefbk{[28]}{Carslaw H S and Jaeger J C 1980}{Heat 
Conduction in Solids}{(Oxford: Oxford University Press) chap. XIV} 
\numrefbk{[29]}{Smythe W R 1939}{Static and
Dynamic Electricity}{(New-York: McGraw-Hill) sect. 5.25}
\numrefbk{[30]}{Jackson J D 1975}{Classical Electrodynamics}{(New-York:
Wiley) sect. 3.4}
\numrefjl{[31]}{Barber M N, Peschel I and Pearce P A
1984}{J. Stat. Phys.}{37}{497}
\numrefjl{[32]}{Peschel I 1985}{Phys. Lett.}{110A}{313}
\numrefjl{[33]}{Kaiser C and Peschel I 1989}{J. Stat. Phys.}{54}{567}
\numrefjl{[34]}{Davies B and Peschel I 1991}{\JPA}{24}{1293}
\numrefjl{[35]}{Peschel I and Truong T T 1987}{Z. Phys. B}{69}{385}
\numrefbk{[36]}{de Gennes P-G 1979}{Scaling Concepts in Polymer
Physics}{(Ithaca and London: Cornell University Press) chap. X}
\numrefjl{[37]}{Guttmann A J and Torrie G M 1984}{\JPA}{17}{3539}
\numrefjl{[38]}{Cardy J L and Redner S 1984}{\JPA}{17}{L933}
\numrefjl{[39]}{Vanderzande C 1990}{\JPA}{23}{563}
\numrefjl{[40]}{Duplantier B and Saleur H 1986}{\PRL}{57}{3179}
\numrefjl{[41]}{Considine D and Redner S 1989}{\JPA}{22}{1621}
\numrefjl{[42]}{Burkhardt T W and Guim I 1987}{Phys.
Rev. B}{36}{2080}
\numrefjl{[43]}{Burkhardt T W and Xue T 1991}{\PRL}{66}{895}
\numrefjl{[44]}{\dash\  1991}{Nucl. Phys. B}{354}{653}
\numrefjl{[45]}{Cardy J L and Peschel I 1988}{Nucl.
Phys. B}{300}{[FS22] 377}
\numrefjl{[46]}{Bl\"ote H W J, Cardy J L and
Nightingale M P 1986}{\PRL}{56}{742}
\numrefjl{[47]}{Affleck I 1986}{\PRL}{56}{746}
\numrefbk{[48]}{Cardy J L 1990}{Fields, Strings and Critical
Phenomena}{ed. E Brezin and J~Zinn-Justin (Amsterdam:
North-Holland) p 169}
\numrefjl{[49]}{Kac M 1966}{Amer. Math. Monthly}{73}{1}
\numrefjl{[50]}{McKean H P and Singer I M 1967}{J. Diff.
Geom.}{1}{43}
\numrefbk{[51]}{Baltes H P and Hilf E R 1976}{Spectra of Finite
Systems}{(Mannheim: Bibliogr. Institut) chap. VI}
\numrefjl{[52]}{Peschel I 1988}{unpublished}{}{}
\numrefjl{[53]}{Gelfand M P and Fisher M E 1990}{Physica A}{166}{1}
\numrefjl{[54]}{Kleban P and Vassileva I 1991}{\JPA}{24}{3407}
\numrefjl{[55]}{Privman V 1988}{Phys. Rev. B}{38}{9261}
\numrefjl{[56]}{Peschel I, Turban L and Igl\'oi F
1991}{\JPA}{24}{L1229}
\numrefjl{[57]}{Blawid S 1993}{Diplomarbeit,
Freie Universit\"at Berlin}{}{}
\numrefbk{[58]}{Morse P M and Feshbach H 1953}{Methods of
Theoretical Physics}{(New-York: McGraw-Hill) chap. X}
\numrefjl{[59]}{van Beijeren H 1977}{Phys. Rev. Lett.}{38}{933}
\numrefjl{[60]}{Davies B and Peschel I 1992}{Ann. Physik}{2}{79}
\numrefjl{[61]}{Binder K and Wang J S 1989}{J. Stat.
Phys.}{55}{87}
\numrefjl{[62]}{Hornreich R M, Luban M and Shtrikman S
1975}{\PRL}{35}{1678}
\numrefbk{[63]}{Privman V and \v Svraki\^c N M 1989}{Directed
Models of Polymers, Interfaces and Clusters}{Lecture Notes
in Physics vol 338 (Berlin: Springer)}
\numrefjl{[64]}{Turban L 1992}{\JPA}{25}{L127}
\numrefjl{[65]}{Igl\'oi F 1992}{Phys. Rev. A}{45}{7024}
\numrefjl{[66]}{Turban L and Berche B 1993}{J. Phys.
I France}{3}{925}
\numrefjl{[67]}{Burkhardt T W 1982}{Phys. Rev. Lett}{48}{216}
\numrefjl{[68]}{Cordery R 1982}{Phys. Rev. Lett}{48}{215}
\numrefjl{[69]}{Hilhorst H J and van Leeuwen J M J 1981}{Phys.
Rev. Lett.}{47}{1188}
\numrefbk{[70]}{Syozi I 1972}{Phase Transitions and Critical
Phenomena}{vol 1 ed. C. Domb and M.S. Green (London:
Academic) p 269}
\numrefjl{[71]}{Burkhardt T W and Guim I 1984}{Phys.
Rev. B}{29}{508}
\numrefjl{[72]}{Burkhardt T W, Guim I, Hilhorst H J and
van Leeuwen J M J 1984}{Phys. Rev. B}{30}{1486}
\numrefjl{[73]}{Bl\"ote H W J and Hilhorst H J 1983}{Phys.
Rev. Lett.}{51}{20}
\numrefjl{[74]}{\dash\  1985}{\JPA}{18}{3039}
\numrefjl{[75]}{Luther A and Peschel I 1975}{Phys. Rev.
B}{12}{3908}
\numrefjl{[76]}{Igl\'oi F and Turban L 1993}{Phys. Rev.
B}{47}{3404}
\numrefbk{[77]}{Fisher M E 1974}{Renormalization Group in
Critical Phenomena and Quantum Field Theory}{Proceedings
of a Conference ed. J D Gunton and M S Green
(Philadelphia: Temple University Press) p 65 }
\numrefjl{[78]}{Abraham DB 1971}{Stud. Appl. Math.}{50}{71}
\numrefjl{[79]}{Peschel I 1984}{Phys. Rev. B}{30}{6783}
\numrefjl{[80]}{Lieb E H, Schultz T D and Mattis D C
1961}{Ann. Phys. (N. Y.)}{16}{406}
\numrefbk{[81]}{Abramowitz M and Stegun I A 1965}{Handbook
of Mathematical Functions}{(New York: Dover)}
\numrefjl{[82]}{Turban L and Berche B 1993}{\JPA}{26}{3131}
\numrefjl{[83]}{Burkhardt T W and Igl\'oi F
1990}{\JPA}{23}{L633}
\numrefjl{[84]}{Burkhardt T W and Guim I 1985}{\JPA}{18}{L25}
\numrefjl{[85]}{Igl\'oi F 1990}{Phys. Rev. Lett.}{64}{3035}
\numrefjl{[86]}{Berche B and Turban L 1990}{\JPA}{23}{3029}
\numrefjl{[87]}{Choi J-Y 1993}{\JPA}{26}{L327}
\numrefjl{[88]}{Burkhardt T W and Guim I 1982}{\JPA}{15}{L305}
\numrefjl{[89]}{Igl\'oi F 1992}{Europhys. Lett.}{19}{305}
\numrefjl{[90]}{Bariev R Z and Peschel I 1991}{Phys.
Lett.}{153A}{166}
\numrefjl{[91]}{Luck J M 1993}{J. Stat. Phys.}{72}{417}
\numrefjl{[92]}{Igl\'oi F 1993}{\JPA}{26}{L703}
\numrefjl{[93]}{Harris A B 1974}{\JPC}{7}{1671}
\numrefjl{[94]}{Turban L and Berche B 1993}{Z. Phys. B}{92}{307}
\numrefjl{[95]}{Hucht A 1992}{Physica}{A183}{223}
\numrefjl{[96]}{Wildpaner V, Rauch H and Binder K
1973}{J. Phys. Chem. Solids}{34}{925}
\numrefjl{[97]}{Binder K, Stauffer D and Wildpaner V
1975}{Acta Met.}{23}{119}
\numrefjl{[98]}{Burkhardt T W and Eisenriegler E
1981}{Phys. Rev. B}{24}{1236}
\numrefjl{[99]}{Eisenriegler E and Burkhardt T W
1982}{Phys. Rev. B}{25}{3283}
\numrefjl{[100]}{Diehl H W, Dietrich S and Eisenriegler E
1983}{Phys. Rev. B}{27}{2937}
\numrefjl{[101]}{Abe R 1981}{Prog. Theor. Phys.}{65}{1237}
\numrefjl{[102]}{\dash\  1981}{Prog. Theor. Phys.}{65}{1835}
\numrefjl{[103]}{Yamamoto K and Abe R 1981}{Prog. Theor.
Phys.}{66}{1947}
\numrefjl{[104]}{Abe R and Yamamoto K 1982}{Prog. Theor.
Phys.}{67}{139}
\numrefjl{[105]}{Abe R, Yamamoto K and Ideura K
1983}{Prog. Theor. Phys.}{69}{464}
\numrefjl{[106]}{Ideura K and Abe R 1984}{Prog.
Theor. Phys.}{71}{27}
\numrefjl{[107]}{Ideura K 1984}{Prog. Theor. Phys.}{71}{474}
\numrefjl{[108]}{Bariev R Z 1979}{Zh. Eksp. Teor.
Fiz.}{77}{1217}
\qquad [{\frenchspacing\sl Sov. Phys. JETP
\bf 50} 613 (1979)]\par
\numrefjl{[109]}{McCoy B M and Perk J H H 1980}{\PRL}{44}{840}
\numrefjl{[110]}{Fisher M E and Ferdinand A E
1967}{\PRL}{19}{169}
\numrefjl{[111]}{Ko L-F, Au-Yang H and Perk J H H
1985}{\PRL}{54}{1091}
\numrefjl{[112]}{Kadanoff L P 1981}{Phys. Rev. B}{24}{5382}
\numrefjl{[113]}{Brown A C 1982}{Phys. Rev. B}{25}{331}
\numrefjl{[114]}{Burkhardt T W and Choi J-Y 1992}{Nucl.
Phys. B}{376}{447}
\numrefjl{[115]}{Abraham D B,Ko L-F and \v Svraki\'c N M
1988}{\PRL}{61}{2393}
\numrefjl{[116]}{Uzelac K, Jullien R and Pfeuty P
1981}{\JPA}{14}{L17}
\numrefjl{[117]}{Turban L 1982}{\JPA}{15}{1733}
\numrefjl{[118]}{Nozi\`eres P and de Dominicis C T
1969}{Phys. Rev.}{178}{1097}
\numrefjl{[119]}{Peschel I and Schotte K D 1984}{Z. Phys.
B}{54}{305}
\numrefjl{[120]}{Cabrera G G and Jullien R 1986}{\PRL}{57}{393}
\numrefjl{[121]}{Zinn-Justin J 1986}{\PRL}{57}{3296}
\numrefjl{[122]}{Cabrera G G and Jullien R 1987}{Phys.
Rev. B}{36}{7062}
\numrefjl{[123]}{Barber M N and Cates M E 1987}{Phys.
Rev. B}{36}{2024}
\numrefjl{[124]}{Igl\'oi F 1989}{Phys. Rev. B}{40}{5187}
\numrefjl{[125]}{Nightingale M P and Bl\"ote H W J
1982}{\JPA}{15}{L33}
\numrefjl{[126]}{Kaufman M and Griffiths R B 1982}{Phys.
Rev. B}{26}{5282}
\numrefjl{[127]}{Turban L 1985}{\JPA}{18}{L325}
\numrefjl{[128]}{Guimar\~aes L G and Drugowich de Felicio
J R 1986}{\JPA}{19}{L341}
\numrefjl{[129]}{Henkel M and Patk\'os A 1987}{\JPA}{20}{2199}
\numrefjl{[130]}{Henkel M, Patk\'os A and Schlottmann M
1989}{Nucl. Phys. B}{314}{609}
\numrefjl{[131]}{Henkel M and Patk\'os A 1988}{\JPA}{21}{L231}
\numrefjl{[132]}{Irving A C, \'Odor G and Patk\'os A
1989}{\JPA}{22}{4665}
\numrefjl{[133]}{Henkel M and Patk\'os A 1987}{Nucl.
Phys. B}{285}{29}
\numrefjl{[134]}{Baake M, Chaselon P and Schlottmann M
1989}{Nucl. Phys. B}{314}{625}
\numrefbk{[135]}{Henkel M 1990}{Finite-Size Scaling and
Numerical Simulations of Statistical Systems}{ed.
V Privman (Singapore: World Scientific) chap. VIII}
\numrefjl{[136]}{Schlottmann M 1988}{Diplomarbeit
Bonn--IR--88--41}{}{}
\numrefjl{[137]}{Grimm U 1990}{Nucl. Phys. B}{340}{633}
\numrefjl{[138]}{Wittlich T 1990}{\JPA}{23}{3825}
\numrefjl{[139]}{Grimm U 1988}{Diplomarbeit
Bonn--IR--88--30}{}{}
\numrefjl{[140]}{Hinrichsen H 1990}{Nucl. Phys. B}{336}{377}
\numrefjl{[141]} {Bariev R Z 1988}{Zh. Eksp. Teor.
Fiz.}{94}{374}
\qquad [{\frenchspacing\sl Sov. Phys. JETP
\bf 67} 2170 (1988)]\par
\numrefjl{[142]} {\dash\  1989}{\JPA}{22}{L397}
\numrefjl{[143]} {Bariev R Z and Malov O A 1989}{Phys.
Lett.}{136A}{291}
\numrefjl{[144]} {Bariev R Z and Ilaldinov I Z
1989}{\JPA}{22}{L879}
\numrefjl{[145]} {Igl\'oi F, Berche B and Turban L
1990}{Phys. Rev. Lett.}{65}{1773}
\numrefbk{[146]}{Burkhardt T W 1984}{Phase Transitions
in Disordered Systems}{Lecture Notes in Physics vol
206 ed. A Pekalski and J Sznajd (Berlin: Springer)}
\numrefjl{[147]}{Peschel I and Wunderling R
1992}{Ann. Physik}{1}{125}
\numrefjl{[148]}{Peschel I and Truong T T
1991}{Ann. Physik}{48}{185}
\numrefjl{[149]}{Bariev R Z and Peschel I 1991}{\JPA}{24}{L87}
\numrefjl{[150]}{Turban L 1991}{Phys. Rev. B}{44}{7051}
\numrefjl{[151]}{Privman V and Fisher M E 1984}{Phys.
Rev. B}{30}{322}
\numrefjl{[152]}{Hamer C J and Barber M N 1981}{\JPA}{14}{241}
\numrefjl{[153]}{Henkel M 1987}{\JPA}{20}{995}
\numrefjl{[154]}{Burkhardt T W and Guim I 1987}{Phys.
Rev. B}{35}{1799}
\numrefjl{[155]}{Reinicke P 1987}{\JPA}{20}{4501}
\numrefjl{[156]}{Kramers H A and Wannier G 1941}{Phys.
Rev.}{60}{252}
\numrefjl{[157]}{Saleur H 1988}{\JPA}{22}{L41}
\numrefjl{[158]}{Saleur H and Bauer M 1989}{Nucl.
Phys. B}{320}{591}
\numrefbk{[159]} {Fisher M E 1971} {Critical
Phenomena}{ed. M S Green (London: Academic) p 1}
\numrefjl{[160]} {Polyakov A M 1970}{Zh. Eksp. Teor.
Fiz. Pis. Red.}{12}{538}
\qquad [{\frenchspacing\sl Sov. Phys. JETP Lett.
\bf 12} 381 (1970)]\par
\numrefjl{[161]}{Belavin A A, Polyakov A M and 
Zamolodchikov A B 1984}{J Stat Phys}{34}{763}
\numrefbk{[162]}{{\rm for a pedagogical introduction to 
the subject, see} Christe P and Henkel M 1993}{Introduction 
to Conformal Invariance and its Applications to 
Critical Phenomena}{Lecture Notes in Physics  vol m16 (Berlin:
Springer)} 
\numrefjl{[163]}{Friedan D, Qiu Z and Shenker S 1984}{Phys. Rev.
 Lett.}{52}{1575}
\numrefjl{[164]}{Cardy J L 1984}{\JPA}{ 17}{L385}
\numrefjl{[165]}{Montroll E W 1941}{J. Chem. Phys.}{9}{706}
\numrefjl{[166]}{Lassetre E N and Howe J P
1941}{J. Chem. Phys.}{9}{747}
\numrefjl{[167]}{Schultz T D, Mattis D C and Lieb E H
1964}{Rev. Mod. Phys.}{36}{856}
\numrefjl{[168]}{Kaufman B 1949}{Phys. Rev.}{76}{1232}
\numrefjl{[169]}{Fradkin E and Susskind L 1978}{Phys. Rev. D}
{17}{2637}
\numrefjl{[170]}{Kogut J B 1979}{Rev. Mod. Phys.}{51}{659}
\numrefjl{[171]}{Jordan P and Wigner E 1928}{Z.
Phys.}{47}{631}
\numrefjl{[172]}{Pfeuty P 1970}{Ann. Phys. (N. Y.)}{57}{79}
\numrefjl{[173]}{Katsura S 1962}{Phys. Rev.}{127}{1508}
\numrefjl{[174]}{Onsager L 1944}{\PR}{65}{117}
\numrefjl{[175]}{Baxter R J 1977}{J. Stat. Phys.}{17}{1}
\numrefjl{[176]}{Baxter R J 1981}{Physica}{106A}{18}
\numrefbk{[177]}{Baxter R J 1985}{Integrable Systems in
Statistical Mechanics}{VII ed. G M d'Ariano, A Monterosi and
M G Rasetti (Singapore: World Scientific)}
\numrefjl{[178]}{Davies B 1988}{Physica}{154A}{1}
\numrefjl{[179]}{Truong T T and Peschel I 1989}{Z. Phys.
B}{75}{119}
\numrefjl{[180]}{Peschel I 1988}{\JPA}{21}{L185}
\numrefjl{[181]}{Davies B and Pearce P A 1990}{\JPA}{23}{1295}
\numrefjl{[182]}{Onsager L 1949}{Nuovo Cim. (Suppl.)}{6}{201}
\numrefjl{[183]}{Polyakov A M 1969}{Zh. Eksp. Teor.
Fiz.}{57}{271}
\qquad [{\frenchspacing\sl Sov. Phys. JETP
\bf 30} 151 (1969)]\par
\numrefjl{[184]}{Kadanoff L P and Ceva H 1971}{Phys.
Rev. B}{3}{3918}
\numrefjl{[185]}{Kadanoff L P and Wegner F J
1971}{Phys. Rev. B}{4}{3989}
\numrefjl{[186]}{Polyakov A M 1972}{Zh. Eksp.
Teor. Fiz.}{63}{24}
\qquad [{\frenchspacing\sl Sov. Phys. JETP
\bf 36} 12 (1972)]\par
\numrefjl{[187]}{Bariev R Z and Turban L 1992}{Phys.
Rev. B}{45}{10761}
\vfill\eject
\bye